\documentclass[useAMS,usenatbib,times]{mn2e}
\usepackage{graphicx}
\usepackage{natbib}

\def\d{{\mathrm{d}}}  
\newcommand{\mathd}{\mathrm{d}}

\newcommand{\mathpi}{\pi}
\def\parn{\par\noindent}
\def\MG#1{{\mbox{\boldmath $ #1$}}} 
\def\M#1{{\mathbf #1}}

\def\R#1{{\mathrm{#1}}}

\def\Eq#1{{Eq.~\ref{e:#1}}}
\def\Ep#1{{~(\ref{e:#1})}}

\def\EQN#1{\label{e:#1}}        

\def\M#1{{\mathbf{#1}}}
\def\T#1{{{#1}^{\top}}}

\newcommand{\tmmathbf}[1]{\mathbf{#1}}

\def\bfr {{\bf r}}
\def\bfv {{\bf v}}
\def\bfk {{\bf k}}
\def\bfw {{\bf w}}
\def\bfI {{\bf I}}

\def\bfx {{\bf x}}

\def\R#1{{\mathrm{#1}}}         
\def\Eq#1{{Eq.~(\ref{e:#1})}}   
\def\Ep#1{{~(\ref{e:#1})}}      

\def\EQN#1{\label{e:#1}}        

\def\be{\begin{equation}}
\def\ee{\end{equation}}
\def\ba{\begin{eqnarray}}
\def\ea{\end{eqnarray}}

\def\M#1{{\mathbf{#1}}} 
\def\T#1{{{#1}^{\top}}}         
\def\d{{\R{d}}}         
\def\MG#1{{\mbox{\boldmath $ #1$}}} 
\def\tmmathbf#1{{\mathbf{#1}}} 

 \def\bv{{\bf v}}
      \def\br{{\bf r}}

 \def\bk{{\bf k}}

\def\bI{{\bf I}}
\def\bw{{\bf w}}
\def\bo{{\MG{\omega}}}
\def\bO{{\MG{\Omega}}}

\def\T#1{{{#1}^{\top}}}

\newcommand{\nicefrac}[2]{\leavevmode\kern.1em
            \raise.5ex\hbox{\the\scriptfont0 #1}\kern-.1em
      /\kern-.15em\lower.25ex\hbox{\the\scriptfont0 #2}}

\usepackage{times}

\begin{document}

   \title[ Dark matter anisotropic cosmic infall on $L_{\star}$ haloes]{ The
    origin and implications  of \\ dark matter anisotropic  cosmic infall on
    $\approx L_{\star}$ haloes }

   \author[D. Aubert, C. Pichon and S. Colombi]{D. Aubert$^{1,3}$\thanks{E-mail:aubert@astro.u-strasbg.fr},
   C. Pichon$^{1,2,3}$ and S. Colombi$^{2,3}$\\
$^{1}$ Observatoire astronomique de Strasbourg, 11 rue de l'Universite, 67000
   Strasbourg, France\\
$^{2}$ Institut d'Astrophysique de Paris, 98 bis boulevard
       d'Arago, 75014 Paris, France \\
$^{3}$ Numerical Investigations in Cosmology (N.I.C.), CNRS, France}



   \date{Typeset \today ; Received / Accepted}
 \maketitle     
\begin{abstract}
We measure the anisotropy of dark matter flows on small scales ($\le 500$ kpc)
in the near  environment of haloes using a large set  of simulations.  We rely
on two different  approaches to quantify the anisotropy  of the cosmic infall:
we  measure  the flows  at  the haloes'  virial  radius  while describing  the
infalling matter via fluxes through  a spherical shell; we measure the spatial
and kinematical  distributions of  satellites and substructures  around haloes
detected  by the  subclump finder  ADAPTAHOP first  described in  the appendix
\ref{s:ADAPT}.   The   two  methods  are   found  to  be  in   agreement  both
qualitatively and  quantitatively via one  and two points statistics.   \\ The
peripheral and  advected momentum is correlated  with the spin  of the embeded
halo at a  level of $30\%$ and $50\%$.  The  infall takes place preferentially
in the  plane perpendicular to the  direction defined by the  halo's spin.  We
computed  the excess  of equatorial  accretion both  through rings  and  via a
harmonic expansion  of the  infall.  \\ The  level of anisotropy  of infalling
matter  is  found  to be  $\sim15  \%$.   The  substructures have  their  spin
orthogonal to  their velocity vector  in the halo's  rest frame at a  level of
about $5\%$,  suggestive of  an image of  a flow along  filamentary structures
which  provides   an  explanation  for  the  measured   anisotropy.   Using  a
`synthetic'  stacked halo,  it is  shown  that the  satellites' positions  and
orientations relative to the direction of  the halo's spin are not random even
in projection. The average ellipticity  of stacked haloes is $10\%$, while the
alignment excess  in projection reaches $2\%$.  All  measured correlations are
fitted by a simple 3 parameters  model.  \\ We conclude that a halo does not see
its environment  as an isotropic perturbation, investigate  how the anisotropy
is  propagated   inwards  using  perturbation  theory,   and  discuss  briefly
implications for weak lensing, warps and the thickness of galactic disks.
\end{abstract}

%

\section{Introduction}

Isotropy  is one of  the fundamental  assumptions in  modern cosmology  and is
widely verified on  very large scales, both in large  galaxies' surveys and in
numerical  simulations.   However   on  scales  of  a  few   Mpc,  the  matter
distribution  is structured  in  clusters  and filaments.   The  issue of  the
anisotropy down to galactic and cluster scales has long been studied, as it is
related to the search for large scale structuration in the near-environment of
galaxies.   For   example,  both  observational   study  (e.g.   \citet{West},
\citet{Plionis},      \citet{Kitz})      and     numerical      investigations
(e.g.  \citet{Falten}) showed  that galaxies  tend  to be  aligned with  their
neighbours  and support the  vision of  anisotropic mergers  along filamentary
structures.  On smaller  scales, simulations of rich clusters  showed that the
shape  and  velocity  ellipsoids  of  haloes  tend  to  be  aligned  with  the
distribution   of   infalling  satellites   which   is  strongly   anisotropic
(\citet{Tormen}).  However the point is still moot and recent publications did
not confirm such an anisotropy  using resimulated haloes; they proposed 20$\%$
as  a  maximum  for  the  anisotropy  level  of  the  satellites  distribution
(\citet{Vitvitska}).

When considering preferential directions  within the large scale cosmic web,
the  picture  that comes  naturally to mind  is  one  involving these  long
filamentary structures  linking large  clusters to one  other. The  flow of
haloes  within these  filaments  can  be responsible  for  the emergence  of
preferential directions and alignments. Previous publications showed that the
 distributions of spin vectors are   not  random.   For  example,  haloes  in
simulations tend to  have their spin pointing orthogonally  to the filaments'
direction  (\citet{Falten}).   Furthermore,  down  to galactic  scales,  the
angular  momentum remains  mainly aligned  within  haloes (\citet{Bullock}).
Combined with the results suggesting that haloes' spins are mostly sensitive to
recent  infall (\citet{Haarlem}),  these  alignment properties  fit well with
accretion  scenarii  along special  directions~:  angular  momentum can  be
considered as a good marker to test this picture.

Most  of  these  previous studies  focused  on  the  fact  that  alignments  and
preferential directions are consequences of the formation process of haloes.
However, the effects of such preferential directions on the inner properties
of  galaxies have  been  less addressed.   It  is widely  accepted that  the
properties  of galaxies  partly result  from their  interactions  with their
environments.   While the  amplitude  of the  interactions  is an  important
parameter, some  issues cannot  be studied without  taking into  account the
spatial  extension  of these  interactions.   For  example,  a warp  may  be
generated  by   the  torque  imposed   by  infalling  matter  on   the  disk
(\citet{OsBin}, \citet{Lopez})~: the direction but also the amplitude of the
warp  are  a  direct  consequence   of  the  spatial  configuration  of  the
perturbation.  Similarly, it is likely  that disks' thickening due to infall
is  not   independent  of  the   incoming  direction  of   satellites  (e.g.
\citet{Quinn}, \citet{Velazquez}, \citet{Huang}).

Is it possible to observe  the small-scale alignment ? In particular,
weak lensing deals with effects as small as the level of detected anisotropy
(if  not smaller)  (e.g.   \citet{Hatton}, \citet{Croft},  \citet{Heavens}),
hence the  importance to  put quantitative constraints  on the  existence of
alignments on small scales. Therefore, the present paper also  addresses the
issue of detecting preferential projected orientations on the sky
of  substructures  within  haloes.

Our  main  aim  is  to   provide  quantitative  measurements  to  study  the
consequences of  the existence of  preferential directions on  the dynamical
properties  of  haloes  and  galaxies,  and on  the  observation  of  galaxy
alignments.    Hence  our  point   of  view   is  more   galactocentric  (or
cluster-centric)  than  previous studies.   We  search  for local  alignment
properties on  scales of  a few hundred  kpc.  Using  a large sample  of low
resolution  numerical simulations,  we aim  to extract  quantitative results
from  a large  number of  halo  environments.  We  reach a  higher level  of
statistical significance while reducing the cosmic variance.  We applied two
complementary approaches  to study the  anisotropy around haloes:  the first
one is particulate and uses a new substructure detection tool ADAPTAHOP, the
other  one  is  the  spherical  galactocentric fluid  approach.   Using  two
methods, we can assess the self-consistency of our results.

 After a brief  description of our set of simulations  ($\S 2$), we describe
the  galactocentric  point of  view  and  study  the properties  of  angular
momentum and infall  anisotropy measured at the virial  radius ($\S 3$).  In
($\S 4$) we  focus on anisotropy in the  distribution of discrete satellites
and substructures and we study the properties of the satellites' proper spin
, which provides an explanation for the detected anisotropy.  In ($\S 5$) we
discuss the  level of anisotropy as seen  in projection on the  plane of the
sky.  We then  investigate how the anisotropic infall  is propagated inwards
and discuss the possible implications of  our results to weak lensing and to
the dynamics  of the disk through  warp generation and  disk thickening ($\S
6$).   Conclusions  and  prospects   follow.   The  appendix  describes  the
substructures detection tool ADAPTAHOP  together with the relevant aspects of
one    point centered  statistics on  the sphere.   We  also formally
derive there the perturbative inward  propagation of infalling fluxes into a
collisionless self gravitating sphere.


\section{Simulations}
\begin{figure} 
\centering \resizebox{7cm}{7cm} {\includegraphics{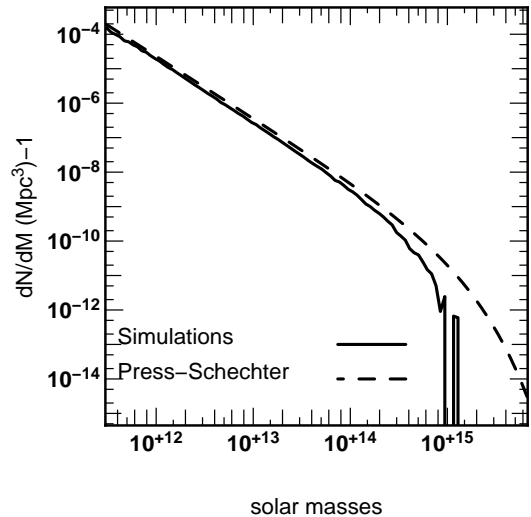}}
\caption{The number  density, $dN/dM$, of haloes  with a mass  M (within dM)
detected in  our simulations at redshift  zero (solid line)  compared to the
distribution expected  by Press-Schechter  (dashed line).  The  agreement is
good  for  masses  between  $3\cdot  10^{11} M_\odot$  and  $3\cdot  10^{14}
M_\odot$. 
}
\label{fmass} 
\end{figure}

In order  to achieve  a sufficient  sample and ensure  a convergence  of the
measurements, we  produced a set  of $\sim 500  $ simulations. Each  of them
consists  of  a  50  $h^{-1}$  Mpc$^3$ box  containing  $128^3$
particles. The mass resolution is $5\cdot10^{9} M\odot$.
A
$\Lambda$CDM  cosmogony ($\Omega_m=0.3$,  $\Omega_\Lambda=0.7$,  $h=0.7$ and
$\sigma_8=0.928$) is  implemented with different  initial conditions.  These
initial conditions were produced with GRAFIC (\citet{Grafic}) where we chose
a  BBKS  (\citet{BBKS})  transfer  function  to compute  the  initial  power
spectrum.   The initial  conditions  were  used as  inputs  to the  parallel
version of the treecode GADGET (\citet{Gadget}). We set the softening length
to 19 $h^{-1}$ kpc.
The halo  detection was performed  using the halo finder  HOP (\citet{Hop}).
We   employed    the   density   thresholds   suggested    by   the   authors
($\Delta_{\mathrm{outer}}=80,
\delta_{\mathrm{saddle}}=2.5\delta_{\mathrm{outer}},
\delta_{\mathrm{peak}}=3.\delta_{\mathrm{outer}}$) 
As a check, we computed the halo
mass  distribution.  It is  shown in  Fig. \ref{fmass}  and compared  to the
Press-Schechter mass function  (\citet{Press}).  The measured distribution
is in agreement with the theoretical curve up to masses $\sim 3\cdot 10^{14}
M_\odot$ (100 000 particles), which validates our completeness in mass.

\begin{figure} 
\centering \resizebox{7cm}{7cm} {\includegraphics{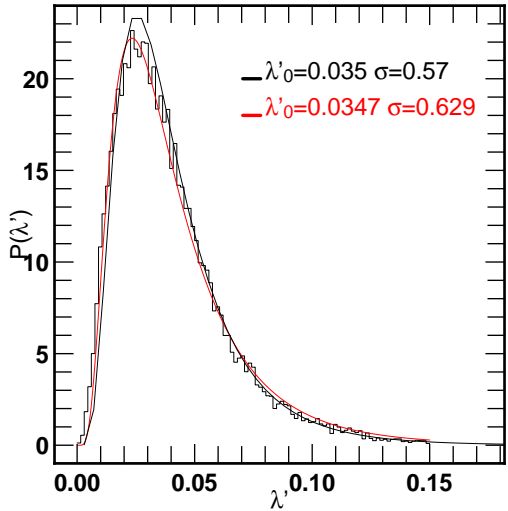}}
\caption{The  distribution of  the  spin parameter  $\lambda  '$ defined  as
$\lambda ' \equiv J/(\sqrt{2}MVR_{200})$  computed using 100 000 haloes with
a mass greater than $5\cdot10^{12}M_\odot$. The distribution can be fit with
a  log-normal function  with parameters  $\lambda_0' =  0.0347\pm0.0006$ and
$\sigma =0.63\pm0.02$ (red line).  The curve parametrized by $\lambda_0' =
0.035$ and  $\sigma =0.57$ is  also shown (black  line).  The two  results are
almost concurrent, indicating that the  value of $\sigma$ is not so strongly
constrained using a log-normal distribution.} 
\label{lambda} 
\end{figure}

As an other  means to check our simulations and  to evaluate the convergence
ensured by our large set of haloes, we computed the probability distribution
of the spin parameter $\lambda '$, defined as (\citet{Bullock}):
\begin{equation}
\lambda ' \equiv \frac{J}{\sqrt{2}MVR_{200}}.
\end{equation}
Here $J$  is the  angular momentum  contained in a  sphere of  virial radius
$R_{200}$ with a mass M  and $V^2=GM/R_{200}$. The measurement was performed
on  100~000  haloes  with  a  mass  larger  than  $5\cdot10^{12}M_\odot$  as
explained in the next section. The resulting distribution for $\lambda '$ is
shown in  Fig.  \ref{lambda}. The distribution $P(\lambda')$  is well fitted
by a log-normal distribution (e.g. \citet{Bullock}):
\begin{equation}
P(\lambda')\d\lambda'=\frac{1}{\lambda  '  \sqrt{2\pi}\sigma}\exp\left(-\frac{\ln^2(\lambda
'/\lambda_0 ')}{2\sigma^2}\right)\d\lambda'.
\end{equation}
We found $\lambda_0' =  0.0347\pm0.0006$ and $\sigma =0.63\pm0.02$ as best-fit
values  and  they are  consistent  with  parameters  found by  \citet{Peirani}
($\lambda ' =0.035$  and $\sigma=0.57$) but our value  of $\sigma$ is slightly
larger.   However,  using  $\sigma=0.57$  does  not lead  to  a  significantly
different result.   The value of $\sigma$  is not strongly  constrained and no
real disagreement  exists between our  and their best-fit values.   The halo's
spin, on  which some  of the following  investigations are based,  is computed
accurately.
\section{A galactocentric point of view}
%
The analysis of exchange processes  between the haloes and the intergalactic
medium will be carried out using two  methods.  The first one can be described as
`discrete'.  The  accreted objects are  explicitly counted as  particles or
particle groups. This  approach will be applied and  discussed later in this
paper.  The other method relies on measuring directly relevant quantities on a
surface  at \textit{the interface}  between the  halo and  the intergalactic
medium.  In  this approach,  the measured quantities  are scalar,  vector or
tensor fluxes,  and we  assign to them  \textit{ flux densities}.   The flux
density representation  allows us to  describe the angular  distribution and
temporal  coherence  of infalling  objects  or  quantities  related to  this
infall.   The  formal relation  between  a  flux density,  $\MG{\varpi}({\bf
\Omega})$, and its associated total flux through a region $S$, $\Phi$, is:
\begin{equation}
\Phi\equiv\int_S \MG{\varpi}({\bf \Omega})\cdot \d {\bf \Omega},
\end{equation}
where ${\bf \Omega}$ denotes the position on the surface where $\MG{\varpi}$
is evaluated  and $\d {\bf  \Omega}$ is the  surface element normal  to this
surface.  Examples of flux densities
are mass flux density, $\rho {\bf v_r}$, or accreted angular momentum, $\rho
{\bf v_r} {\bf L}$.  In particular, this description in terms of a spherical
boundary  condition is  well-suited  to study  the  dynamical stability  and
response of  galactic systems.   In this section,  these fields are  used as
probes of the environment of haloes.

\subsection{Halo analysis}
\begin{figure} 
\centering
\resizebox{7cm}{7cm}{\includegraphics{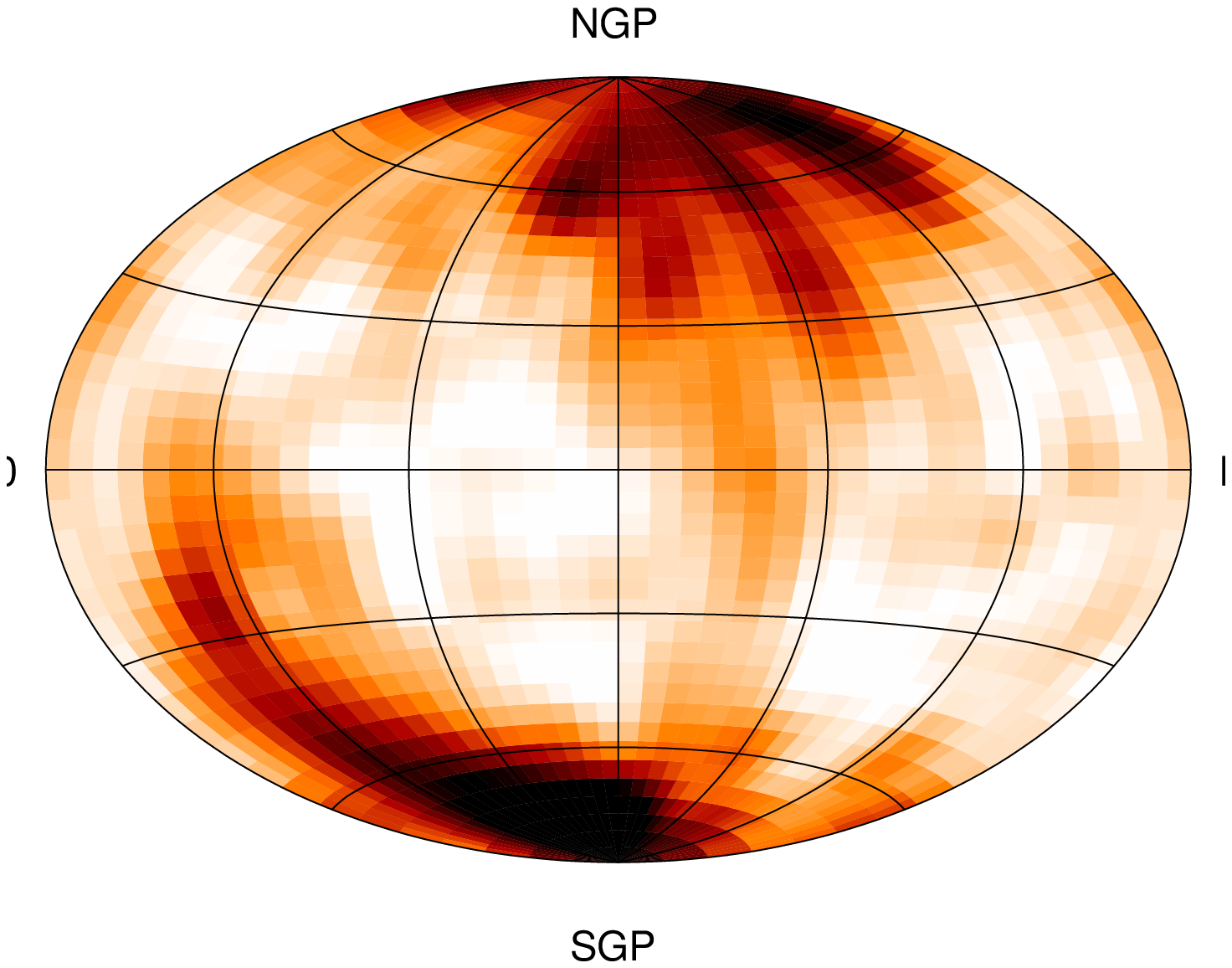}}
\resizebox{7cm}{7cm}{\includegraphics{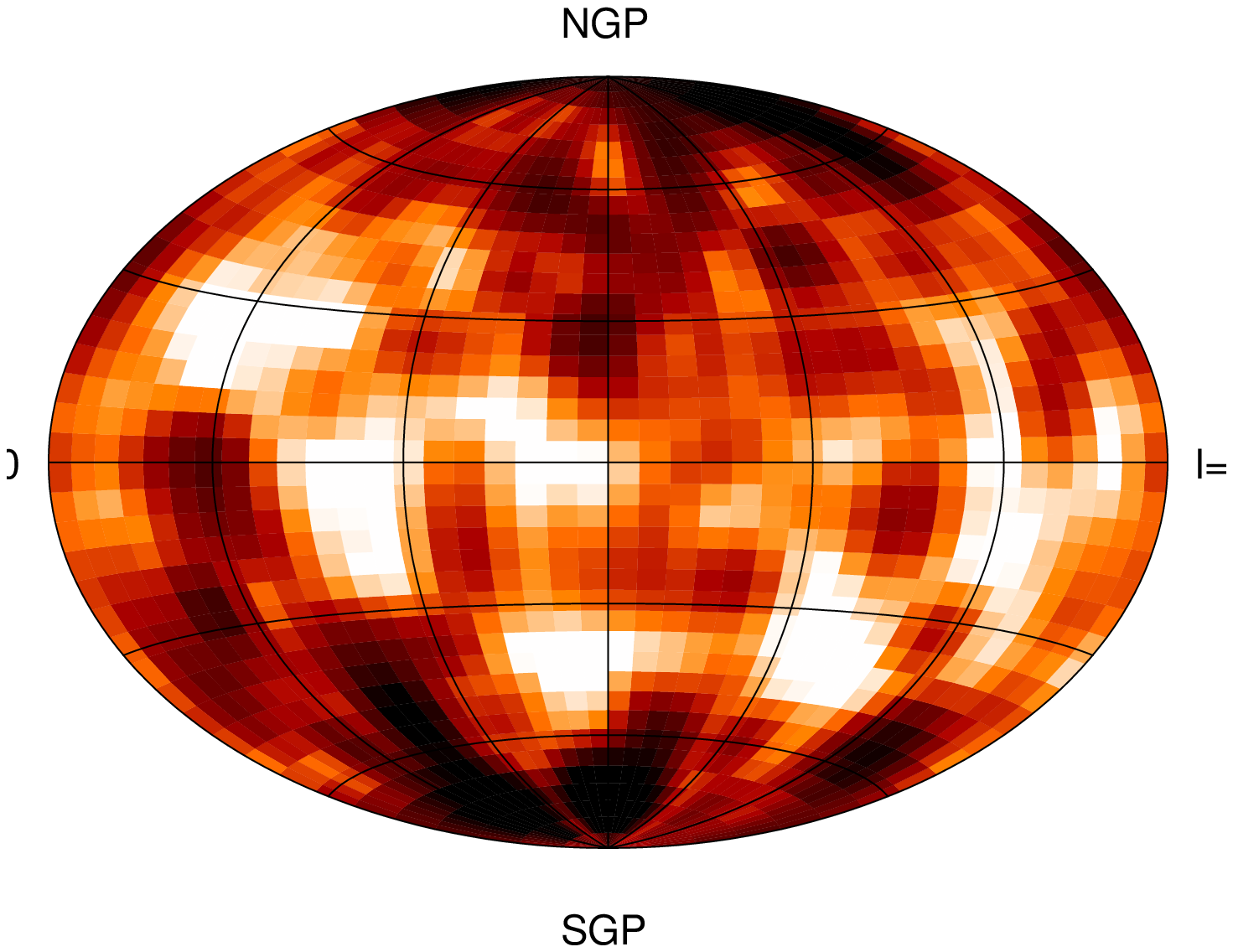}}
\caption{A  galactocentric point of  view of  the density  field, $\rho({\bf
\MG{\Omega}})$  ({\sl top})  and of  the flux  density of  mass,  $\rho {\bf
v_r}({\bf \MG{\Omega}})$,  surrounding the  same halo ({\sl  bottom}).  This
measurement was  extracted from a $\Lambda$CDM  cosmological simulation. The
considered halo  contained about $10^{13}  M_\odot$ or 2000  particles.  The
high density zones are darker.  The density's spherical field shows a strong
quadrupolar component with high density  zones near the two poles while this
component is less important for the  mass flux density field measured on the
sphere.  This  discrepancy between  the two spherical  fields is  common and
reflects the shape of the halo as discussed in the main text.}  \label{Mapsphere}
\end{figure}

Once a  halo is  detected, we study  its environment using  a galactocentric
point  of view.   
The  relevant  fields ${\MG\varpi}({\bf\MG{\Omega}})$  are  measured on  the
surface of a  sphere centered on the  halo's centre of mass with  a $ R_{200}$
radius  (where  $3M/(4\pi  R_{200}^3)\equiv 200\overline{\rho}$)  (cf.   fig
.\ref{Mapsphere}).     
There is  no
exact, nor unique, definition of the halo's outer boundary and our choice of a
$ R_{200}$ (also  called the virial radius) is the  result of a compromise
between a large distance to the halo's center and a good signal-to-noise ratio
in the spherical density fields determination.

We  used $40\times  40$ regularly  sampled  maps in  spherical angles  ${\bf
\Omega}=(\vartheta,\phi)$, allowing for an  angular resolution of 9 degrees.
We take into account haloes with  a minimum number of $1000$ particles, which
gives a  good representation of high  density regions on  the sphere. This minimum corresponds    to
$5\cdot10^{12}M_\odot$ for a halo, and allows us to reach a total number of 10~000 haloes
at  z=2 and  50~000  haloes at  z=0. This  range  of mass  corresponds to  a
somewhat  high  value  for a  typical  $L_*$  galaxy  but results  from  our
compromise  between  resolution  and  sample  size.   Detailed  analysis of
the effects of resolution  is postponed to \citet{Aubert1}.

The density, $\rho({\bf \MG{\Omega}}$), on  the sphere is computed using the
particles located in  a shell with a radius of $R_{200}$  and a thickness of
$R_{200}/10$  (this  is  quite  similar  in  spirit to  the  count  in  cell
techniques widely used  in analyzing the large scale  structures, but in the
context of  a sphere the cells  are shell segments).   Weighting the density
with quantities such as the radial  velocity or the angular momentum of each
particle contained within the  shell, the associated spherical fields, $\rho
{\bf  v_r}({\MG{\Omega}})  $  or  $\rho  {\bf L}({\MG{\Omega}})  $,  can  be
calculated for each  halo. Two examples of spherical maps  are given in Fig.
\ref{Mapsphere}. They  illustrate a frequently  observed discrepancy between
the two types of spherical  fields, $\rho({\bf \MG{\Omega}}$) and $\rho {\bf
v_r}({\MG{\Omega}}$).     The    spherical    density   field,    $\rho({\bf
\MG{\Omega}}$), is strongly quadrupolar, which is due to the intersection of
the halo  triaxial 3-dimensional density field by  our 2-dimensional virtual
sphere.    By   contrast   the   flux   density  of   matter,   $\rho   {\bf
v_r}({\MG{\Omega}}$),  does  not have  such  quadrupolar distribution.   The
contribution of halo particles to the  net flux density is small compared to
the contribution of particles coming from the outer intergalactic region.

\subsection{Two-points statistics:   advected  momentum
   and halo's spin}
\label{s:two-points}
 \begin{figure}         
  \centering
\resizebox{7cm}{7cm}{\includegraphics{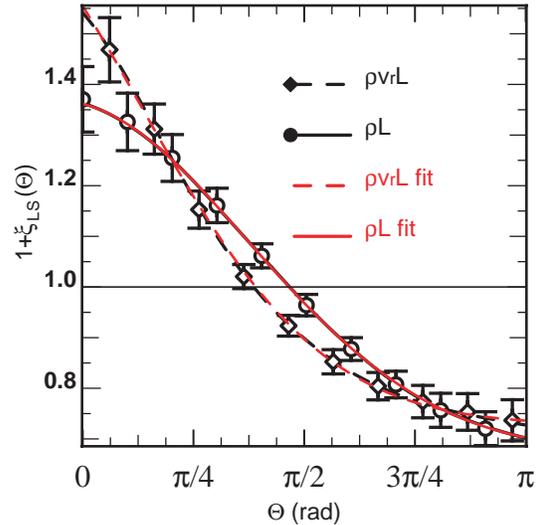}}
\caption{Excess  probability,  $1+\xi_\mathrm{LS}(\theta)$,   of  the  angle,  $\theta$,
between the halo's spin ($\bf S$)  and the angular momentum (${\bf L}_T$ for
total, or ${\bf L}_A$ for accreted)  measured on the virial sphere using the
fluid located at the virial radius.
Here ${\bf L}_T$ represents  the \textit{total} angular momentum measured on
the  virial sphere (corresponding  curve~: solid  line, circle  symbols) and
${\bf  L}_A$ the \textit{total  accreted} angular  momentum measured  on the
sphere (corresponding curve: dashed  line, diamond symbols).  The error bars
represent the $3 \sigma$ dispersion measured on subsamples of 10~000 haloes.
The   correlation  takes  into   account  the   uncertainty  on   the  angle
determination due  to the  small number of  particles at the  virial radius.
Here $\xi_\mathrm{LS}(\theta)\equiv0$ would be expected for an isotropic distribution of
angles  between $\bf  S $  and  $\bf L  $ while  the measured  distributions
indicate that  the aligned configuration ($\theta \sim  0$) is significantly
more  likely. The  two excess  probability distibutions  are  well-fitted by
Gaussian functions (red curves: see main text). }  \label{LS}
\end{figure}

The influence of infalling matter on the dynamical state of a galaxy depends
on whether  or not the infall  occurs inside or outside  the galactic plane.
If  the infalling  matter is  orbiting in  the galactic  plane,  its angular
momentum is aligned with the angular  momentum of the disk.  Taking the halo's
spin as  a reference for the direction  of the `galactic' plane,  we want to
quantify  the  level  of  alignment  of  the  orbital  angular  momentum  of
peripheral structures (i.e.  as measured  on the virial sphere) relative to
that spin.   The inner
spin   $\bf{S}$  is   calculated  using   the   positions-velocities  $({\bf
r_{\mathrm{part}}},{\bf  v_{\mathrm{part}}})$ of  the  particles inside  the
$R_{200}$   sphere    in   the   centre   of   mass    rest   frame   $({\bf
r_{\mathrm{0}}},{\bf v_{\mathrm{0}}})$:
\begin{equation}
{\bf                      S}=\sum_{\mathrm{part}}                      ({\bf
r_{\mathrm{part}}-r_{\mathrm{0}}})\times                              ({\bf
v_{\mathrm{part}}-v_{\mathrm{0}}}).
\end{equation}
${\bf r_{\mathrm{0}}}$  is the  position of the  halo centre of  mass, while
${\bf  v_{\mathrm{0}}}$  stands  for  the  average velocity  of  the  halo's
particles. This  choice of  rest frame is  not unique; another  option would
have been to  take the most bounded particle  as a reference.  Nevertheless,
given the considered mass range, no significant alteration of the results is
to be expected.
The  total angular  momentum, ${\bf  L}_T$ (measured  at  the virial  radius,
$R_{200}$) is  computed for each halo  using the spherical  field $\rho {\bf
L}({\MG{\Omega}})$:
\begin{equation}
{\bf L}_T=\int_{4 \pi}  \rho {\bf L}({\MG{\Omega}})\d{\MG{\Omega}}. 
\end{equation}
The angle, $\theta$,  between the spin of the inner particles  $\bf S$ and the
total orbital momentum ${\bf L}_T$  of `peripheral' particles is then easily
computed:
\begin{equation}
\theta=\cos^{-1} (\frac{{\bf L}_T\cdot{\bf S}}{|{\bf L}_T||{\bf S}|}).
\end{equation}
Measuring this angle $\theta$ for all the haloes of our simulations allow us
to  derive  a  raw  probability  distribution of  angle,  $d_r(\theta)$.  An
isotropic  distribution  corresponds to  a  non-uniform probability  density
$d_{\mathrm{iso}}(\theta)$. Typically $d_{\mathrm{iso}}$ is smaller near the
poles (i.e.  near  the region of alignment) leading  to a larger correction
for  these angles  and  to larger  error  bars in  these  regions (see  fig.
\ref{LS}):  this is the  consequence of smaller  solid angles in  the polar
regions (which  scales like $\sim  \sin \theta$) than in  equatorial regions
for a given $\theta$ aperture. The true anisotropy is estimated by measuring
the ratio:
\begin{equation}
d_r(\theta)/d_{\mathrm{iso}}(\theta)\equiv1+\xi_{\mathrm {LS}}(\theta),
\end{equation} 
Here, $1+\xi_{\mathrm{LS}}(\theta)$  measures the excess probability  of finding $\bf
S$ and  ${\bf L}_T$  away from each  other, while $\xi_{\mathrm{LS}}(\theta)$  is the
cross correlation  of the angles  of $\bf S$  and ${\bf L}_T$.   Thus having
$\xi_\mathrm{LS}(\theta)>0$ (resp.   $\xi_\mathrm{LS}(\theta)<0)$ implies an excess  (resp.  a lack)
of  configurations with  a $\theta$  separation relative  to  an isotropic
situation.

To take into account the error  in the determination of $\theta$, each count
(or Dirac distribution) is replaced with a Gaussian distribution
and contributes to several bins:
\begin{equation}
\delta(\theta-\theta_0)     \rightarrow     {\cal     N}(\theta_0,\sigma_0)=
\frac{1}{\sigma_0\sqrt{2\pi}}\exp\left(-\frac{(\theta-\theta_0)^2}{2\sigma_0^2}\right),
\end{equation} 
where $\cal N$  stands for a normalized Gaussian  distribution and where the
angle  uncertainty  is  approximated  by  $\sigma_0  \sim  (4\pi/N)^{1/2}$ using
$N$ particles as
suggested by  \citet{Hatton}. If $N_v$ is  equal to the  number of particles
used to compute  $\rho {\bf L}({\bf \MG{\Omega}})$ on  the virial sphere and
if $N_h$ is the number of particles used to compute the halo spin, the error
we associated to the angle between the angular momentum at the virial sphere
and the halo spin is:
\begin{equation}
\sigma_0=\sqrt{(4\pi/N_v)+(4\pi/N_h)}\sim \sqrt{(4\pi/N_v)},
\end{equation}
because  we  have  $N_v  \ll  N_h$. Note  that  this  Gaussian  correction
introduces  a  bias  in mass:  a  large  infall  event (large  $N_v$,  small
$\sigma_0$)  is weighted more  for a  given $\theta_0$  than a  small infall
(small $N_v$,  large $\sigma_0$).  All  the distributions are added  to give
the final distribution:
\begin{equation}
d_r(\theta)=\sum_p^{N_p} {\cal N}(\theta_p,\sigma_p),
\end{equation}
where $N_p$  stands for  the total number  of measurements (i.e.   the total
number of  haloes in our set  of simulations).    The
corresponding isotropic angle distribution is  derived using the same set of
errors randomly redistributed:
\begin{equation}
d_{\mathrm{iso}}(\theta)=\sum_p^{N_p}                                   {\cal
N}(\theta_p^{\mathrm{iso}},\sigma_p).
\end{equation}

Fig. \ref{LS} shows  the excess probability, $1+ \xi_\mathrm{LS}(\theta)$,  of the angle
between the total  orbital momentum of particles at  the virial radius ${\bf
L}_T$ and the halo spin $\bf S$.  The solid line is the correlation deduced
from 40~000 haloes at redshift  $z=0$.  The error bars were determined using
50 subsamples  of 10 000  haloes extracted from  the whole set  of available
data. 
An average Monte-Carlo correlation  and a Monte-Carlo dispersion $\sigma$ is
extracted.  In Fig.  \ref{LS}, the symbols stand for the average Monte-Carlo
correlation,  while  the  vertical   error  bars  stand  for  the  $3\sigma$
dispersion.  

The correlation  in Fig. \ref{LS} shows  that all angles  are not equivalent
since $\xi_\mathrm{LS}(\theta)\ne0$.  It  can be fitted with a  Gaussian curve using the
following parametrization:
\begin{equation}
1+\xi_\mathrm{LS}(\theta)=\frac{a_1}{\sqrt{2\pi}a_3}\exp\left[-(\theta-a_2)^2/(2a_3^2)\right]+a_4.
\label{gaussfit}
\end{equation}
The  best  fit  parameters  are  $a_1=2.351\pm0.006$,  $a_2=-0.178\pm0.002$,
$a_3=1.343\pm0.002$,  $a_4=0.6691\pm0.0004$.  The  maximum being  located at
small angles, the aligned configuration, $\widehat{{\bf L}_T{\bf S}}\sim 0$,
is the most enhanced configuration  (relative to an isotropic distribution of
angle $\theta$).  The  aligned configuration of ${\bf L}_T$  relative to $\bf
S$ is $35\%$  ($\xi_\mathrm{LS}(0) =0.35$) more frequent in our  measurements than for a
random   orientation  of  ${\bf   L}_T$.   As   a  consequence,   matter  is
preferentially  located  in  the  plane  perpendicular  to the spin,  which  is
hereafter referred to as the `equatorial' plane.

The  angles, $(\vartheta,\phi)$,  are  measured relative  to the  simulation
boxes z-axes and x-axes and not relative to the direction of the spin.  Thus
we  do no  expect artificial  ${\bf L}_T$-$\bf  S$ correlations  due  to the
sampling procedure.  Nevertheless it  is expected on geometrical ground that
the  aligned configuration  is more  likely since  the contribution  of {\sl
recent} infalling dark matter to the halo's spin is important. As a check, the
same  correlation was  computed  using the  total \textit{advected}  orbital
momentum:
\begin{equation}
{\bf L}_A=\int_{4 \pi}  {\bf L} \rho {v_r}({\MG{\Omega}}) \cdot\d{\MG{\Omega}}\,.
\end{equation}
 The resulting correlation  (see Fig.  \ref{LS}) is similar  to the previous
one  but the  slope toward  small  values of  $\theta$ is even  stronger and  for
example  the excess  of aligned  configuration reaches  the level  of $50\%$
($\xi_\mathrm{LS}(0)   \sim0.5$).     The   correlation   can    be   fitted   following
Eq.~\ref{gaussfit}   with    $a_1=3.370\pm0.099$,   $a_2=-0.884\pm0.037$,
$a_3=1.285  \pm 0.016$ and  $a_4=0.728\pm0.001$.  This  enhancement confirms
the relevance of advected momentum for the build-up of the halo's spin, though
the increase in amplitude is limited  to 0.2 for $\theta=0$.  The halo's inner
spin is  dominated by  the orbital momentum  of infalling clumps  (given the
larger lever arm  of these virialised clumps) that  have just passed through
the   virial   sphere,  as   suggested   by   \citet{Vitvitska}  (see   also
appendix~\ref{s:convergence}).   It  reflects a  temporal  coherence of  the
infall of matter  and thus of angular momentum, and  a geometrical effect: a
fluid clump  which is just being  accreted can intersect  the virtual virial
sphere, being in  part both ``inside'' and ``outside''  the sphere.  Thus it
is expected that the halo spin $\bf S$ and the momenta ${\bf L}_T$ and ${\bf
L}_A$ at the virial radius are  correlated since the halo's spin is dominantly set
by the properties of the angular momentum in its outer region.
The anisotropy of the two fields  ${\bf L}_T$ and ${\bf L}_A$ do not have
the same  implication. The spatial distribution of  advected angular momentum,
${\bf L}_A$,  contains stronger  dynamical information.  In  particular, the
variation of the halo+disk's angular momentum is induced by tidal torques but also
by accreted momentum for an open system. For example the anisotropy
of ${\bf L} \rho {\bf v_r}$  should be reflected in the statistical properties
of  warped   disks  as  discussed   later  in  sections   \ref{s:kalnajs}  and
\ref{s:warp}.
\subsection{One-point statistics: equatorial infall anisotropy }
\label{s:one-point}
\begin{figure}         
  \centering            \resizebox{7cm}{7cm}{\includegraphics{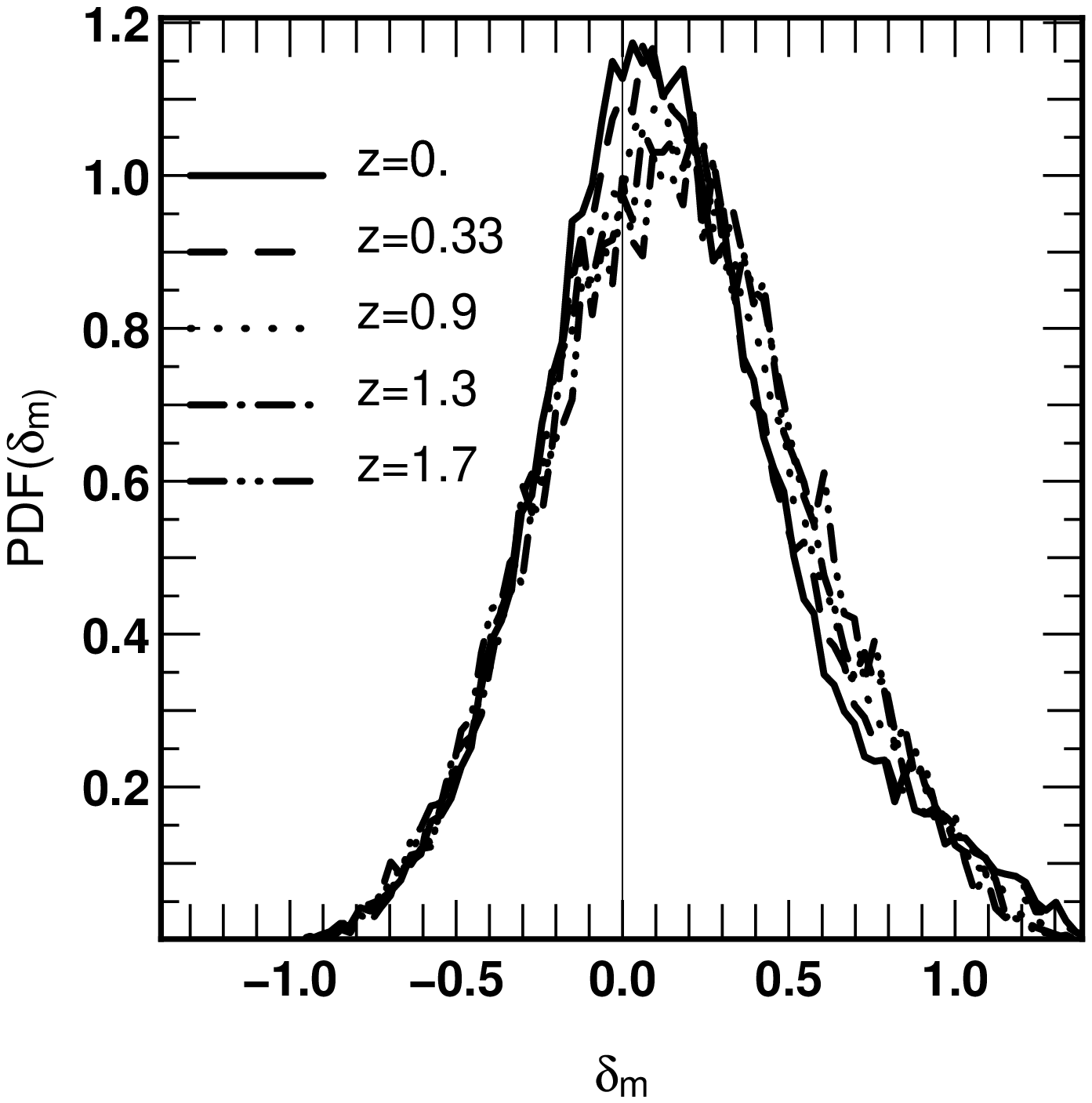}}
\resizebox{7cm}{7cm}{\includegraphics{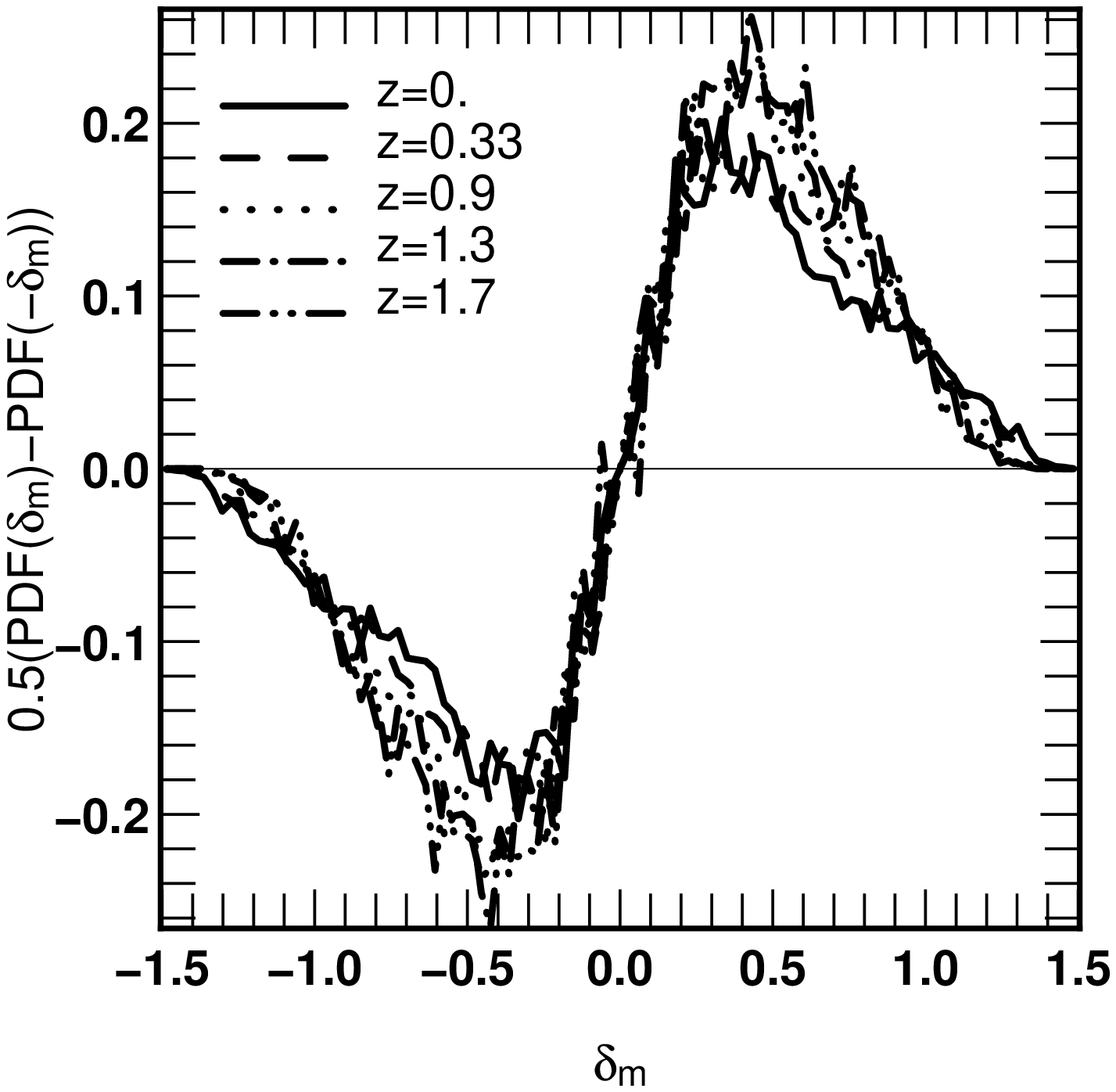}}
\caption{ \textit{Top: } Normalized  probability distributions (PDF) of the
excess of equatorial infall,  $\delta_m$,  measured  at the  virial radius.
The quantity  $1+\delta_m$ stands for the  ratio between the  flux of matter
through the  equatorial sub-region of  the $R_{200}$ sphere and  the average
flux  of  matter  through   the  whole  $R_{200}$  sphere.   The  equatorial
sub-region is defined as being  perpendicular to the direction of the halo's
spin.   It formally corresponds  to the  top-hat-smoothed mass  flux density
contrast. $\delta_m=0$ is expected for an isotropic infall of matter through
the virial sphere. The average value  of $\delta_m$ is always greater than 0
indicating that the  infall of matter is, on average,  more important in the
direction orthogonal  to the  halo's spin vector  than in  other directions.
\textit{Bottom:}  the antisymmetric  part of  the  $\delta_m$ distribution.
Being positive for positive values  of $\delta_m$, the antisymmetric part of
the $\delta_m$ distributions shows that accretion in the equatorial plane is
in  excess relative  to  the one  expected  from an  isotropic accretion  of
matter.}
\label{NI}          
\end{figure}          
\begin{figure}          \centering
\resizebox{7cm}{7cm}{\includegraphics{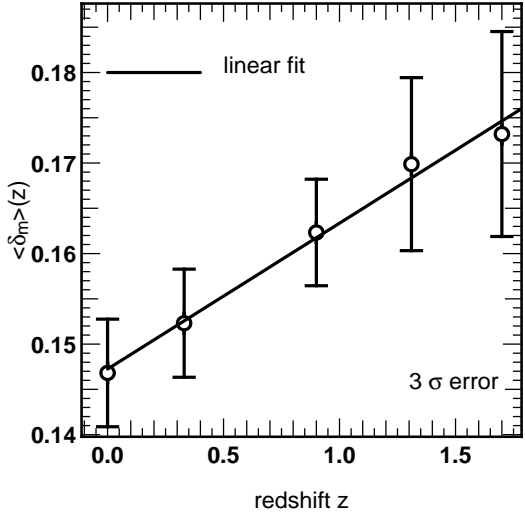}}
\caption{The redshift  evolution of $\langle \delta_m  \rangle$. The average
$\langle \cdot \rangle$  is performed on a set of 40~000  haloes at $z=0$ and
10  500  haloes at  $z=1.8$.  The  error bars  stand  for  the  error on  the
estimation       of        $\langle       \delta_m       \rangle$       with
$\Delta=\sigma(\delta_m)/\sqrt{N}$.  N  is the  number  of  haloes needed  to
compute  $\langle \delta_m  \rangle$. $\langle  \delta_m \rangle$  is always
positive  and indicates  an excess  of accretion  in  the equatorial
plane. This  redshift evolution can  be fitted as $\langle  \delta_m \rangle
(z) =0.0161(\pm 0.0103)  z +0.147 (\pm 0.005)$. This  excess is detected for
every redshift smaller than $z=2$, which indicates an excess of accretion in
the equatorial region. We  applied a mass threshold of $5\cdot10^{12}M_\odot$
to our haloes for every redshift. Then, the halo population is different from
one redshift to another. This  selection effect may dominate the observed time
evolution.}  \label{IvsZ}
\end{figure}
The previous  measurement doesn't account  for dark matter falling  into the
halo with a  very small angular momentum (radial orbits). 
We  therefore measured the excess
of equatorial accretion, $\delta_m$, defined  as follow.  We can measure the
average flow density of matter, $\Phi_r$, in a ring centered on the equatorial
plane:
\begin{equation}
\Phi_r\equiv\frac{1}{S_r}\int_{-\pi/8<\theta -\pi/2<\pi/8} \rho {\bf v_r}( 
\MG{\Omega}) \cdot \d{\MG{\Omega}},
\end{equation} 
where   $S_r=\int_{-\pi/8<\theta    -\pi/2<\pi/8}   \d{\MG{\Omega}}$.    The
ring-region  is  defined  by  the  area  where  the  polar  angle  satisfies
$\theta_{pol}=\pi/2\pm\pi/8$ which corresponds to about 40 $\%$ of the total
covered solid angle.  The larger  this region is, the better the convergence
of the  average value of $\Phi_r$,  but the lower the  effects of anisotropy,
since averaging over  a larger surface leads to a  stronger smoothing of the
field.   This  value  of  $\pm\pi/8$  is  a  compromise  between  these  two
contradictory trends.  In  the next section and in  the appendix, we discuss
more general filtering involving  spherical harmonics which are related
to the  dynamical evolution  of the  inner component of  the halo.   We also
measure the flow averaged on all the directions $\overline{\Phi}$:
\begin{equation}
\overline{\Phi}\equiv\overline{\rho {\bf
v_r}}\equiv \frac{1}{4\pi}\int_{4\pi} \rho {\bf v_r}( \MG{\Omega}) 
\cdot\d{\MG{\Omega}}.
\end{equation}
Since   we  are   interested  in   accretion,  we   computed   $\Phi_r$  and
$\overline{\Phi}$  using only  the infalling  part  of the  density flux  of
matter, where $\rho {\bf v_r} (\MG{\Omega})\cdot\d{\MG{\Omega}}<0$, ignoring
the  outflows.
We  therefore define $\delta_m$ as
\begin{equation}
 \delta_m\equiv\frac{\Phi_r-\overline{\Phi}}{\overline{\Phi}}.
\end{equation}   
This number  quantifies the  anisotropy of the  infall. It is  positive when
infall  is in  excess in  the  galactic-equatorial plane, while for  isotropic
infall $\delta_m\equiv 0$.  The  quantity $\delta_m$ can be regarded  as being the
`flux density' contrast of the infall of matter in the ring region (formally
it is the  centered top-hat-filtered mass flux density  contrast as shown in
appendix~\ref{s:statistics}). This measurement,  in contrast to those of the
previous section does  not rely on some  knowledge on the  inner region of the  halo but
only on the properties of the environment.

Fig.~\ref{NI} displays the normalized distribution of $\delta_{m }$ measured
for 50~000 haloes  with a mass in excess of  $5\cdot10^{12} M_\odot$ and for
different redshifts ($z=1.8, 1.5, 0.9,  0.3, 0.0$).  The possible values for
$\delta_m$  range between $\delta_m\sim  -1$ and  $\delta_m \sim  1.5$.  The
average value $\langle \delta_m \rangle$ of the distributions is statistically
larger than
zero (see also Fig.~\ref{IvsZ}).
Here $\langle  \quad \rangle$ stands for the  statistical expectation, which
in this paper is approximated by  the arithmetic average over many haloes in
our simulations. The antisymmetric part of the distribution
of $\delta_m$ is  positive for positive $\delta_m$.  The  PDF of $\delta_m$
is skewed,  indicating an excess  of accretion through the  equatorial ring.
The median  value for $\delta_m$ is  $\delta_{\mathrm{med}}=0.11$, while the
first $25  \%$ haloes  have $\delta_m<\delta_{25}\equiv-0.11$ and  the first
$75 \%$  haloes have  $\delta_m<\delta_{75}\equiv 0.37$.  Therefore  we have
$(\delta_{75}-\delta_{\mathrm{med}})/(\delta_{25}-\delta_{\mathrm{med}})=1.13$,
which quantifies how the  distribution of $\delta_{m}$ is positively skewed.
The              skewness             $S_3=\langle             (\delta-{\bar
\delta})^3\rangle/\langle(\delta-{\bar \delta})^2\rangle^{3/2}$  is equal to
$0.44$.  Combined  with the  fact that the  average value  $\langle \delta_m
\rangle$ is always positive, this shows  that the infall of matter is larger
in the equatorial plane than in the other directions.

This  result  is   robust  with  respect  to  time   evolution  (see  Fig.
\ref{IvsZ}).  At redshift $z=1.8$,  we have $\langle \delta_m \rangle =0.17$
which  falls down to $\langle  \delta_m \rangle=0.145$  at redshift  $z=0$.  This
redshift  evolution   can  be  fitted  as  $\langle   \delta_m  \rangle  (z)
=0.0161(\pm 0.0103) z +0.147 (\pm 0.005)$. 
This trend  should be taken with  caution. For \textit{every}  redshift z we
take  in account  haloes with  a mass  bigger than  $5\cdot10^{12} M_\odot$.
Thus the population of haloes studied at  z=0 is not exactly the same as the
one studied  at z=2.  Actually,  at z=0, there  is a strong  contribution of
small haloes (i.e.  with a mass close to $5\cdot10^{12} M_\odot$) which just
crossed the mass threshold.  The accretion on small haloes is more isotropic
as shown  in more details in appendix \ref{sec:appmasdep}.  One possible
explaination  is   that  they  experienced  less   interactions  with  their
environment  and  have had  since  time to  relax  which  implies a  smaller
correlation with the spatial distribution of the infall.  Also bigger haloes
tend to lie  in more coherent regions, corresponding  to rare peaks, whereas
smaller haloes are more evenly  distributed.  The measured time evolution of
the anisotropy  of the  infall of  matter therefore seems  to result  from a
competition between  the trend for haloes  to become more  symmetric and the
bias corresponding to a fixed mass cut.

In  short,  the infall  of  matter  measured at  the  virial  radius in  the
direction orthogonal to  halo spin is larger than  expected for an isotropic
infall.

\subsection{Harmonic  expansion of anisotropic infall}
\label{s:expansion}
\begin{figure}          
    \centering \resizebox{8cm}{9cm}{\includegraphics{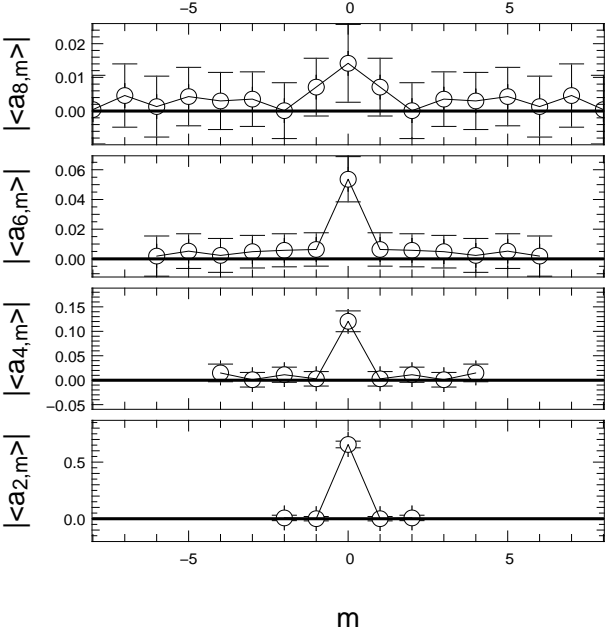}}
\caption{
The convergence of modulus of  the real part of $\langle {\tilde a_{\ell}^{m}}
\rangle$,  for $\ell=2,4,6,8$. The  ${\tilde a_{\ell}^{m}}$  decomposition was
computed for 25  000 haloes and each coefficient has  been normalized with the
corresponding  $C_0$  (see text  for  details).  {Here,  $\langle\,\,\,\rangle$
stands for the median while the  error bars stand for the distance between the
5th   and   the   95th   centile.}    The  median   value   of   $\langle\tilde
a_{\ell}^{m}\rangle$    is     zero    except    for     the    $\langle\tilde
a_{\ell}^{0}\rangle$ coefficient: this is a  signature of a field invariant to
azimuthal rotations.}
\label{peakalm}
\end{figure} 
\begin{figure}          
    \centering
\resizebox{8cm}{8cm}{\includegraphics{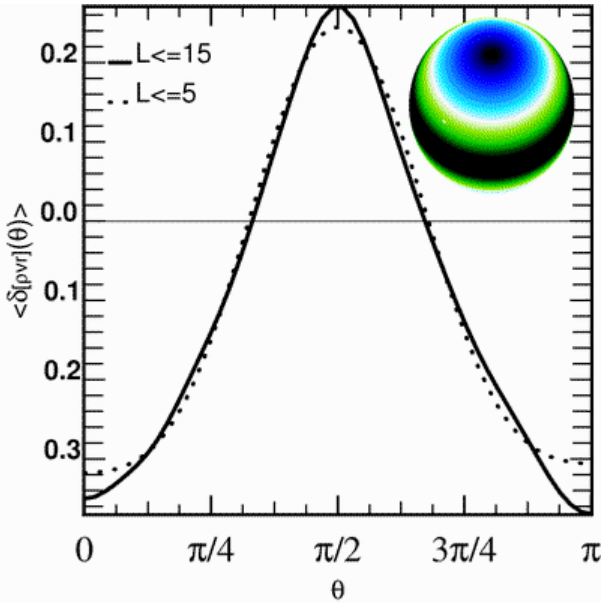}}
\caption{An illustration  of  the  convergence   of  the  $\tilde  a_{\ell  m}$
presented in Fig. \ref{peakalm}.  The  solid line stands for the azimuthal
average of the spherical  contrast of accretion computed using \Eq{defalm2}.
The dotted  line for  the spherical field  reconstructed with $\ell  \le 5$.
The insert represents the  reconstructed spherical field using the expansion
of the $\tilde  a_{\ell m}$ of the mass flux measured  at the virial sphere.
The sphere presents an excess  of accretion in the equatorial region because
of the non-zero average value  of $\tilde a_{\ell 0}$ coefficients (for even
values of $\ell$).}
\label{profilalm}
\end{figure}

As mentioned  earlier (and demonstrated  in appendix~\ref{s:response}), the
dynamics of  the inner halo and  disk is partly governed  by the statistical
properties  of the  flux  densities  at the  boundary.   Accounting for  the
gravitational  perturbation  and the  infalling  mass  or momentum  requires
projecting  the perturbation  over a  suitable basis  such as  the spherical
harmonics:
\begin{equation}
 \varpi  (  \MG{\Omega} )  =  \sum_{\ell,  m}  \alpha^m_{\ell} Y_{\ell}^m  (
 \MG{\Omega} ) \, , \EQN{defalm}
\end{equation} 
Here, $\varpi$ stands  for, {\sl e.g.}  the mass  flux density, the advected
momentum  flux density, or  the potential  perturbation.   The  resulting $\alpha_\ell^m$
coefficients  correspond  to  the  spherical harmonic  decomposition  in  an
\textit{arbitrary} reference frame.
The different $m$ correspond  to the different fundamental orientations for
a given  multipole $\ell$.  A spherical field  with no  particular orientation
gives rise to a field  averaged over the different realisations which appear
as a monopole, i.e.  $\langle \alpha_{\ell  m} \rangle =0$ for $\ell \ne 0$.
Having constructed  our virial sphere in  a reference frame  attached to the
simulation  box, we effectively  performed a  randomization of  the spheres'
orientation.
However, since  the direction of the halo's spin is associated  to a  general preferred
orientation for  the infall, it  should be traced through  the $\alpha_{\ell
m}$ coefficients.   Let us define the rotation  matrix, $\mathcal{R}$, which
brings the z-axis  of the simulation box along the direction of the halo's spin. The
spherical  harmonic  decomposition  centered   on  the  spin  of  the  halo,
$a_\ell^{m}$, is given by (e.g. \citet{Moment}):
\begin{equation} 
a^m_{\ell}  = \mathcal{R}  [  \alpha^{m'}_{\ell'} ]  \equiv \sum_{\ell'  m'}
   \mathcal{R}_{\ell,    \ell'}^{m,    m'}    (   \vartheta,    \varphi    )
   \alpha^{m'}_{\ell'} \,.
\end{equation} 
If  the  direction of the  spin  defines  a  preferential  plane of  accretion,  the
corresponding $  a_\ell^{m}$ will be systematically  enhanced.  We therefore
expect  the  equatorial direction  (which  corresponds  to  $m=0$ for  every
$\ell$) not to converge to zero.

We computed  the spherical harmonic  decomposition of $\rho  v_r(
\vartheta, \varphi  )$ given by  \Eq{defalm} for
the mass flux density of 25 000 haloes at $z=0$, up to $\ell=15$.  For each spherical
field of  the mass density flux,  we performed the rotation  that brings the
halo's spin along the z-direction to obtain a set of `centered' $a_{\ell}^{m}$
coefficients.  We also computed the related angular power spectrums $C_\ell$
:
\begin{equation}
C_\ell   \equiv   \frac{1}{4\pi}   \frac{1}{2\ell+1}   \sum_{m=-\ell}^{\ell}
|a_{\ell}^{m}  |^2=  \frac{1}{4\pi} \frac{1}{2\ell+1}  \sum_{m=-\ell}^{\ell}
|\alpha_\ell^m|^2.  \EQN{defCl}
\end{equation}
Let us  define the normalized  $\tilde a_{\ell}^{m}$ (or  harmonic contrast,
see appendix~\ref{s:statistics}),
\begin{equation}
{\tilde a_{\ell}^{m} }\equiv  \sqrt{4\pi}\frac{ a_{\ell}^{m} }{a_0^0} = \frac{
a_{\ell}^{m}}{\mathrm{sign}(a_0^0)\sqrt{C_0}}. \EQN{tildealm}
\end{equation}
This compensates for  the variations induced by our range  of masses for the
halo.
For  each $\ell$,  we  present in  Fig.   \ref{peakalm} the  median value,  $|
\langle   \mathrm{Re}\left\{\tilde   a_{\ell}^{   m}   \right\}\rangle|$   for
$\ell=2,4,6,8$ computed for 25 000 haloes.  All the $\tilde a_{\ell}^{m}$ have
converged toward zero,  except for the $\tilde a_{\ell  0}$ coefficients.  The
imaginary parts of  $\tilde a_{\ell}^{m}$ have the same  behaviour, except for
the ${\rm  Im}\{\tilde a_{\ell 0}\}$  coefficients which vanish  by definition
(not shown here).  The $m=0$ coefficients are statistically non-zero.  We find
$\langle{\tilde    a_{2}^0}\rangle=   -0.65    \pm    0.04$,   $\langle{\tilde
a_{4}^0}\rangle= 0.12  \pm 0.02$, $\langle{\tilde  a_{6}^0}\rangle= -0.054 \pm
0.015$ and  $\langle{\tilde a_{8}^0}\rangle= 0.0145 \pm  0.014$. {Errors stand
for the  distance between the 5th  and the 95th centile}.   The typical pattern
corresponding  an  $m=0$  harmonic  is  a  series of  rings  parallel  to  the
equatorial plane.  This confirms that the accretion occurs preferentially in a
plane perpendicular to the direction halo's spin.

The  spherical  accretion
contrast    $\langle\delta_{\rho   v_r}    (    {\vartheta},\phi   )\rangle$
can be reconstructed  using the  $\langle\tilde  a_{\ell}^{m}\rangle$
coefficients (as shown in the appendix): 
\begin{equation}
\delta_{[\rho  v_r]}(  \vartheta, \varphi  )=\sum_{\ell,  m}
 {\tilde  a^m_{\ell}}  Y_{\ell}^m  (  \vartheta, \varphi  )  -1.
\end{equation}
In  Fig.~\ref{profilalm}, the polar profile:
\begin{equation}
\langle  \delta_{[\rho  v_r]} (  {\vartheta}  )\rangle\equiv \sum_{\ell,  m}
 \langle{\tilde  a^m_{\ell}}\rangle  Y_{\ell}^m  (  \vartheta, 0  )  -1\,  ,
 \EQN{defalm2}
\end{equation}
of this  reconstructed spherical contrast  is shown.  This  profile has
been  obtained using  the  $\langle\tilde a_{\ell}^{m}\rangle$  coefficients
with $\ell \le 5$ and $\ell \le 15$. The contrast is large and positive near
$\vartheta =  \pi/2$ as expected  for an equatorial accretion.   The profile
reconstructed using $\ell \le 5$ is quite similar to the one using $\ell \le
15$.   This indicates that  most of  the energy  involved in  the equatorial
accretion is contained in a typical angular scale of 36 degrees (a scale
which is significantly larger than $\pi/20$ corresponding to the cutoff
frequency in our sampling of the sphere as mentionned earlier).

 Using  a spherical  harmonic expansion  of the  incoming mass  flux density
(\Eq{defalm}), we confirmed the excess  of accretion in the equatorial plane
found  above.  This  similarity was  expected since  these  two measurements
(using a ring or using a  spherical harmonic expansion) can be considered as
two different filterings of the  spherical accretion field as is demonstrated
in  the  appendix~\ref{s:statsphere}.   The   main  asset  of  the  harmonic
filtering resides in its relevance for the description of the inner dynamics
as is discussed in section~6.

\subsection{Summary}
To sum  up, the  two measurements of  section~3.2 and  3.3 (or 3.4)  are not
sensitive to the same effects.  The first measurement (involving the angular
momentum $\rho  {\bf L}$ at  the virial radius)  is mostly a measure  of the
importance of infalling matter in building the halo's proper spin.
The second and the third  measurements (involving the excess of accretion in
the equatorial  plane, $\delta_m$, using  rings and harmonic  expansion) are
quantitative measures of coplanar accretion.  The equatorial plane of a halo
is favoured  relative to  the accretion of  matter (compared to  an isotropic
accretion) to a level of $\sim 12  \%$ between $z=2$ and $z=0$.  Down to the
halo scale ($\sim 500$ kpc), anisotropy  is detected and is reflected in the
spatial configuration of infalling matter.
 

\section{Anisotropic infall of substructures}
                                                
To confirm and assess the detected anisotropy of the matter infall on haloes
in our simulations,  let us now move on to a  discrete framework and measure
related quantities  for satellites  and substructures.  In  the hierarchical
scenario, haloes are built up  by successive mergers of smaller haloes. Thus
if an anisotropy  in the distribution of infalling matter  is to be detected
it seems reasonable that this anisotropy should also be detected in the distribution
of satellites. The  previous galactocentric approach for the  mass flow does
not  discriminate between  an infall  of  virialised objects  and a  diffuse
material accretion  and therefore is  also sensitive to  satellites merging:
one would  need to consider, say,  the energy flux density.   However, it is
not   clear  if   satellites  are   markers  of   the  general   infall  and
\citet{Vitvitska}  did not  detect any  anisotropy at  a level  greater than
20$\%$.

The detection  of substructures and  satellites is performed using  the code
ADAPTAHOP which is described in  details in the appendix.  This code outputs
trees of  substructures in our  simulations, by analysing the  properties of
the local dark matter density in terms of peaks and saddle points.  For each
detected halo we can extract  the whole hierarchy of subclumps or satellites
and  their  characteristics. Here  we  consider  the  leaves of  the  trees,
i.e.  the  most elementary  substructures  the  haloes  contain.  Each  halo
contains a  `core' which is the  largest substructure in  terms of particles
number  and `satellites'  corresponding to  the smaller  ones.  We  call the
ensemble core +  satellites the `mother' or the  halo.  Naturally the number
of  substructures is  correlated with  the  mother's mass.   The bigger  the
number of substructures, the bigger  the total mass.  Because the resolution
in mass of our simulations is  limited, smaller haloes tend to have only one
or two satellites.
Thus in  the following  sections we will  discriminate cases where  the core
have less than  4 satellites.  A total of 50~000  haloes have been examined
leading to a total of about 120~000 substructures.


\subsection{Core spin - satellite orbital momentum correlations}
\label{s:Corespin_satmomentum}
\begin{figure}          
 \centering
\resizebox{7cm}{7cm}{\includegraphics{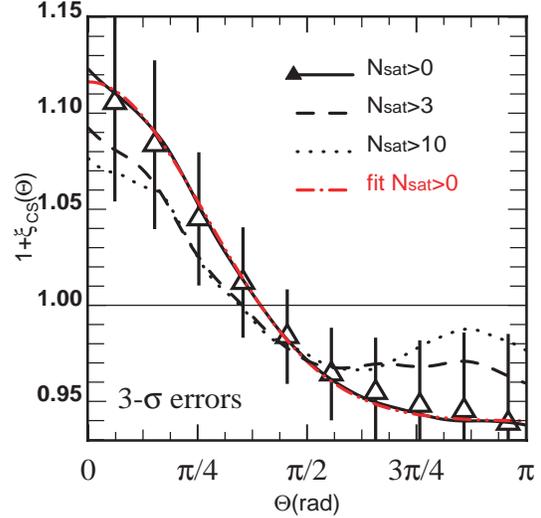}}
\caption{Excess probability, $1+\xi_\mathrm{CS}(\theta_{\rm  cs})$, of the angle between
the core's spin and the orbital momentum of satellites.  Cores have at least
1  satellite (solid  line), 4  satellites  (dashed line)  and 10  satellites
(dotted line).  These curves have  been normalized by the expected isotropic
distribution and the  Gaussian correction was applied to  account for errors
on  the  angle  determination.   Here  $\xi_\mathrm{CS}(\theta)=0$ is  expected  for  an
isotropic distribution  of angles between the  core 's spin  and the orbital
momentum.   All  satellites are  external  to the  core,  yet  an excess  of
alignment is  present.  The triangles represent the  angle distribution, the
error bars  stand for the  $3\sigma$-dispersion for 50 subsamples  of 10~000
satellites (out of 35 000) while  the red curve stands for the best gaussian
fit  of  the distribution  for  systems with  at  least  one satellite  (see
Eq.~\ref{gaussfit}  for parameterization).   The best  fit  parameters are:
$a_1=0.3993\pm0.0038,        a_2=0.0599\pm0.0083,       a_3=0.8814\pm0.0055,
a_4=0.9389\pm0.0002$.  The isotropic case is excluded with a good confidence
level, even for systems with a large number of satellites. }
\label{CosSub_refcore}
\end{figure} 
In the  mother-core-satellite picture, it is  natural to regard  the core as
the  central galactic  system, while  satellites  are expected  to join  the
halo from  the intergalactic medium. One way to  test the effect of
large scale  anisotropy is  to directly compare  the angle between  the core's
spin, ${\bf S}_c$, and the satellites' angular momentum, ${\bf L}_s$, relative to
the core.   These two angular momenta  are chosen since they  should be less
correlated  with  each other  than  e.g. the  haloes'  spin  and the  angular
momentum of  its substructures.  Furthermore,  particles that belong  to the
cores  are strictly  distinct from  those  that belong  to satellites,  thus
preventing any `self contamination'  effect.  As a final safeguard, we
took in account only satellites with  a distance relative to the core larger
than the mother's radius. The core's spin is:
\begin{equation}
{\bf S}_c=\sum_{p} ({\bf r_p-r_c})\times ({\bf v_p-v_c}),
\end{equation}
where $\bf r_p$ and $\bf v_p$ (resp.  $\bf r_c$ and $\bf v_c$) stands for particles' position
and velocities (resp. the core's centre of mass position and velocity) and where:
\begin{equation}
r_p<d_c,
\end{equation}
where $d_c$  is the core's radius.  The angular  momentum for a  satellite is
computed  likewise,  with  a  different selection  criterion  on  particles,
namely:
\begin{equation}
|{\bf r_p - r_s}| <d_s,
\end{equation}
where $\bf r_s$  stands for the satellite's centre of  mass position and $d_s$
is its radius.

Fig.  \ref{CosSub_refcore}  displays the reduced distribution  of the angle,
$\theta_{\rm  cs}$, between  the  core's spin  and  the satellites'  orbital
momentum, where $\theta_{\rm cs}$ is defined by:
\begin{equation}
\theta_{\rm cs}=\cos^{-1}   (\frac{{\bf  L}_s\cdot{\bf   S}_c}{|{\bf  L}_s||{\bf
S}_c|}).
\end{equation}
The Gaussian correction  was  applied as described in section 3.2, to take into
account the uncertainty on the determination of $\theta_{\rm cs}$.

The  correlation of  $\theta_{\rm cs}$ indicates  a preference  for  the aligned
configuration  with an  excess of  $\sim  12 \%$  of aligned  configurations
relative  to the  isotropic  distribution. We  ran Monte-Carlo  realisations
using  50  subsamples  of  10~000  haloes  extracted from  our  whole  set  of
substructures to  constrain the  error bars. We  found a $3\sigma$  error of
$6\%$: the detected  anisotropy exceeds our errors, i.e. $\xi_\mathrm{CS}(\theta_{\rm cs})$
is not uniform with a good confidence level.
The variations with the fragmentation  level (i.e.  the number of satellites
per system) remains within the error  bars.  The best fit parameters for the
measured   distributions  of  systems   with  at   least  1   satellite  are
$a_1=0.3993\pm0.0038,        a_2=0.0599\pm0.0083,       a_3=0.8814\pm0.0055,
a_4=0.9389\pm0.0002$  (see  Eq.~\ref{gaussfit}  for  parametrization).   Not
surprisingly, a  less structured  system shows a  stronger alignment  of its
satellites' orbital  momentum relative to  the core's spin.  In  the extreme
case of a  binary system (one core  plus a satellite), it is  common for the
two bodies to  have similar mass. Since the two  bodies are revolving around
each other, a  natural preferential plane appears.  The  core's spin will be
likely to be orthogonal to  this plane.  Increasing the number of satellites
increases  the  isotropy  of   the  satellites'  spatial  distribution  (the
distributions  maxima  are  lower  and   the  slope  toward  low  values  of
$\theta_{\rm cs}$ is  gentler), but switching from at  least 4 satellites to
at  least  10  satellites  per  system does  not  change  significantly  the
overwhole shape  distribution.  This suggests that  convergence, relative to
the  number  of satellites,  has  been  reached  for the  $\theta_{\rm  cs}$
distribution.

 As  the  measurements  of   the  anisotropy  factor  $\delta_m$  indirectly
suggested, satellites  have an anisotropic distribution  of their directions
around  haloes.   Furthermore  the  previous  analysis  of  the  statistical
properties of  $\delta$ (section  \ref{s:one-point}) indicated an  excess of
aligned configuration of $15\%$ which  is consistent with the current method
using substructures.  While  the direction of the core's  spin should not be
influenced  by the  infall  of matter,  we  still find  the  existence of  a
preferential plane for this infall, namely the core's equatorial plane.

\subsection{ Satellite velocity - satellite spin correlation}
\label{s:satvel}
The  previous sections compared  haloes' properties  with the  properties of
satellites.   In   a  galactocentric   framework,  the  existence   of  this
preferential plane could only be local.  In the extreme each halo would then
have its own  preferential plane without any connection  to the preferential
plane of the next halo.
Taking the satellite itself as a reference, we have analyzed the correlation
between  the  satellite average  velocity  in  the  core's rest frame  and  the
structures' spin.  Since part of  the properties of these two quantities are
consequences of  what happened outside the galactic  system, the measurement
of their alignment should provide  information on the structuration on scales
larger than the haloes scales, while sticking to a galactocentric point of view.

\begin{figure}          
 \centering
\resizebox{7cm}{7cm}{\includegraphics{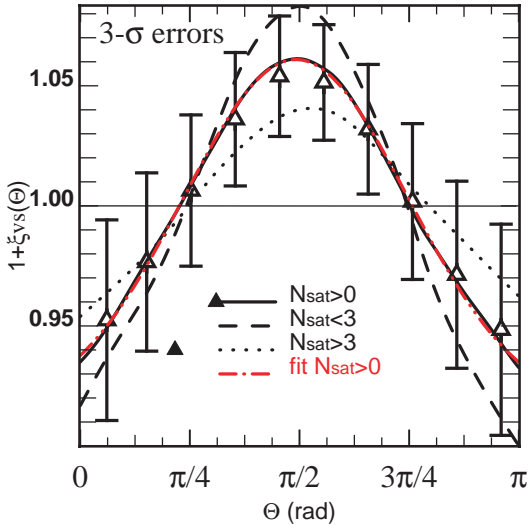}}
\caption{Excess probability, $1+\xi_{\mathrm{vs}}$, of the angle between the
substructures'  spin and  their velocities  in the  mother's rest  frame.  The
Gaussian  correction was  applied to  take into  account uncertainty  on the
angle  determination.   The  distributions  was  measured  for  all  mothers
(solid), mothers  with at  least 4 substructures  (dot) and mothers  with at
most  3  substructures  (dash).   The  triangles  represent  the  mean  angle
distribution.  The error bars represent the Monte Carlo $3\sigma$-dispersion
for  50 subsamples of  10~000 haloes  (out of  35~000).
The red dash-dotted curve stands for
the best fit  of the distribution with a Gaussian  function for systems with
at least 1 satellite (see Eq.~\ref{gaussfit} for parametrization).  The best
fit     parameters    are:     $a_1=0.2953\pm0.0040,    a_1=1.5447\pm0.0015,
a_2=0.8045\pm0.0059,  a_3=0.9144\pm0.0010$.  In the  core's rest  frame, the
satellites' motion  is orthogonal to  the direction of the satellites' spin. 
 This
configuration would fit in a picture where structures move along filamentary
directions.}
\label{spinvel}
\end{figure}

For each satellite, we extract  the angle, $\theta_{\rm vs}$, between the velocity
and  the  proper spin  and  derive  its  distribution using  the  Gaussian
correction  (see fig.  \ref{spinvel}).   The satellite's  spin ${\bf  S}_s$ is
defined by:
\begin{equation}
{\bf S}_s=\sum_{p} ({\bf r_p-r_s})\times ({\bf v_p-v_s}),
\end{equation}
where $\bf r_s$ and $\bf v_s$ stands for the satellite's position and velocity in the halo
core's rest frame. 
 The angle, $\theta_{\mathrm{vs}}$, between the satellite's spin and the
satellite's velocity is:
\begin{equation}
\theta_{\mathrm{vs}}=\cos^{-1}   (\frac{{\bf  S}_s\cdot{\bf   v}_s}{|{\bf  S}_s||{\bf
v}_s|}).
\end{equation}

Only satellites  external to the  mother's radius are considered while computing
the distribution of angles.  This leads to a sample  of about 40~000 satellites,
at redshift  $z=0$.  The  distribution $\xi(\theta_{\rm vs})$ was  calculated as
sketched in section 2. An isotropic distribution of $\theta_{\rm vs}$ would
as usual lead to a
uniform  distribution $\xi(\theta_{\rm vs})=0$.   The  result is  shown in  Fig.
\ref{spinvel}. The error bars were computed using the same Monte-Carlo simulations
described before with 50 subsamples of 10~000 satellites.

We obtain  a peaked  distribution with a  maximum for $\theta_{\rm vs}  = \pi/2$
corresponding to an excess of  orthogonal configuration of $5\%$ compared to
a random distribution of satellite  spins relative to their velocities.  The
substructures' motion  is preferentially  perpendicular to their  spin.  This
distribution of angles  for systems with at least 1  satellite can be fitted
by a Gaussian function with  the following best fit parameters (see Eq
\ref{gaussfit}):      $a_1=0.2953\pm0.0040,
a_1=1.5447\pm0.0015,    a_2=0.8045\pm0.0059,    a_3=0.9144\pm0.0010$.    The
variation  with the  mother's fragmentation  level is  within the  error bars.
However  the effect  of an  accretion orthogonal  to the  direction of the spin  is
stronger for satellites which belong to less structured systems. This may be
again related  to the case where  two comparable bodies  revolve around each
other, but from  a satellite point-of-view. The satellite  spin is likely to
be  orthogonal to  the revolution  plane  and consequently  to the  velocity's
direction.

This  result was  already known  for haloes  in  filaments (\citet{Falten}),
where  their motion  occurs along  the filaments  with their  spins pointing
outwards. The current results show  that the same  behaviour is measured
down  to the  satellite's scale.   However this  result should  be  taken with
caution since  Monte-Carlo tests  suggest that the  error (deduced  from the
$3\sigma$   dispersion)  is   about   $4\%$.   

This configuration where  the spins of haloes and  satellites are orthogonal
to  their motion  fit  with the  image of  a  flow of  structures along  the
filaments.  Larger structures are formed  out of the merging of smaller ones
in  a hierarchical  scenario.  Such small  substructures  should have  small
relative  velocities in order  to eventually  merge while  spiraling towards
each other. 
The  filaments correspond  to  regions  where most  of  the flow  is
  laminar, hence the merging between satellites is more likely to occur when
  one satellite  catches up with  another, while both satellites  move along
  the  filaments. During  such encounter,  shell crossing  induces vorticity
  perpendicular to the flow as was demonstrated in \cite{pichon}. This
  vorticity is then converted to momentum, with a spin orthogonal to the 
direction of the filament.


Finally,  the  flow  of matter  along  the  filaments  may also  provide  an
explanation for  the excess of  accretion through the equatorial  regions of
the virial sphere.  If a sphere is embedded in a `laminar' flow, the density
flux detected near  the poles should be smaller than  that detected near the
`equator'  of the  sphere. The  flux  measured on  the sphere  is larger  in
regions  where the normal  to the  surface is  collinear with  the `laminar'
flow, i.e.  the `equator'.  On the other  hand, a nil flux  is expected near
the poles since the vector normal  to the surface is orthogonal to direction
of the  flow.  The same effect is  measured on Earth which  receives the Sun
radiance: the temperature is larger on the Tropics than near the poles.  Our
observed excess of accretion through the equatorial region supports the idea
of a  filamentary flow  orthogonal to  the direction of  the halo's  spin
down to scales $\leq 500$ kpc. 

\section{Projected anisotropy}

\subsection{Projected satellites population }
\label{s:projsatpop}
\begin{figure}          
\centering
\resizebox{6cm}{6cm}{\includegraphics{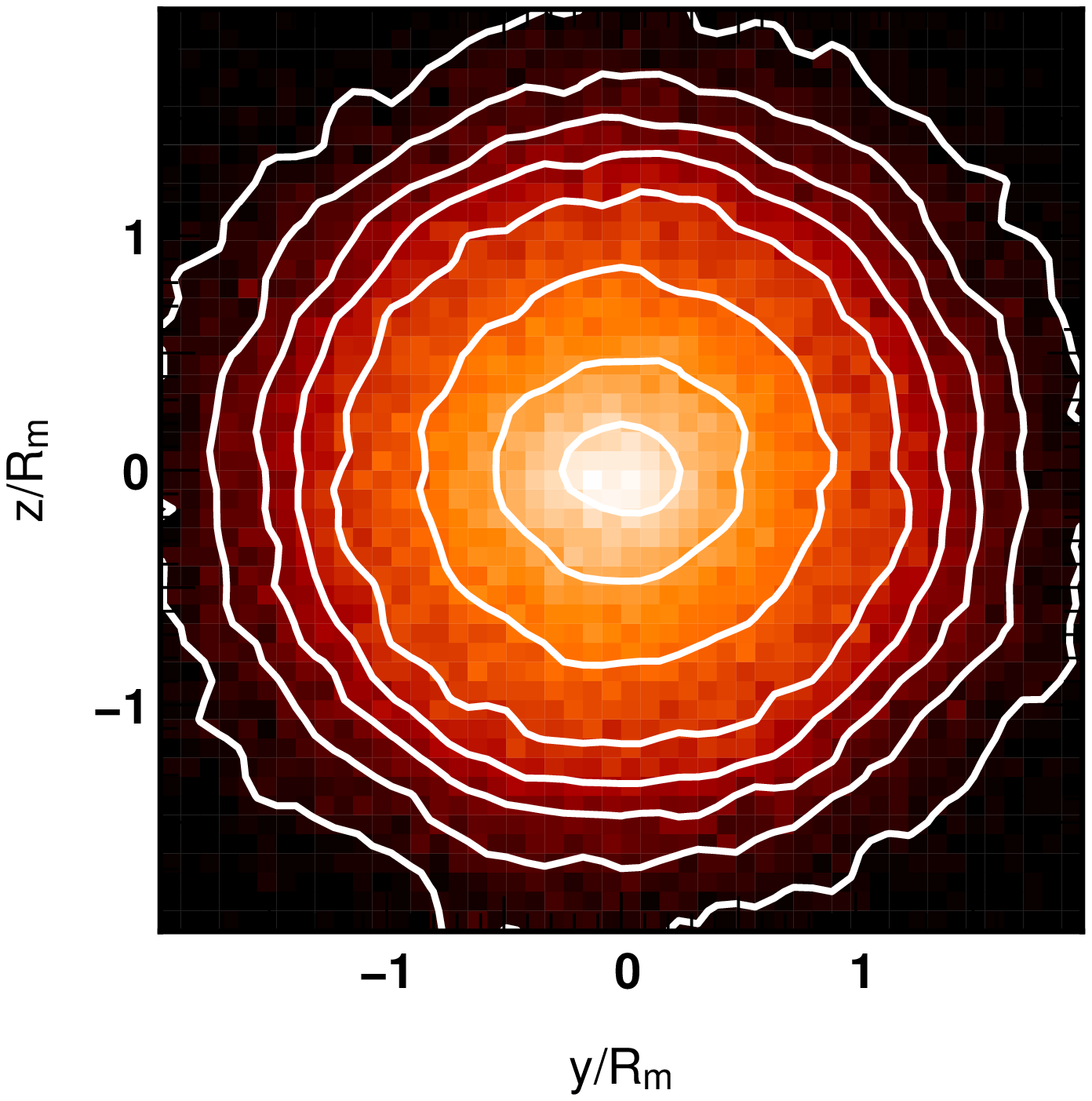}}
\resizebox{6cm}{6cm}{\includegraphics{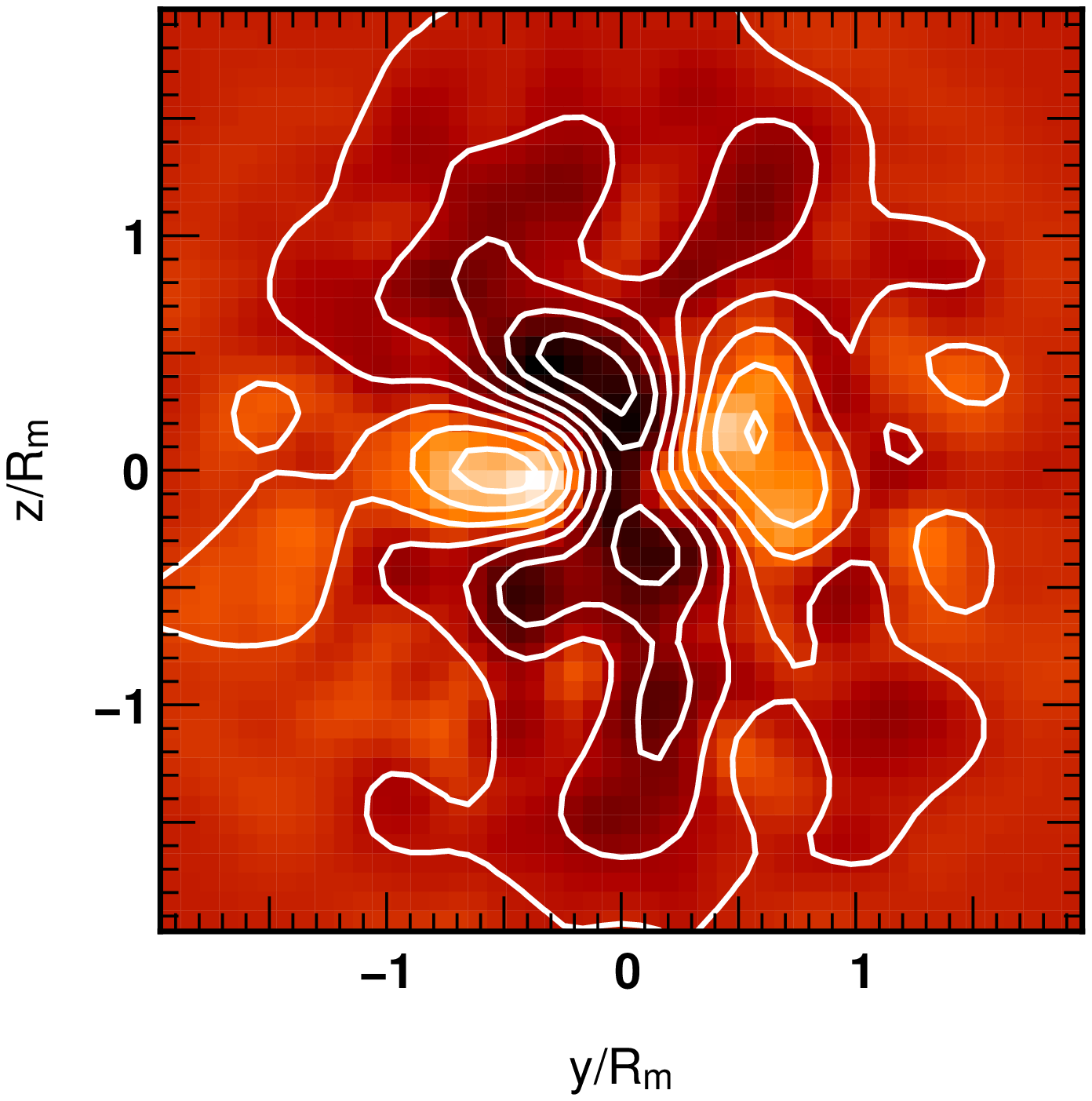}}
\caption{The projected  distribution of satellites around  the core's centre
of mass. We  used the position of 40~000  satellites around their respective
core  to produce a  synthetic halo+satellites  (a `mother')  system. The
projection is performed along the x-axis.  The y and z coordinates are given
in units of the  mother's radius.  The  z-axis  is colinear to  the direction
of the core's spin. \textit{Top:  }The isocontours of the number density
of satellites around the core's centre  of mass present a flattened shape. The
number of  satellites is  lower in  darker bins than  in lighter  bins.  The
flattened isocontours  indicate that  satellites lie preferentially  in the
plane orthogonal to   direction of the spin.  \textit{Bottom:} The excess number of
satellites surrounding the core.  We compared the distribution of satellites
measured  in our  simulations to  an isotropic  distribution  of satellites.
Light zones stand for an excess  of satellites in these regions (compared to
an isotropic distribution)  while dark zones stand for  a lack of
satellites.  
The satellites are more numerous in the equatorial region
than expected  in an isotropic  distribution of satellites around  the core.
Also, there are fewer satellites along the spin's axis than expected for an isotropic
distribution of satellites. }
\label{Triax1}
\end{figure}

We looked directly into  the spatial distributions of satellites surrounding
the haloes  cores to confirm the  existence of a preferential  plane for the
satellites  locations in  projection.  In  Fig.  \ref{Triax1},  we  show the
compilation  of the  projected positions  of satellites  in the  core's rest
frame.  The result is a synthetic galactic system with 100 000 satellites in
the same rest frame.  We performed suitable rotations to bring the spin axis
collinear to  the z-axis for  each system of  satellites, then we  added all
these systems to  obtain the actual synthetic halo  with 100~000 satellites.
The  positions were normalized  using the  mother's radius  (which is  of the
order of  the virial radius).   A quick analysis  of the isocontours  of the
satellite  distributions indicates  that satellites  are more  likely  to be
found in the equatorial plane,  even in projection.  The axis ratio measured
at  one mother's  radius is  $\epsilon(R_m)\equiv a/b-1=0.1$  with  $a>b$.  We
compared  this  distribution to  an  isotropic  `reference' distribution  of
satellites surrounding  the core.  This reference distribution  has the same
average  radial profile  as the  measured satellite  distributions  but with
isotropically distributed directions.  The  result of the substraction of the
two  profiles is  also  shown in  Fig.~\ref{Triax1}.   The equatorial  plane
(perpendicular to the z-axis) presents an excess in the number of satellites
(light regions).  Meanwhile, there  is a lack  of satellites along  the spin
direction (dark regions).  This  confirms our earlier results obtained using
the alignment of  orbital momentum of satellites with  the core's spin, {\it
i.e.}  satellites lie  more likely in the plane orthogonal  to the halo spin
direction.   Qualitatively,  these results  have  already  been obtained  by
\citet{Tormen},  where  the major  axis  of  the  ellipsoid defined  by  the
satellite's distribution  is found  to be aligned  with the  cluster's major
axis.  This synthetic halo is more directly comparable to observables since,
unlike the dark matter halo  itself, the satellites should emit light.  Even
though $\Lambda$CDM  predicts too many satellites,  its relative geometrical
distribution might still be correct.   In the following sections, our intent
is to quantify more precisely this effect.

The propension of satellites to lie in the plane orthogonal to the direction
of the  core's spin appears as an  `anti-Holmberg' effect.  \citet{Holmberg}
and  more  recently \citet{Zaritsky}  have  found  observationnally that  the
distribution of satellites around disks  is biased towards the pole regions.
Thus if the orbital momentum vector  of galaxies is aligned with the spin of
their parent haloes, our result  seems to contradict these observations. One
may  argue  that  satellites  are  easier  to detect  out  of  the  galactic
plane. Furthermore our measurements are  carried far from the disk while its
influence  is not  taken  in  account.  \citet{Huang}  have  shown that  the
orbital  decay and  the  disruption  of satellites  are  more efficient  for
coplanar orbits  near the disk. It  would explain the lack  of satellites in
the  disk plane.   Thus our  distribution of  satellites can  still  be made
consistent with the `Holmberg effect'.

\subsection{Projected satellite orientation and spin }
\label{s:projsat}
In addition to  the known alignment on large scales, we  have shown that the
orientation of structures on smaller scales should be different from the one
expected for a random distribution  of orientations.  Can this phenomenon be
observed ?  The  previous measurements were carried in 3D while this latter
type of observations  is performed in projection on  the sky. The projection
`dilutes'   the   anisotropy   effects  detected   using   three-dimensional
information.  Thus an effect of $15 \%$  may be lowered to a few percents by
projecting on  the sky. However, even  if the deviation from  isotropy is as
important as a few percents, as we will suggest, this should be relevant for
measurements involved in extracting a signal just above the noise level, such
as weak lensing.

To see the effect of projection  on our previous measurements, we proceed in
two steps.  First, every mother (halo core + satellites) is rotated to bring
the direction of the core's spin to the  z-axis. Second, every  quantity is computed
using only  the y and z components  of the relevant vectors, corresponding to a
projection  along  the  x-axis.   

The  first  projected measurement  involves  the  orientation of  satellites
relative to  their position  in the core's  rest frame. The  spin of a  halo is
statistically orthogonal to  the main axis of the distribution of  matter of that halo
(\citet{Falten}),  and   assuming  that  this  property   is  preserved  for
satellites, their spin $\bf S_s$  is an indicator of their orientation.  The
angle, $\theta_{\rm P}$ (\textit{in projection}),  between the satellites' spin and
their position vector (in the core's rest frame) is computed as follow:
\begin{equation}
\theta_{\mathrm{P}}=\cos^{-1}  (\frac{{\bf  S}_s^{y,z}\cdot{\bf  r}_{sc}^{y,z}}{|{\bf
S}_s^{y,z}||{\bf r}_{sc}^{y,z}|}),
\end{equation}
with
\begin{equation}
{\bf r}_{sc}={\bf r}_{s}-{\bf r}_{c},
\end{equation}
where ${\bf  r}_{s}$ and ${\bf  r}_{c}$ stand respectively for  the position
vector  of  the  satellite  and  the  core's centre of  mass.   Two  extreme
situations  can be imagined.   The `radial'  configuration corresponds  to a
case where  the satellite's main axis  is aligned with the  radius joining the
core's centre of  mass to the satellite centreof  mass (spin perpendicular to
the radius, or $\theta_{\rm P} \sim  \pi/2$).  The `circular' configuration is the
case  where  the satellite  main  axis is  orthogonal  to  the radius  (spin
parallel to the radius, $\theta_{\rm P}  \sim 0 [\pi]$).  These reference configuration will be
discussed in what follows.

\begin{figure}          
\centering
\resizebox{7cm}{7cm}{\includegraphics{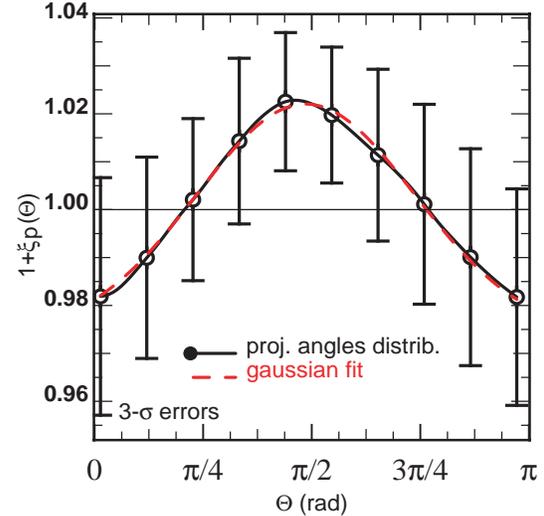}}
\caption{ Excess  probability, $1+\xi_P$, of the projected angles  between the direction
of the  spin of substructures and  their position vector in  the core's rest
frame  .  The  projection  is made  along  the x-axis  where  the z-axis  is
concurrent with  the core's spin  direction.  The solid line  represents the
average  distribution  of  projected  angles  of  50  subsamples  of  50~000
substructures  (out of  100~000  available substructures).   The error  bars
represents the  $3\sigma$ -dispersion relative  to these 50  subsamples.  An
isotropic distribution of  orientation would correspond to a  value of 1 for
$1+\xi_{\rm P}$.  The  projection plus the reference to  the position vector
instead of the  velocity's direction lowers the anisotropy  effect.  The red
curve  stands for  the  best Gaussian  fit  of the  excess probability  (see
Eq.~\ref{gaussfit}  for  parameterization).  The  best  fit parameters  are:
$a_1=0.0999\pm   0.0030,   a_2=1.5488\pm   0.0031,   a_3=0.8259\pm   0.0131,
a_4=0.9737\pm 0.0007$. It seems that on average the projected orientation of
a substructure is orthogonal to its projected position vector.}
\label{proj}
\end{figure}

The  resulting distribution,  $1+\xi_{\rm P}(\theta_{\rm  P})$, is  shown in
Fig.  \ref{proj}. As before, an isotropic distribution of orientations would
lead to $\xi_{\rm P}(\theta_{\rm  P})=0$.  The distribution is computed with
100  000 satellites, without  the cores,  while the  error bars  result from
Monte-Carlo simulations on 50 subsamples of 50~000 satellites each. As compared to
the   distribution  expected   for  random   orientations,   the  orthogonal
configuration is present in excess of $\xi_{\rm P}(\pi/2) \sim 0.02$. If the spin of
satellites is  orthogonal to their  principal axis, the direction  vector in
the core's  rest frame  is more aligned  with the satellites  principal axes
than  one   would  expect  for   an  isotropic  distribution   of  satellites'
orientations.  This   configuration is  `radial'.  The peak  of the
distribution  is  slightly  above  the error  bars:  $\Delta  \xi_{\rm P}(\theta_{\rm
P}\sim\pi/2) \sim  0.02$.  The  distribution can be  fitted by  the Gaussian
function  given   in  Eq.~\ref{gaussfit}  with   the  following  parameters:
$a_1=0.0999  \pm  0.0030,  a_2=1.5488  \pm 0.0031,  a_3=0.8259  \pm  0.0131,
a_4=0.9737  \pm 0.0007$.   The alignment  seems to  be difficult  to detect  in
projection.  With 50 000 satellites, we barely detect the enhancement of the
orthogonal configuration  at the 3-$\sigma$ level,  thus we do  not expect a
detection  of  this  effect at  the  1-$\sigma$  level  for less  than  6000
satellites.  Nevertheless,  the distribution of  the satellites' orientation
in projection seems to be `radial'
on dynamical grounds, without reference to a lensing potential.

\begin{figure}          
\centering
\resizebox{9cm}{9cm}{\includegraphics{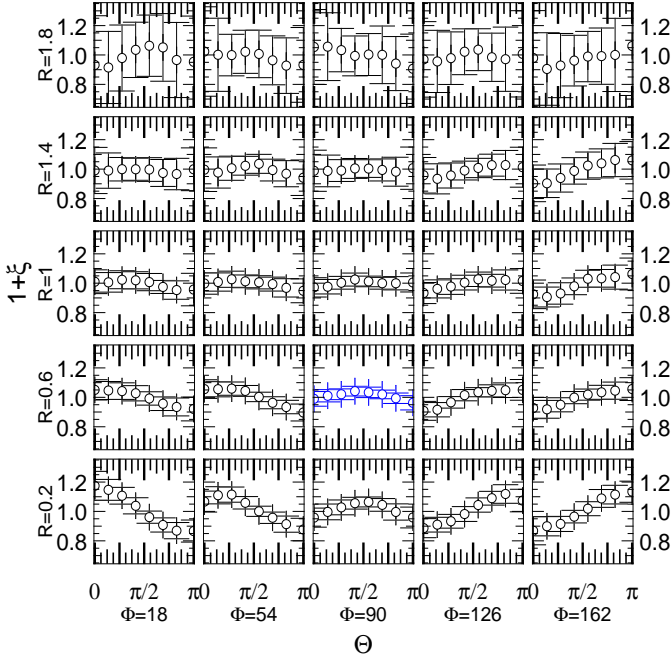}}
\caption{ Radial  and azimuthal grid of the  excess probability, $1+\xi_{\rm
P}$,  of  the  projected  angles  between  the  direction  of  the  spin  of
substructures and  their direction relative  to the central position  of the
core (as shown on average  in Fig~\ref{proj}).  The projection is made along
the x-axis  where the z-axis is  concurrent with the direction of the  core's spin.
Each  row represents a  distance relative  to central  core in  the mother's
radius units  (from bottom to  top): $R\in[0,0.4[$, $R\in  [0.4,0.8[$, $R\in
[0.8,1.2[$, $R \in [1.2,1.6[$ and $R\in [1.6,2[$.  Each column represents an
angular distance (in  degrees) relative to the direction  of the core's spin
(z-axis)  : $\phi_s\in  ]0,36]$,  $\phi_s\in[36,72[$, $\phi_s\in  [72,108[$,
$\phi_s\in[108,144[$  and $\phi_s\in[144,180[$.   The  isotropic orientation
distribution  corresponds  to  a  value   of  1.   Each  sector  presents  a
preferential direction  that depends  on its position  relative to  the spin
direction  of the  central core.   The distributions  are computed  using 50
samples of  50~000 satellites each.   In each sector, the  points represents
the distribution  averaged over the  50 samples.  The error  bars represent
the  $3\sigma$-Monte Carlo  dispersion  of the  distribution  over these  50
samples.}
\label{lensmap}
\end{figure}

\begin{figure}          
    \centering
\resizebox{8cm}{8cm}{\includegraphics{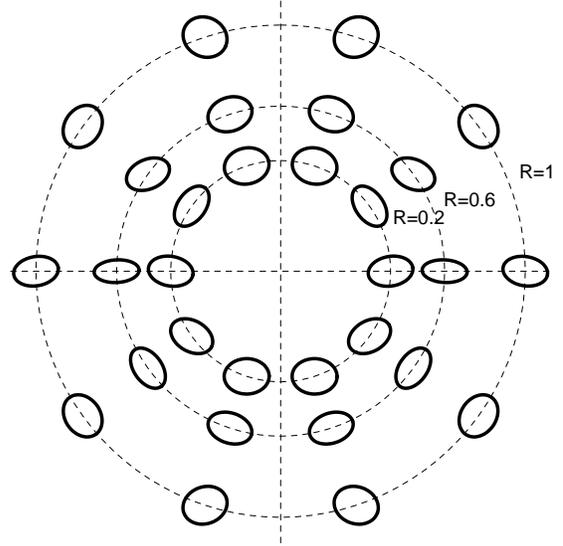}}
\caption{  Geometric  configuration of  mean  satellites  around their  core
galaxy; each panel of Fig.~\ref{lensmap}  is represented by an ellipse at its
log radius and angle around the core galaxy. The axis ratio of the ellipses
is  proportional  to  the  peak-to-peak  amplitude  of  the  corresponding
correlation  (accounting for  the relative  SNR), while  its  orientation is
given by the orientation of the maximum of $1+\xi_{\rm P}$. }
\label{lensmap2}
\end{figure}

Our previous  measurement was `global' since  it does not  take into account
the possible  change of orientation with  the relative position of  the satellites in
the core's rest frame.  In  Fig.  \ref{lensmap}, we explore the evolution of
$1+\xi_{\rm P}$ with the radial distance relative to the core's centre of mass
and with the angular distance relative  to the z-axis, i.e.  relative to the
direction of  the core's spin.  The  previous synthetic halo  was divided in
sectors and for  each sector, $1+\xi_{\rm P}$ can  be computed.  The sectors
are thus defined by their radius (in the mother's radius units): $R\le0.4$,
$0.4<R\le0.8$,  $0.8<R\le1.2$, $1.2<R\le1.6$ and  $1.6<R\le2$ and  by their
polar  angle relative  to the  direction of  the core's  spin  (in degrees):
$\phi_s\le36$,  $36<\phi_s\le72$, $72<\phi_s\le108$,  $108<\phi_s\le144$ and
$144<\phi_s\le180$.  Each of the previous Monte-Carlo subsamples can also be
divided into  sectors in  order to compute  the dispersion $\sigma$  for the
distributions  within the subsamples.   The error  bars still  represent the
$3\sigma$ dispersions.

The  Fig. \ref{lensmap2}  is  a qualitative  representation  of the  results
presented  in   Fig.  \ref{lensmap}.   Each  sector  with   $R\le1$  in  Fig.
\ref{lensmap}  is represented  by an  ellipse at  its actual  position.  The
orientation  of the ellipse  is given  by the  angle of  the maximum  of the
corresponding $1+\xi_{\rm P}(\theta_p)$  function.  We  chose to represent  the spin's
direction perpendicular to  the ellipse's major axis.
We also chose to  scale the ellipse axis ratio with  the signal-to-noise ratio of
$1+\xi_{\rm P}(\theta_p)$.   Indeed large  errors leads  to  weak  constrains  on the  spin
orientation and the galaxy would  be seen as circular on average. Conversely
a strongly constrained orientation leads to a typical axis ratio of 0.5.

Two  effects seem  to emerge  from this investigation.  For  some  sectors, the
orthogonal configuration is in  excess compared to an isotropic distribution
of satellites' orientation relative to the radial vector. This seems to be
true especially for radii smaller than  the mother's radius but the effect is
still present at larger distances,  especially near $\phi_s \sim \pi/2$. Switching from  low values to high  values of $\phi_s$ changes  the slope of
the  $1+\xi_{\rm P}(\theta)$  distribution. This  may  be  a marker  of  a  `circular
configuration'  of the orientation  of  satellites.

The existence of  a `radial' component in the  orientation of the satellites
was expected,  both from the  unprojected measurements made in  the previous
sections  and from  the  global distribution  extracted  from the  projected
data. The fact that the `radial' signature is stronger around the equatorial
plane ($72<\phi_s<108$ in Fig. \ref{lensmap}) may be an another evidence for
a filamentary  flow of  satellites, even in  projection.  It seems  that the
existence  of a  `circular'  component  was mostly  hidden  in the  previous
measurements by the dominant signature of the `radial' flow. Nevertheless, the
dominance of 'circular' orientations near the poles fits with the
picture of a halo surrounded by satellites with their spin pointing
orthogonally to the filament directions. 

The `circular' flow  may alternatively be related to  the flow of structures
around  clusters located  at  the connection  between  filaments. There  are
observations of  such configurations  (Kitzbichler $\&$ Saurer  2003), where
galaxies have  their spin  pointing along their  direction of  accretion and
these observations could be consistent with our `circular' component.



\section{Applications}
\label{s:applications}

Let  us give  here a  quick  overview of  the implications  of the  previous
measurements for the inner dynamics of the halo down to galactic scales.  In
particular let us see how the self-consistent dynamical response of the halo
propagates anisotropic  infall inwards,  and then briefly  and qualitatively
discuss implications  of anisotropy to  galactic warps, disk  thickening and
lensing.

\subsection{Linear response of galaxies}
\label{s:kalnajs}
In  the spirit  of  e.g.   \citet{Kalnajs} or  \citet{Tremaine}  we show  in
appendix~\ref{s:linearresponse}  and   elsewhere  (\citet{Aubert1})  how  to
propagate dynamically the perturbation from  the virial radius into the core
of  the  galaxy  using  a  self consistent  combination  of  the  linearized
Boltzmann and  Poisson equations under the  assumption that the  mass of the
perturbation  is small compared  to the  mass of  the host galaxy.   Formally, we
have:
\begin{equation}
{\mathbf                      r}({\mathbf                     x},t)={\mathbf
R}[F,\MG{\Omega},\M{x},t-\tau]\left({{\varpi}} (  \MG{\Omega},\tau ) \right)
\,,
\label{decomp}
\end{equation}
where  $\mathbf R$ is  a linear  operator which  depends on  the equilibrium
state  of  the  galactic  halo  (+disk) characterized  by  its  distribution
function  $F$, and  $\M{r}({\mathbf  x},t)$ represents  the self  consistent
response of the inner halo at time t due to a perturbation $ {{\varpi}} (
\MG{\Omega},\tau )$ occuring at time $\tau$.  Here $\varpi$ represents formally the perturbed
potential on  the virial sphere and  the flux density  of advected momentum,
mass and kinetic  energy at $R_{200}$. A `simple'  expression for $\M{R}$ is
given in  Appendix~\ref{s:response} for the self  consistent polarisation of
the halo.   The linear operator, $\M{R}$,  follows from \Eq{defexp},\Ep{ap3}
and \Ep{sourceexpr}.  These equations generalize the work of Kalnajs in that
it accounts for a consistent infall  of advected quantities at the
outer   edge    of   the   halo.     It   is   shown   in    particular   in
appendix~\ref{s:response}  that self-consistency  requires the  knowledge of
all    ten    (scalar,   vector    and    symmetric    tensor)   fields    $
\varpi_{\rho}(\bO,\tau),\varpi_{\rho \bfv }(\bO,\tau), \varpi_{\rho \sigma_i
\sigma_j }(\bO,\tau) $.

\begin{figure}          
    \centering
\resizebox{8cm}{8cm}{\includegraphics{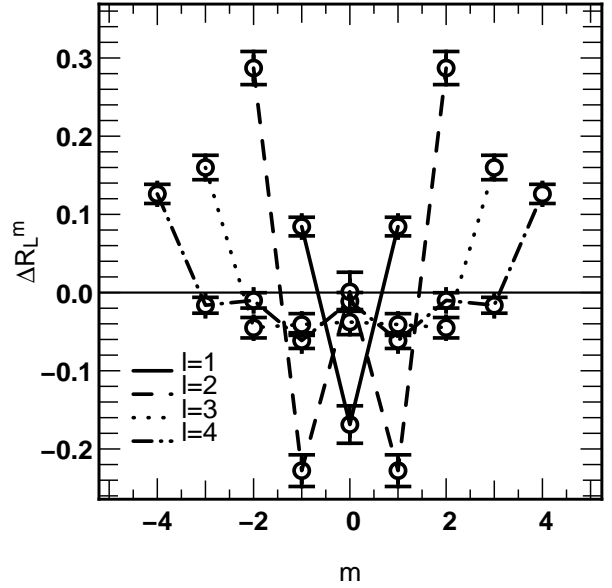}}
\caption{ The  residual anisotropic  harmonic power spectra,  $\Delta \tilde
R_{\ell}^{m}$,  introduced  in  \Eq{defDeltaR}  as  a function  of  $m$  for
$\ell=1,2,3,4$. These  residuals will serve  as input to the  computation of
the dynamical response of the halo.  }
\label{f:YLM}
\end{figure}

When dealing with disk broadening,  $\M{r}$ could be the velocity orthogonal
to the plane of the disk, or,  for the warp, its amplitude, as a function of
the position  in the disk, $\M{x}$ (or  the orientation of each  ring if the
warp is described as concentric rings).  More generally, it could correspond
to  the  perturbed  distribution  function  of  the  disk+halo.   The  whole
statistics of $\M{r}$ is  relevant. The average response  $\langle{\mathbf r}({\mathbf
x},t)\rangle$ can be written as:
\begin{equation}
\langle{\mathbf r}({\mathbf x},t)\rangle={\mathbf R}\langle \varpi(\MG{\Omega},\tau
)\rangle = \sum_{\ell m} {\mathbf R}  Y_{\ell}^{  m}(\MG{\Omega}) \langle a^m_{\ell} \rangle
\end{equation}
Since the accretion is anisotropic,  $\langle a^m_{\ell}\rangle$ do not
converge toward zero (see section \ref{s:expansion}) inducing a non-zero
average response.   
Most  importantly the  two  point
correlation of  the response  since it will  tell us qualitatively  what the
correlation length and  the root mean square amplitude  of the response will
be.  For  the purpose of  this section, and  to keep things simple,  we will
ignore   temporal  issues  (discussed   in  Appendix~\ref{s:linearresponse})
altogether,  both  for  the  mean  field and  the  cross-correlations.   The
two-point correlation  of $\M{r}({\mathbf x})$ then depends  linearly on the
two-point correlation of ${\varpi}$:
\begin{equation}
\langle    {{\mathbf     r}({\mathbf     x})} \cdot  \T{\mathbf     r}({\mathbf
y})\rangle= {\mathbf R}  \langle   {{\varpi}}(\MG{\Omega}) \cdot  \T{{\varpi}}
(\MG{\Omega}') \rangle \T{{\mathbf   R}} \,,
\end{equation}
where $\T{}$ stands for the transposition.  Clearly, if the infall, $\varpi(
\MG{\Omega}$),  is anisotropic  the response  will be  anisotropic.   As was
discussed in section~3.4  when the infall is not isotropic, we have
\begin{equation}
\langle    |\tilde    a^m_{\ell}|^2    \rangle    \neq    \langle    |\tilde
 \alpha^m_{\ell}|^2   \rangle=   \frac{1}{2\ell+1}   \sum_{m=-\ell}^{m=\ell}
 \langle |\tilde \alpha^m_{\ell}|^2 \rangle \,,
\end{equation}
Let us therefore introduce:
\begin{equation}
    \Delta  \tilde  R_{\ell}^{m}  \equiv \langle  |\tilde a^{m}_{\ell}|^2  \rangle-
\langle |\tilde \alpha^{m}_{\ell}|^2 \rangle, \EQN{defDeltaR}
\end{equation} 
which   would  be   identically  zero   if  the   field   were  stationary
on the sphere.
Here $\Delta\tilde  R_{\ell}^{m}$  represents the  anisotropic  excess for  each
harmonic  correlation.   In  particular,  the  excess  polarisation  of  the
response induced by the anisotropy reads
\begin{equation}
\Delta  \langle  {{\mathbf  r}({\mathbf  x})} \cdot  \T{\mathbf  r}({\mathbf
y})\rangle   =  \sum_{\ell   m}  {\mathbf   R}   Y_{\ell}^{  m}(\MG{\Omega})
\Delta\tilde    R_{\ell}^{m}    Y_{\ell}^{    m}(\MG{\Omega}')    \T{\mathbf
R}\,.\EQN{defcorR}
\end{equation}
Fig.~\ref{f:YLM}    displays    $\Delta    \tilde    R_{\ell}^{m}$,    for
$\ell=1,2,3,4$.  The different $\Delta \tilde R_{\ell}^{m}$ clearly converge
toward different  non-zero values. Consequently the  response should reflect
the anisotropic nature of the external perturbations.

 It is beyond the scope of this paper to pursue the quantitative exploration
of the response of the inner  halo to a given anisotropic infall, since this
would require an explicit expression  of the response operator, $\M{R}$, for
each dynamical problem investigated. 
%

\subsection{Implication for warps, thick disks and lensing}
\label{s:warpthicklens}

In this paper,  the main emphasis is on measured  anisotropies. It turns out
that it never  exceeds 15 $\%$ in accretion. For a  whole class of dynamical
problems  where anisotropy  is  not the  dominant  driving force  it can  be
ignored at that level.  Here we now discuss qualitatively the implication of
the previous measurements to galactic warps, the thick disk and weak lensing
where anisotropy is essential.

\subsubsection{Galactic warps}
\label{s:warp}
The action of  the torque applied on  the disk of a galaxy  is different for
different angular and radial  position of the perturbation. Consequently the
warp's   orientation  and  its   amplitude  are   functions  of   the  spatial
configuration of  the external potential.  For  example, \citet{Lopez} found
that the  warp's amplitude due  to an intergalactic  flow is dependent  on the
direction  of  the  incoming  `beam'  of matter.  
Having modelled the intergalactic flow  applied to the Milky Way, they found
that the warp amplitude rises steeply as the beam leaves the region coplanar
to the disk and this warp  amplitude reaches a maximum for an inclination of
30  degrees  relative  to the disk's   plane.   As  the  beam  direction  becomes
perpendicular to  the galactic plane,  the warp amplitude  decreases slowly.
In this  context, the existence of  a typical spatial  configuration for the
incoming  intergalactic matter  or satellites  infall may  induce a  kind of
'typical' warp in the disk of galaxies.

The existence of a preferential  plane for the accretion of angular momentum
also implies  that the  recent evolution  of the halo's  spin has  been rather
smooth. \citet{Bullock} have shown that the angular momentum tends to remain
aligned within  haloes.  Furthermore, the  accretion of matter by  haloes is
preferentially performed on plunging radial  orbits, thus the inner parts of
haloes  are  aware  of  the  properties of  the  recently  accreted  angular
momentum.
Therefore, a  disk embedded in the  halo would also  `feel' this anisotropic
accretion. \citet{OsBin}  have shown that  the misalignment of  the accreted
angular momentum and the disk's spin forces the latter to slew the symmetry
axis of its inner parts.
The warp  line of nodes  is also found  to be aligned  with the axis  of the
torque applied  to the  disk.  As stressed  by \citet{Binney},  non straight
line  of nodes  can  be associated  with  changes in  the  direction of  the
accreted angular  momentum.  Using a  sample of 12  galaxies, \citet{Briggs}
established rules  of thumb for galactic  warps, one of them  being that the
line of nodes is  straight in the inner region of disk  while it is wound in
the outer  parts.  If the angular  momentum is accreted  along a stationary
preferential direction, as we suggest,  the warp line of nodes should remain
mostly straight.  However, if the  accretion plane differs slightly from the
disk  plane,  more than  one  direction  of  accretion become  possible  (by
symmetry around the  vector defining the disk plane)  and, as a consequence,
different directions are possible for the torque induced by accreted matter.
We  may then  consider  a varying  torque  along accretion  history with  an
accreted angular momentum `precessing' around the halo's spin but close to its
direction.  In  this scenario, the difference  in the behaviour  of the warp
line of  nodes between  the inner and  outer regions  of the galaxies  may be
explained. 
\subsubsection{Galactic disk thickening}
Thin galactic disks put serious constraints on merging scenarii, since their
presence implies a  fine-tuning between the cooling  mechanisms ({\sl e.g.}
coplanar  infall  of gas),  and  the  heating  processes (merging  of  small
virialised objects, deflection of spirals  on molecular clouds). It has been
shown  that small  mergers can  produce a  thick disk  (e.g.  \citet{Quinn},
\citet{Walker} ).  However,  the presence of old stars  within the thin disk
cannot  be explained  in  the framework  of  the merging  scenario unless  a
fraction of  the accretion  took place within  the equatorial plane  of the
galaxy. Furthermore, the geometric  characteristic of the infall is essential
in the formation  process of a thick disk.   In \citet{Velazquez}, numerical
simulations of interactions between  galactic disks and infalling satellites
show  that  the heating  and  thickening  is  more efficient  for  coplanar
satellites.   They  also stressed  the  differences  between  the effect  of
prograde  or  retrograde orbits  of  infalling  satellites  (relative to  the
rotation of the disk): prograde  orbits induce disk heating while retrograde
orbits  induce disk tilting.   Our results  indicate that  the  infall is
preferentially  prograde and  coplanar relative  to  the halo's  spin: if  we
consider an alignment between the halo's spin and the galaxy's angular momentum,
the thickening  process may be  more efficient than  the one expected  in an
isotropic configuration  of infalling matter.  Furthermore,  our estimate of
the fraction of coplanar accretion at  the virial scale may be considered as
a lower bound  near the disk since the presence of
a  disk will  focus  the infall  closer  to the  galactic  plane.  In  fact,
\citet{Huang} found that the disk tends  to tilt toward the orbital plane of
infalling  prograde  low-density  satellites. This  effect  would  also
contribute  to enhance  the excess  of  coplanar accretion down to   galactic
scales.

However the nature of infalling virialised objects was shown to affect their
ability to heat or destroy  the disk.  \citet{Huang} found that the presence
of low density satellites should induce preferentially a tilting of the disk
instead of  a thickening: one needs  to enhance the relative  mass of the
satellite ($\sim 30\%$ of the disk mass) to produce an observable thickening
in the inner parts of the galaxy. Unfortunately such a massive satellite has
a  destructive impact  on the  outer parts  of the  disk.   The relationship
between the excess of accretion and the satellite mass should be constrained
but  our  limited  mass  resolution  prevents  us  from  performing  such  a
quantitative analysis.  We should  therefore aim at achieving higher angular
resolution  of the  virial sphere  and higher  mass resolution  in  order to
describe well compact virialised objects.

\subsubsection{Gravitational lensing}


The  first detection  of  cosmic shear  was reported  by  four different
groups in 2000 (\citet{Bacon}, \citet{Kaiser}, \citet{VW}, \citet{Wittman}).
One  of the  basic assumptions  made  by cosmic  shear studies  is that  the
intrinsic  ellipticities of galaxies  are expected  to be  uncorrelated, and
that  the observed correlations  are the  results of  gravitational lensing
induced  by  the large  scale  structures  between  those galaxies  and  the
observer.   Hence,   the  detection  of   weak  lensing  signal   assumes  a
gravitationally  induced  departure  from  a  random  distribution  of  the
galactic  shapes.  Consequently,  if  there exists  intrinsic alignments  or
preferential patterns in galactic orientations, this would potentially affect
the  interpretation from  weak  lensing measurements.   Several papers  have
already  considered  the  `contamination'  of  the weak  lensing  signal  by
intrinsic galactic alignment. Using analytic arguments, \citet{Catelan} have
shown that such alignments should exist.   The issue of the amplitude of the
intrinsic  correlations compared to  the correlation  induced by  the cosmic
shear  has also  been explored  by \citet{Croft}  and  \citet{Heavens}.  The
`intrinsic' correlations  may overcome  the shear-induced signal  in surveys
with a narrow redshift range.
We have shown that the orientation of satellites around haloes is not randomly
distributed, which  is a  clear indication of  intrinsic correlations  for our
considered  scales ($\sim$  500  kpc).   Taking $z_m=1$  as  a typical  median
redshift  for large  lensing surveys,  the  corresponding angular  scale is  1
arcminute  in  our  simulations'  cosmogony.   Furthermore,  the  prospect  of
studying  the  redshift  evolution   of  gravitational  clustering  via  shear
measurements will  require investigating narrower  redshift bins and  as such,
small  scale  dynamically induced  polarisation  might  become  an issue.   As
recommended  by  \citet{Catelan},  our  measurement  may also  be  used  as  a
`numerical' calibration of the  relation between ellipticity and tidal fields.
Interestingly, they suggested to compensate  for the finite number of galaxies
around  clusters  by  `stacking'  several  clusters, which  is  precisely  the
procedure we  followed to extract  signal from our simulations.  Finally, Weak
lensing predicts  no `curl' component  in the shear field  (e.g.  \citet{Pen})
and such  `curl' configurations would  serve to extract the  intrinsic signal.
Even though satellites exhibit  both `circular' and `radial' configurations in
our simulations, we do not observe  a clear signature of a 'curl' component of
orientations at our level of detection.

\section{Conclusion $\&$ Prospects}

\begin{figure*}          
    \centering
\resizebox{15cm}{10cm}{\includegraphics{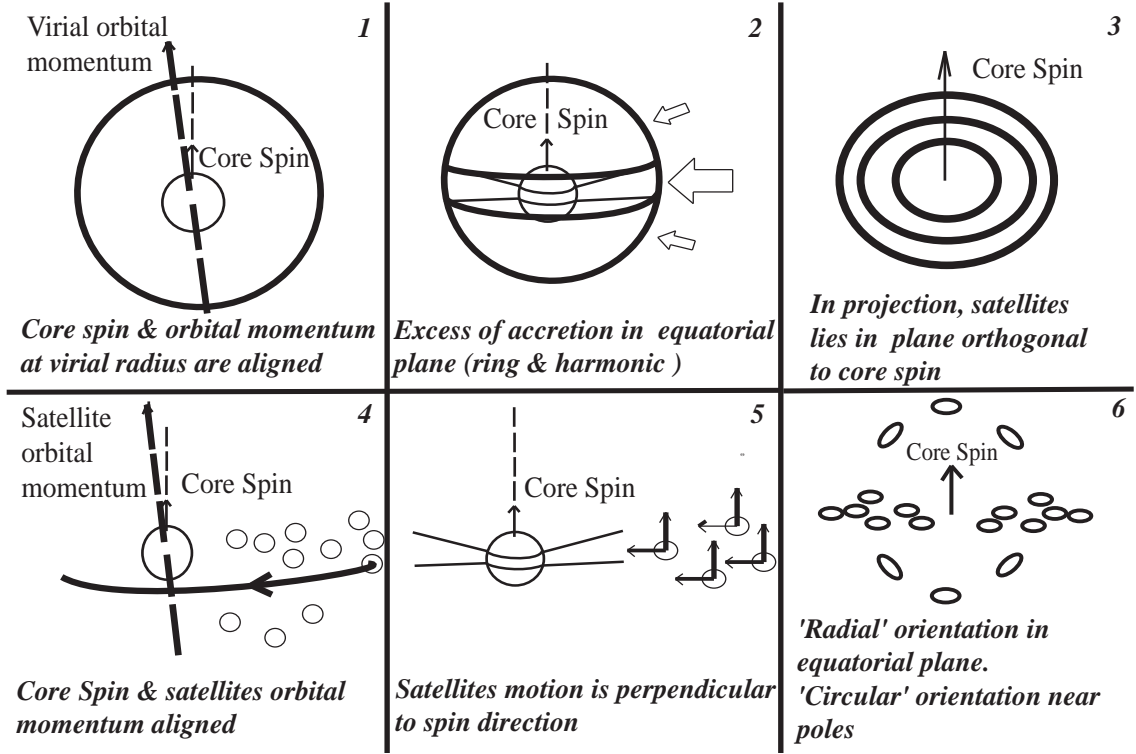}}
\caption{A schematic representation of all estimate of anisotropic accretion
considered in  this paper.   {\bf 1)}: we  measured the distribution  of the
angle  between the  orbital momentum  on the  virial sphere  and  the halo's
spin. The average  orbital momentum measured on the  virial sphere is mostly
aligned with the  spin of the halo embedded in  the virial sphere (discussed
in section \ref{s:two-points}).  {\bf 2)}:  we compared the accretion in the
plane  orthogonal to  the  direction of  the  halo's spin  with the  average
accretion on  the sphere.  On  the virial sphere,  we detected an  excess of
ring-like  or  harmonic accretion  in  the  equatorial  plane (discussed  in
section \ref{s:one-point}  and \ref{s:expansion}). {\bf  3)}: in projection,
we  used  a `synthetic'  halo  to look  at  the  distribution of  satellites
detected  with ADAPTAHOP  and at  the orientation  of their  spin  around the
direction of the spin.  In  projection, satellites lie preferentially in the
projected equatorial plane  (discussed in section \ref{s:projsatpop}).  {\bf
4)}: We measured the angle between  the halo's spin and the orbital momentum
of  each satellite.  The  orbital momentum  of satellites  is preferentially
aligned  with  the  spin  of   their  hosting  core  (discussed  in  section
\ref{s:Corespin_satmomentum}).   {\bf 5)}:  We compared  the  orientation of
each  satellite  velocity  vector  (in  the  core's  rest  frame)  with  the
orientation of  their own spin.  The  velocity vector of  satellites (in the
core's rest frame)  is orthogonal to the direction  of their spin (discussed
in  section  \ref{s:satvel}).   {\bf  6)}:  In  the  equatorial  plane,  the
projected  orientation  of  satellites  is  more 'radial',  while  near  the
direction of  the spin  a 'circular' configuration  of orientation  seems to
emerge (discussed in section \ref{s:projsat}).}
\label{sketch}
\end{figure*} 

\begin{table}
\caption{Summary of the fitting parameters for the angular correlations}
\begin{tabular}{@{}lcccc@{}}
\hline angle & $a_1$ & $a_2$  & $a_3$ & $a_4$\\ \hline $\theta_{\rm \rho L}$
&    2.351$\pm$0.006     &    -0.178$\pm$0.002    &     1.343$\pm$0.002    &
0.669$\pm$0.000\\\hline   $\theta_{\rm  \rho   vrL}$  &   3.370$\pm$0.099  &
-0.884$\pm$0.037 &  1.285$\pm$0.016 & 0.728$\pm$0.001\\  \hline $\theta_{\rm
cs}$   &    0.399$\pm$0.003   &   0.059$\pm$0.008    &   0.881$\pm$0.005   &
0.938$\pm$0.000\\   \hline    $\theta_{\rm   vs}$   &    0.295$\pm$0.004   &
1.544$\pm$0.001    &     0.804$\pm$0.005    &    0.914$\pm$0.001\\    \hline
$\theta_{p}$&0.099 $\pm$  0.003 &  1.548 $\pm$ 0.003  & 0.825 $\pm$  0.013 &
0.973 $\pm$ 0.000\\ \hline
\end{tabular}

Here $\theta_{\rm  \rho L}$  is the  angle between the  halo's spin  and the
angular momentum measured  on the virial sphere; $\theta_{\rm  \rho vrL}$ is
the angle between the halo's spin and the accreted angular momentum measured
on the virial sphere; $\theta_{\rm cs}$ is the angle between the core's spin
and the satellite orbital momentum;   $\theta_{\rm vs}$ is the angle between
the satellite velocity  in the core 's rest frame  and the satellite's spin;
$\theta_p$  is the  projected angle  between  the satellite's  spin and  its
direction relative  to the core s'  position.  The fitting model  we used is
$1+\xi(\theta)=a_1/(\sqrt{2\pi}a_3)\exp\left[-(\theta-a_2)^2/(2a_3^2)\right]+a_4$.
\end{table}

\begin{table}
\caption{Summary of other quantities related to anisotropic accretion}
\begin{tabular}{@{}lc@{}}
\hline $\langle  \delta_m\rangle(z)$ & 0.0161($\pm$ 0.0103)  z +0.147 ($\pm$
0.005)\\  \hline $S_3(\delta_m)$  &  0.44\\ \hline  $\epsilon(R_m)$ &  0.1\\
\hline ${\tilde  a_{2 0}}$  & -0.65 $\pm$  0.04\\ \hline${\tilde  a_{40}}$ &
0.12  $\pm$  0.02\\  \hline  ${\tilde  a_{6  0}}$  &  -0.054  $\pm$  0.015\\
\hline${\tilde  a_{80}}$ &  0.0145 $\pm$  0.0014\\ \hline  
\end{tabular}
\medskip

$\langle \delta_m\rangle(z)$ is the redshift evolution of the average excess
of   accretion  in   the  plane   orthogonal   to  the   direction  of   the
spin.  $S_3(\delta_m)$ is  the skewness  of  the distribution  of excess  of
accretion.  $\epsilon(R_m)$ is  the axis  ratio  $a/b-1$ with  $a>b$ of  the
projected  satellite distribution.  ${\tilde a_{2  0}}$,  ${\tilde a_{40}}$,
${\tilde  a_{6 0}}$  and  ${\tilde  a_{8 0}}$  are  the normalized  harmonic
coefficients of the `equatorial' modes.
\end{table}

\subsection{Conclusion}
                                  
Using a set of 500 $\Lambda$CDM simulations, we investigated the properties of
the spatial configuration of the  cosmic infall of dark matter around galactic
$\approx L_{\star}$  haloes.  The aim of the  present work was to  find out if
the existence  of preferential  directions existing on  large scales  (such as
filaments) is reflected in the behaviour of matter accreted by haloes, and the
answer is a clear quantitative yes.  

Two important assumptions  were made in the present  paper.We did not consider
different class of haloes' masses  (except for Fig.  \ref{Imass}), but instead
applied normalisations  to includes all haloes in  our measurements (considering
e.g. the statistical average of constrasts).   We also did not take in account
outflows and focused on accreted quantities.

First  we looked  at the  angular distribution  of matter  at  the interface
between the  intergalactic medium and the  inner regions of  the haloes.  We
measured the accreted  mass and the accreted angular  momentum at the virial
radius, describing these quantities as spherical fields.
\begin{itemize}
\item The  total (resp.  advected)  angular momentum measured at  the virial
radius is strongly aligned with the inner spin of the halo with a proportion
of aligned  configuration $30\%$ (resp.  $50\%$) more frequent than  the one
expected  in   an  isotropic  distribution  of   accreted  angular  momentum
($1+\xi_\mathrm{LS}(0)  \sim  1.5$).   This  result reflects  the  importance  of
accreted angular momentum in the building of the haloes' inner spin.
\item
The accretion of mass measured at  the virial radius in the ring-like region
perpendicular to the direction of the halo's spin  is $\sim 15 \%$ larger than the one
expected in the case of an isotropic infall of matter.\\
We also detected the excess of  accretion at the same level in the equatorial  plane using a
spherical  harmonic  expansion  of  the  mass density  flux.
\item  In  the  spin's frame,  the  average  of  the harmonic  $a_{\ell  0}$
coefficients  does not  converge toward  zero,  indicating that  there is  a
systematic accretion
structured in rings parallel to the equatorial plane.

 Using  the substructure  detection  code ADAPTAHOP,  we  confirmed that  the
existence of a preferential plane for the infalling mass is reflected in the
distribution of satellites around  haloes. \item Investigating the degree of
alignment between the orbital momentum of satellites and the central spin of
the halo, it is shown that the aligned configuration is present in excess of
$\sim  12\%$. Satellites  tend to  revolve in  the plane  orthogonal  to the
direction of the halo's spin.   The two methods (using spherical fields and
satellites' detection) yield consistent results and
suggest  that  the  image  of   a  spherical  infall  on  haloes  should  be
reconsidered at the  quoted level.  
\\ We studied the distribution of the angle between the direction
of accretion of satellites and their own spin.
\item An  orthogonal configuration is $5  \%$ more frequent  than what would
one  expect  for  an  isotropic  distribution  of  spin  and  directions  of
accretion.  Satellites  tend to be  accreted in the direction  orthogonal to
their own spin.

These findings  are interpreted as the  results of the  filamentary flows of
structures,  where  satellites  and  haloes  are  accreted  along  the  main
direction  of filaments  with their  spins orthogonal  to  this preferential
direction. The flow along filaments also explains why the matter is accreted
preferentially in the equatorial plane at the virial radius. The halo points
its spin  perpendicular to the  flow and sees  a larger flux in  the regions
normal to the flow direction, i.e.  near the equator.  Thus, it appears that
the existence of preferential directions on large scales is still
relevant on galactic  scales and  should have  consequences for  the inner
dynamics of the halo.

We  addressed the issue  of  observing  these  alignments in
projection.
\item The distribution  of satellites projected onto the  sky is flattened,
with an axis ratio of $1.1$ at the virial radius.
\item  It seems  that the orientation of satellites  around their haloes
is not random, even  if the  two dimensional representation  dilutes the  effects of
alignments.  The `radial' orientation, where the  satellites main axis is
aligned with  the line joining  the satellite to  the halo centre,  is $\sim
5\%$ more frequent  than the one expected in  a completely random distribution
of  orientation.  The  `circular' configuration,  where the  satellites main
axis is perpendicular to the line  joining the satellite to the halo centre,
is also present  in excess compared to an random  distribution near the pole
of  the  host  galaxy. 
\end{itemize}
 All  corresponding  fits  are  summarized  in  Table~1  and~2,  while  Fig.
\ref{sketch} gives a schematic view of the measurements we carried out.

We investigated how the self-consistent dynamical response of the halo would
propagate anisotropic infall down to  galactic scales. In particular we gave
the corresponding polarisation operator in  the context of an opened system.
We  have  shown in  appendix~\ref{s:response} that  accounting  for  dark matter  infall
required the knowledge  of the first three moments of  the flux densities, $
\varpi_{\rho}(\bO,\tau),\varpi_{\rho \bfv }(\bO,\tau), \varpi_{\rho \sigma_i
\sigma_j }(\bO,\tau) $.

It is suggested  that the existence of a preferential  plane of accretion of
matter,  and thus  of angular  momentum, should  have an  influence  on warp
generation and disk thickening.   If the anisotropic properties of infalling
matter measured  in the  outer parts  of haloes are  conserved in  the inner
region  of galaxies, there  may exist  a 'typical'  warp amplitude  and this
anisotropic accretion of  matter may explain the properties  of warp line of
nodes.  In the same spirit, the efficiency of the thickening of the disk may
be enhanced  or reduced  by equatorial accretion.   Finally, our  finding of
intrinsic alignments  on small  scales as well  as specific  orientations of
structures should be  relevant for cosmic shear studies  on wide and shallow
surveys.

\subsection{Prospects}
\label{s:prospect}
                             
The  main   purpose  of  our  investigation  was   to  provide  quantitative
measurements of  the level  of anisotropy involved  in the infall  on scales
$\leq  500$  kpc.   The  next   step  should  clearly  involve  working  out
quantitatively  their implications  for warp,  disk heating  etc...   as was
discussed in section~\ref{s:applications}.

Our measurements were carried out at $R_{200}$, which on galactic scales is a
long way from the inner region  of the galaxy.  One should clearly propagate
the infall (and  its anisotropy) towards the centre of  the galaxy, and more
radial infalling  components will play a  more important role  and should be
weighted accordingly.  It should also be  stressed that we did not take into
account the extra polarisation induced  by the presence of an embedded disk,
which will undoubtedly reinforce the  polarisation and the anisotropy of the
infall.  We also concentrated on  mass accretion, as the lowest order moment
of the underlying ``fluid''  dynamics.  Clearly higher moments involving the
anisotropically accreted momentum, the  kinetic energy etc.  are dynamically
relevant  for  the  evolution of  the  central  object  as is  discussed  in
section~6 and in the appendix. The time evolution of the statistics of these
flux  densities is  also essential  for the  inner dynamic  of the  halo and
should  be addressed  systematically  as  well.  It  will  be worthwhile  to
explore  different  cosmologies  and   their  implications  on  small  scale
dynamics, and  on the  characteristics of infalling  clumps, though  we hope
that the qualitative results sketched here should persist.

It should be emphasized that some  aspects of the present work are exploratory
only,  in that  the  resolution achieved  ($M_{\mathrm{halo}} >  5\cdot10^{12}
M_\odot$) is  somewhat high  for $L_{\star}$ galaxies.   In fact, it  would be
interesting  to  see  if the  properties  of  infall  changes for  lower  mass
($M_{\mathrm{halo}}  <5\cdot10^{10}  M_\odot$)  together  with  the  intrinsic
properties of galaxies. In addition a systematic study of biases
induced  by  the  estimators  of  angular correlations  should  be  conducted,
e.g. the mass weighted errors we introduced in section~\ref{s:two-points}. 

Observationaly, the synthetic  halo described in section \ref{s:projsatpop}
 could be  compared to stacked  satellite distributions relying  on galactic
 surveys such as  the SDSS.  Once the anisotropy has  been propagated to the
 inner  regions  of the  galactic  halo  following  the method  sketched  in
 section~\ref{s:applications},  we should  be  in a  position  to compile  a
 synthetic edge-on  galactic disk and compare  the flaring of  the disk with
 the  corresponding  predictions.   The  residual  preferred  orientation  of
 galactic    disks    around    more    massive   objects    discussed    in
 section~\ref{s:projsat}  should be observed  on the  scales $\leq  500 {\rm
 kpc}$.

Using larger simulations  will allow us to combine  high resolution with the
statistics required to detect the  anisotropic accretion of mass and angular
momentum. A  wide range of halo  masses will become accessible  and the halo
mass dependency of our findings will be constrained without suffering from the
lack of statistics. Better angle determinations will naturally follow from a
better resolution and will improve the accuracy of our quantitative results.
Resimulations  (zoom simulations)  should give  access to  a larger  range of
satellite  masses,  while we  were  here  mostly  sensitive to  the  biggest
substructures.  Large  infalling objects are likely to  feel differently the
effects  of tidal  forces  or dynamical  friction  than smaller  satellites.
Resimulated haloes  allow us  to investigate the  dependency on  the spatial
distribution  of  satellites with  their  masses  corresponding  to a  given
cosmological environment. However using only  a few resimulations may not be
sufficient to overcome cosmic variance and, given the difficulty to produce a
large number of high resolution haloes, such a project remains challenging.

The inclusion  of gas  physics in  these simulations and  their impact  on the
results is the natural following step. For example, gas filaments are known to
be narrower than dark matter filaments,
thus we  expect to  see a higher  level of  anisotropy in the  distribution of
accreted gas by the haloes.  Furthermore, the transmission of angular momentum
from one parcel  of gas to another  (or to the underlying dark  matter) may be
highly effective and would lead to higher homogeneity of the properties of the
accreted   angular  momentum   direction,   enhancing  the   effect  of   spin
alignments. The loss of angular momentum from the gaz to the halo will lead to
a modification  of our pure dark  matter findings.  Yet, the  inclusion of gas
physics  in  simulations  would  force  us  to  address  issues  such  as  the
over-cooling,  the requirement  to take  star formation  and  related feedback
processes into account.  It remains that  in the longer term, the inclusion of
gas physics cannot be avoided and  will give new insights into the anisotropic
accretion of matter by haloes.

\parn{\bf Acknowledgements}

{\sl We  are grateful to J.~Devriendt,  J.~Heyvaerts, A.~Kalnajs, D.~Pogosyan,
E.~Scannapieco, F.~Stoehr, R.~Teyssier, E.~Thi\'ebaut, for useful comments and
helpful  suggestions.  DA  would  like  to thank  C.~Boily  for reading  early
versions of  this paper. CP would  like to thank  F.~Bernardeau for stimulating
discussions during the premises of this work.  We would like to thank D.~Munro
for freely distributing his  Yorick programming language (available at {\em\tt
ftp://ftp-icf.llnl.gov:/pub/Yorick}),   together   with   its   {\em\tt   MPI}
interface, which we used to implement our algorithm in parallel.}

\bibliographystyle{mn2e}
\bibliography{astroph_moment}


\appendix

\onecolumn
\section{Linear response of a spherical halo  to infalling dark matter 
fluxes }
\label{s:response}
\label{s:linearresponse}     
In the  following section, we extend  to open spherical  stellar systems the
 formalism developed by \citet{Tremaine} and e.g. \citet{Murali} by adding a
 source  term  to  the  collisionless Boltzmann  equation.\footnote{This  is
 formally equivalent to summing the response
 of the halo  to a point-like particle for all  entering particles.}  For an
 open  system, the  dark  matter  dynamics within  the  $R_{200}$ sphere  is
 governed by  the collisionless Boltzmann equation coupled  with the Poisson
 equation:
\begin{equation}
 \frac{{dF}}{{dt}}\equiv \frac{\partial F}{\partial t} + \{
  F ,H \}=s^e(\bfr,\bfv,t),
\quad
{\rm and}
\quad
 \nabla^2 \Psi = 4 \pi G \int \d^3 v F ( {\bf v} ).
\label{e:boltz}
\label{e:poisson}
\end{equation}
where $\{ \,\,\}$  is the standard Poisson bracket,  $F(\bfr,\bfv,t)$ is the
system's  distribution   function  submitted  to   $\Psi(\bfr,t)$,  the  total
gravitational  potential  (self-gravity   +  external  perturbation).   The
r.h.s. of (\ref{e:boltz})  is non-zero because of infalling  fluxes from the
environment which  require adding a source term,  $s^e(\bfr,\bfv,t)$, to the
Vlazov  equation.   We  may  now  discriminate  between  a  stationary  part
corresponding  to   the  unperturbed  state  from  a   weak  time-dependent
perturbation induced  by the environnement.  Thus  the DF can  be written as
$F=F_0+f$. Provided the mass of the incoming flux of dark matter is small
compared to the mass  of the halo, we may assume that  $f$ is small compared
to $F_{0}$. Similarly, the Hamiltonian $H$  of the system can be expanded as
$H_{0}+\Delta H$,  with $\Delta H=\psi^e  + \psi$ where $\psi^e$  and $\psi$
stand respectively for the external perturbative potential and for the small
response in potential of the open system.

\subsection{The Boltzmann equation in action-angle }
 
 Given the periodicity of the  system, the most adequate representation of a
spherical  halo corresponds  to action-angle  variables (\citet{Goldstein}).
The linearized Boltzmann equation in such a representation is:
\begin{equation}
\frac{\partial f_\bfk({\bf I},t)}{\partial t}+\imath {\bf k} 
\cdot {\bo}  f_\bfk({\bf I},t)  =
\imath {\bf k} \cdot \frac{\d{}F_0}{\d \bI} \Delta H_\bfk({\bf I},t) +  s^{e}_\bfk({\bf I},t).
\label{e:boltzsphere}
\end{equation}
   The new variables  are the actions $\bfI$ and  the angles $\bfw$ together
with the angular rates  $\bo\equiv\d \bfw/\d t$.  In equation \ref{e:boltzsphere}
we  have Fourier  expanded the  linearized equation  \ref{e:boltz}  over the
periodic angles:
\begin{equation}
X(\br,\bv,t)=\sum_{\bk} X_\bfk({\bf I},t)\exp{(\imath \bk\cdot\bw)},
\quad
{\rm with}
\quad
 X_{\bfk}({\bf I},t) =\frac{1}{(2\pi)^3}\int \d^{3} {\bw} \exp(-\imath \bk 
\cdot \bw ) X(\br,\bv,t),
\end{equation}
where $X$  is any function  of $(\br,\bv,t)$ with  $\bk$ being the Fourier  triple index
corresponding to the three degrees of freedom on the sphere.  The equilibrium
state $F_0$  does not depend on  time nor angles  since it is assumed  to be
stationary.
Then the solution to (\ref{e:boltzsphere}) can be written as:
\begin{equation}
f_\bfk({\bf I},t) =\int_{-\infty}^{t}
\d \tau\exp(\imath {\bf k} 
\cdot {\bf \omega} (\tau-t))
\left[
\imath {\bf k} \cdot \frac{\d F_0}{\d \bI } 
\left[\psi_\bfk(\bI,\tau)+\psi^{e}_\bfk(\bI,\tau)\right] +
s^{e}_\bfk({\bf I},\tau)\, \right],
\label{e:DFsphere}
\end{equation}
where        we        have        written        $\Delta        H_\bfk({\bf
 I},\tau)=\psi_\bfk(\bI,\tau)+\psi^{e}_\bfk(\bI,\tau)$.   We  can  integrate
 (\ref{e:DFsphere}) over velocities  and sum over $\bk$ to recover the density
 perturbation:
\begin{equation}
\rho(\br,t) =\sum_{\bk}\int_{-\infty}^{t} \d \tau
\int  \d^{3}\bv\left( \exp(\imath {\bf k} 
\cdot {\bo} (\tau-t)+\imath \bk \cdot \bw)
\left[
\imath {\bf k} \cdot \frac{\d F_0}{\d \bI } 
\left[\psi_\bfk(\bI,\tau)+\psi^{e}_\bfk(\bI,\tau)\right] +
s^{e}_\bfk({\bf I},\tau)\, \right]\right) .
\label{e:denssphere}
\end{equation} 
Let us expand the potential and the density
 over a bi-orthogonal complete basis 
function $\{ \psi^{[\M{n}]}, \rho^{[\M{n}]}\}$such that 
\begin{equation}
    \psi(\br,t) =\sum_{\M{n}} a_{\M{n}}(t) \psi^{[\M{n}]}(\br) \, ; \quad  
    \rho(\br,t) =\sum_{\M{n}} a_{\M{n}} (t)
    \rho^{[\M{n}]}(\br) \, ; \quad \nabla^{2}\psi^{[\M{n}]}= 4 \pi G \rho^{[\M{n}]}
    \, ; \quad \int \d^{3} \br\psi^{[\M{n}]*}(\br)  \rho^{[\M{p}]}(\br)  = 
\delta^{n}_{p}.
    \label{e:defexp}
\end{equation}
The external potential can be expanded along the same basis as:
\begin{equation}
\quad  \psi^{e}(\br,t) =\sum_{\M{n}} b_{\M{n}}(t) 
    \psi^{[\M{n}]}(\br).  \label{e:defexpe}
\end{equation}  
Note  that   in  \Eq{defexp}  the   expansion  runs  over  a   triple  index
$\M{n}\equiv(n,  \ell,  m)$  corresponding  to  the  radial,  azimuthal  and
alt-azimuthal   degrees  of   freedom,  while   in  \Eq{defexp}   the  three
coefficients are not independent since  the radial variation of the external
potential is  fixed by its boundary  value on the  sphere $R_{200}$.  Making
use   of   the    biorthogonality,   multiplying   (\ref{e:denssphere})   by
$\psi^{[\M{p}]*}(\br)$  for some  given $\M{p}$  and integrating  over $\br$
yields:
\begin{equation}
  a_{\M{p}}(t) =  \sum_{\bk} 
    \int_{-\infty}^{t} \d \tau
\int \int   \d^{3}\bv \d^{3}\br \exp(\imath {\bf k} 
\cdot {\bo} (\tau-t) +\imath \bk \cdot \bw)
\psi^{[\M{p}]*}(\br) \left[\sum_{\M{n}}
\imath {\bf k} \cdot \frac{\d F_0}{\d \bI } 
\left[ a_{\M{n}}(\tau)+ b_{\M{n}}(\tau) \right]  \psi^{[\M{n}]}_\bk(\bI) +
{s}^{e}_\bfk({\bf I},\tau)\, \right]
    \label{e:ap1}
\end{equation}
\subsection{Self-consistency of the response}

We may now swap from position-velocity to  angle-action variables since
$ \d^{3}\bv \d^{3}\br= \d^{3} \bw \d^{3}\bI$.
In (\ref{e:ap1}) only $\psi^{[\M{p}]}(\br)$ depends on $\bw$ so we may carry the 
$\bw$ integration over $\psi^{[\M{p}]*}$, yielding $\psi^{[\M{p}]*}_\bk(\bI)$ which
leads to
\begin{equation}
     a_{\M{p}}(t) =  \sum_{\bk} 
    \int_{-\infty}^{t} \d \tau
\int  \d^{3}\bI  \exp(\imath {\bf k} 
\cdot {\bo} (\tau-t))
\left[\sum_{\M{n}}
\imath {\bf k} \cdot \frac{\d F_0}{\d \bI } 
\left[ a_{\M{n}}(\tau)+ b_{\M{n}}(\tau) \right]\psi^{[\M{p}]*}_\bk(\bI)  
\psi^{[\M{n}]}_\bk(\bI) +
{s}^{e}_\bk({\bf I},\tau)\, \psi^{[\M{p}]*}_\bk(\bI)  \right] 
    \label{e:ap2}
\end{equation}
Note that the  last term of \Eq{ap2} corresponds  to the modulated potential
along  the  unperturbed trajectories  weighted  by  the  number of  entering
particles with $(\M{v},\MG{\Omega})$ at  time $\tau$. This is expected since
it just reflects the fact that  could have linearly summed over all incoming
individual  particles   (since  the  interaction  between   particles  in  a
collisionless fluid  is purely gravitationnal).   In this sense,  this terms
corresponds to a  ray tracing problem in a variable  index medium. Note also
that \Eq{ap2}  does not  account for dynamical  friction since  we integrate
over the unperturbed trajectories.  At this point, we expand the source term
over  a complete  basis; this  basis should  also describe  (known) velocity
space  variations. We assume  that such  a basis  $ \phi^{[\M{n}]}(\br,\bv)$
exists.  We write:
\begin{equation}
   s^e(\br,\bv,t) =\sum_{\M{n}} c_{\M{n}}(t) \phi^{[\M{n}]}(\br,\bv) \, \,\,
{\rm  so }  \,\, {s}^{e}_\bk({\bf  I},\tau)\,=  \sum_{\M{n}} c_{\M{n}}(\tau)
{\sigma}^{[\M{n}]e}_\bk({\bf       I})       \,\,      {\rm       where}\,\,
{\sigma}^{[\M{n}]e}_\bk({\bf  I})\equiv \frac{1}{(2\pi)^3}\int  \d^{3} {\bw}
\exp(-\imath \bk \cdot \bw ) \phi^{[\M{n}]}(\br,\bv).\EQN{defsek}
\end{equation}
Calling  ${\bf a}(\tau) 
=[a_{\M{1}}(\tau),\cdots, 
a_{\M{n}}(\tau)]$,  ${\bf b}(\tau) =[b_{\M{1}}(\tau),\cdots, b_{\M{n}}(\tau)]$, 
  ${\bf  c}(\tau)   =[c_{\M{1}}(\tau),\cdots,  {c}_{\M{n}}(\tau)]$,
and $\Theta(\tau)$ the Heaviside function, we  define  two
tensors:
\begin{equation}
  {K}_{\M{p}\M{n}}(\tau) =(1-\Theta(\tau))\cdot
   \sum_{\bk} 
\int   \d^{3}\bI\exp(\imath {\bf k} 
\cdot {\bo} \tau)
\imath {\bf k} \cdot \frac{\d F_0}{\d \bI } \psi^{[\M{p}]*}_\bk(\bI)  
\psi^{[\M{n}]}_\bk(\bI),
    \label{e:}
\end{equation}
which depends only on the halo equilibrium state via $F_0$ and
\begin{equation}
  {H}_{\M{p}\M{n}}(\tau)  = (1-\Theta(\tau))\cdot \sum_{\bk}  \int \d^{3}\bI
   \exp(\imath {\bf k}  \cdot {\bo} \tau) {\sigma}^{[\M{n}]e}_\bk({\bf I})\,
   \psi^{[\M{p}]*}_\bk(\bI), \label{e:}
\end{equation}
which depends only on the expansion basis,  equation (\ref{e:ap2}) becomes:
\begin{equation}
 {\bf a}(t) = \int_{-\infty}^{\infty} \d \tau   \left\{
{\bf K}(\tau-t) \cdot 
    \left[{\bf 
    a}(\tau)+ {\bf b}(\tau) \right]+ {\bf H}(\tau-t)\cdot {\bf 
c}(\tau) \right\}.
    \label{e:ap3}
\end{equation}
We now perform a Fourier transform with respect to time, hence convolutions become
multiplications and we get:
\begin{equation}
  {\bf \hat a}(p) = (\bI -{\bf \hat K}(p) )^{-1} \cdot 
  \left[{\bf \hat K}(p) \cdot  {\bf \hat  b}(p)+ {\bf \hat  H}(p) \cdot {\bf 
\hat c}(p) \right],
    \label{e:}
\end{equation}
where $p$ stands for the frequency conjugate to time. The computation of the
variance-covariance matrix is straightforward:
\begin{equation}
\langle {\tmmathbf{  \hat  a}}\cdot  {\tmmathbf{ \hat  a}}^{*\T{}}
\rangle= \langle 
(\bI -{\bf \hat K} )^{-1} \cdot 
  \left[{\bf \hat K} \cdot  {\bf \hat  b}+ {\bf \hat  H} \cdot {\bf 
\hat c}\right]\cdot\left[{\bf \hat 
    K} \cdot  {\bf \hat  b}+ {\bf \hat  H} \cdot {\bf 
\hat c}\right]^{\T{} *} \cdot  (\bI 
-{\bf \hat 
    K} )^{-1 * \T{}}  
\rangle\,,
    \label{e:correl}
\end{equation}
where $\bI$ the identity matrix.  Note that $\langle {\tmmathbf{ \hat a}}\cdot
{\tmmathbf{ \hat a}}^{*\T{}} \rangle  $ involves autocorrelation like $\langle
{\tmmathbf{ \hat b}}\cdot {\tmmathbf{ \hat b}}^{* \T{}} \rangle $ and $\langle
{\tmmathbf{ \hat  c}} \cdot {\tmmathbf{ \hat  c}}^{* \T{}} \rangle  $ but also
cross correlation terms such as $\langle {\tmmathbf{ \hat b}}\cdot {\tmmathbf{
\hat  c}}^{* \T{}}  \rangle  $. In  other  words, recalling  that $\M{b}$  and
$\M{c}$  stand respectively  for  the expansion  coeficients  of the  external
potential,   \Eq{defexpe},  and   the   parametrized  velocity   distribution,
\Eq{defsek},  their   cross-correlation  will  \textit{in   fine}  modify  the
correlation  of the  response of  the  inner halo.  Two-points statistics  are
sufficient to caracterize stationnary  perturbations and therefore the induced
response.  Nevertheless,  higher statistics  of  the  response  can be  easily
expressed  in  terms  of  higher  order correlations  of  the  pertubation  if
needed. For example, it can be shown that the three-point correlation function
of the  response's coefficients  can be  written as function  of the  two- and
three-points correlation  of the  perturbations' coefficients.  There  are yet
quite a few  caveats involved; for instance, it is  not completely clear today
that  we  have a  good  understanding  of  what the  unperturbed  distribution
function of a halo+disk should be.

\subsection{The source term}
A  possible choice\footnote{ An alternative choice is made
in \citet{Aubert1} to account for the bimodality of the velocity distribution.
} for the source term consistent  with the first two velocity moments of the
entering matter,  involves constructing $s^e(\bfr,\bfv,t)$  in the following
manner:
\begin{eqnarray}
  s^e(  \bfr, {\bf v} ,t)  &=&
  \sum_{\bf m} Y_{\M{m}}(\bO)
\frac{\delta_{\rm D}(r-R_{200})\,{\hat \varpi}_{\rho, \M{ m}}(t) (2 \pi)^{-3/2}}
{{\rm det}| {{\hat \varpi}_{\rho \sigma_{i}\sigma_{j},\M{m}}(t)}/{{\hat \varpi}_{\rho, \M{m}}} (t) 
 | \,\,} \exp
  \left[ - \frac{1}{2} \left( {\bf v} - \frac{
  {\hat \varpi}_{\rho {\bf v}, \M{m}}(t )}{{\hat \varpi}_{\rho,  
  \M{m}}(t )} \right)^{\T{}}\left(\frac{ {\hat \MG{\varpi}}_{\rho 
  \sigma_{i}\sigma_{j},\M{m}} (t)}{{\hat \varpi}_{\rho,  \M{m}} 
  (t)}\right)^{-1} \left( {\bf v} - \frac{
  {\hat \varpi}_{\rho {\bf v}, \M{m}}(t )}{{\hat \varpi}_{\rho,  
  \M{m}}(t )} \right) \right] \nonumber \\
  &\equiv&   \sum_{\bf m} Y_{\M{m}}(\bO)\delta_{\rm D}(r-R_{200}) {\cal C}_{\M{m}}(\bv,t) \,,
  \label{e:sourceexpr}
 \end{eqnarray}
where  $\M{m}$   stands  for  the   two  harmonic  number,   $(\ell,m)$  and
$Y_{\M{m}}(\bO)\equiv   Y_{\ell}^{m}(\bO)$.    Here   the   Dirac   function
$\delta_{\rm D}(r-R_{200})$ is introduced  since we measure the source terms
at the virial radius.  The global  form is Gaussian and is constructed using
${\hat \varpi}_{\rho,\M{m}}$, $  {\hat \varpi}_{\rho {\bf v},\M{m}}$, ${\hat
\varpi}_{\rho  \sigma_{i}\sigma_{j},\M{m}}$,   the  harmonic  components  of
respectively the mass flux density field, velocity flux density vector field
and the  specific kinetic energy flux  density tensor field  measured on the
$R_{200}$  sphere.    When  taking  the  successive  moment   of  this  flux
distribution over velocity, we get:
\begin{equation}
  \int   \d{}^3    \M{   v}s^e(   {\bfr},    {\bf   v}   )   =      {
    \varpi}_{\rho} (\bfr) \,, \quad \int \d{}^3\M{ v} {\bf v} s^e ( {\bfr}, {\bf
    v} )  = { \varpi}_{\rho \M{v}} (\bfr) \,, \label{e:champs}
\end{equation}
while
\begin{equation}
  \quad  \int  \d{}^3  \M{v}{( v_{i}  -\frac{  {  \varpi}_{\rho  {\bf v},i}  }{  {
  \varpi}_{\rho}  })}\,{(  v_{j}  -  \frac{{  \varpi}_{\rho {\bf  v},j}  }{  {
  \varpi}_{\rho}}    )}   s^e(    \bfr,    {\bv}   )    =   {    \varpi}_{\rho
  \sigma_{i}\sigma_{j}}      (\bfr)    +    \left(
       \sum_{\bf           m}
  Y_{\M{m}}(\bO)\delta(r-R_{200})\frac{ {\hat  \varpi}_{\rho {\bf v}, \M{m}}(t
  )^2}{{\hat \varpi}_{\rho, \M{m}}(t )}
 -\frac{{      \varpi}_{\rho     \M{v}}
  (\bfr)^2}{{\varpi}_{\rho}            (\bfr)} 
\right) \approx{    \varpi}_{\rho
  \sigma_{i}\sigma_{j}}      (\bfr) ,  \label{e:approx2ndmoment}
\end{equation}
so that the Ansatz, \Eq{sourceexpr}, satisfy the first two moments, and 
approximatively the third moment of the fluid equations.
Let us
now  expand ${\cal  C}_{\M{m}}(\bv,t) $  over a  linear complete  basis, say
b-splines covering the radial velocity component and spherical harmonics for
the angle distribution of the velocity vector:
\begin{equation}
    {\cal C}_{\M{m}}(\bv,t)=\sum_{\alpha} C_{\M{m},\alpha}(t) b_{\alpha}(\bv).
\end{equation}
%
The particular choice of \Eq{sourceexpr} has led to the 
parametrisation:
\begin{equation}
 c_{\M{n}}(t)   = { C}_{\M{m},\alpha}(t) \quad {\rm and} \quad  
 \phi^{[\M{n}]}(\bfr,\bfv)=b_{\alpha}(\bv) \, Y_{\M{m}}(\bO) \delta_{\rm D}(r-R_{200})\,,
    \label{e:defexp2}
\end{equation}
while  \Eq{defsek} becomes
\begin{equation}
{\sigma}^{[\M{n}]e}_\bk({\bf   I})=   \frac{1}{(2\pi)^3}\int  \d^{3}   {\bw}
\exp\left(-\imath   \bk  \cdot   \bw   \right)  Y_{\M{m}}[   \bO(\bfI,\bfw)]
b_{\alpha}(\bv[\bfI,\bfw])\delta_{\rm D}(r(\bfI,\bfw)-R_{200}) \, .\EQN{defsedelta}
\end{equation}
Note  that   we  can  make  use   of  the  $\delta_{\rm D}$   function  occurring  in
 \Eq{defsedelta}   since    $w_r\equiv   {\tilde   w}_r(r,\bI)$.   Therefore
 \Eq{defsedelta} reads:
\begin{eqnarray}
{\sigma}^{[\M{n}]e}_\bk({\bf I})&=&\int  \frac{\d^{2} {\bw}}{(2\pi)^3}\int \d w_r
 \exp\left(-\imath   \bk  \cdot   \bw  \right)   Y_{\M{m}}[  \bO(\bfI,\bfw)]
 b_{\alpha}(\bv[\bfI,\bfw])\frac{1}{|\partial {\tilde w}_r/\partial r|^{-1}}
 \delta_{\rm D}(w_r-{\tilde   w}_r[R_{200},\bI])   \,   ,   \nonumber   \\   &=&\int
 \frac{\d^{2}   {\bw}}{(2\pi)^3}\exp\left(-\imath  \bk  \cdot   \bw  \right)
 Y_{\M{m}}[             \bO(\bfI,\bfw,{\tilde            w}_r[R_{200},\bI])]
 b_{\alpha}(\bv[\bfI,\bfw,{\tilde    w}_r(R_{200},\bI)])\frac{\omega_r(\bI)}{
 |{\dot    r}(R_{200},\bI)|}    \exp\left(-\imath    k_r    \cdot    {\tilde
 w}_r[R_{200},\bI]\right)\EQN{Newinteg}
\end{eqnarray}
In \Eq{Newinteg} we  sum over all intersections of the  orbit $\bI$ with the
$R_{200}$  sphere, at the  radial phase  corresponding to  that intersection
(with a weight corresponding to $\omega_r/|{\dot r}|$).

%
Given  \Eq{defexp}, \Eq{sourceexpr},  together with  \Eq{Newinteg}, \Eq{ap3}  can be
recast formally as
\begin{equation}
 \rho(\M{r},t) = 
 \M{R}\{ F_{0},t,\tau,\bO\}
 \left[\psi_{e}(\bO,\tau),
 \varpi_{\rho}(\bO,\tau),\varpi_{\rho \bfv }(\bO,\tau),
 \varpi_{\rho \sigma_i\sigma_j }(\bO,\tau)\right]
    \label{e:ap3n}
\end{equation}
which  corresponds  to  the  form   given  in  the  main  text  in  Equation
\ref{decomp}.  It should be emphasized  once again that the splitting of the
gravitationnal field into two components,  one outside of $R_{200}$, and one
inside, via point particles obeying the distribution $s_e(\M{r},\M{v},t)$ is
completely arbitrary  from the point of  view of the dynamics.  In fact, one
should  account  that $\psi_e(\MG{\Omega},t)$  should  be  switched on  long
before any  particles enter  $R_{200}$ since no  particle is created  at the
boundary. This last constraint is clearly satisfied by our simulations.

\twocolumn

\section{ADAPTAHOP: a substructure finder based on saddle point handling}
\label{s:ADAPT}
Dark matter haloes can contain a hierarchy of subhaloes, which can be viewed
as  a tree  of structures  and sub-structures.  Given a  mass  resolution (a
finite number of particles such as  in our $N$-body simulations), there is a
limit to this hierarchy, which can be formalised as an ensemble of leaves in
a  tree. The  goal here  is to  draw this  tree by  applying  the simplest
principles  of  Morse theory  (e.g.  \citet{Jost}).  Morse theory  basically
involves relating the  topology of an excursion, e.g.,  the regions of space
with density above  a given threshold, $\rho > \rho_{\rm t}$,  to the set of
critical  points it contains,  $\{\bfx, \nabla\rho(\bfx)=0  \}$, and  to the
field lines  connecting these points  together, i.e. the curves  obtained by
following the gradient of the density field.  In that approach, the smallest
substructures, which are  the leaves of the tree, can  be identified as peak
patches,  i.e.  ensembles  of  field  lines converging  to  the  same  local
maximum.  The connectivity  between  substructures is  ruled  by the  saddle
points, which are local maxima in  the surfaces defining the contours of the
peak patches: from the knowledge of these saddle points and the local maxima
they connect, it is possible to extract the full tree of structures (haloes)
and sub-structures (subhaloes) in four steps:

\begin{enumerate}
\item In order  to eliminate, at least partly, the  effects of Poisson noise
and to have an estimate of the local density as close as possible to a Morse
function,\footnote{i.e. a smooth function such that the ensemble of critical
point is discrete and the matrix of second derivatives in their neighborhood
is non  degenerate.}  while  conserving as much  as possible details  of the
distribution,  we  perform  adaptive  smoothing of  this  distribution  with
standard      SPH     technique     (Smooth      Particle     Hydrodynamics,
e.g.  \citet{Monaghan}). This  smoothing  assumes that  each  particle is  a
smooth spherical cloud of given radius  $R$, e.g. a spline $S(r)$.  For each
particle,  the  list  of  its  $N_{\rm SPH}$  closest  neighbors  is  found,
typically $N_{\rm SPH}$  of a few tens (here we  take $N_{\rm SPH}=64$). The
distance from the furthest neighbor fixes  $R$, while the SPH density at the
particle of  interest is  estimated by a  summation over its  neighbors with
weight $S(r)$.  To  find rapidly the closest neighbors  of each particle, we
use a standard Oct-tree  algorithm, which decomposes hierarchically space in
subcells untill they contain zero or one particle.
\item The leaves of the  tree of structures and substructures are identified
while associating  each particle to  the peak patch  it belongs to.  This is
performed by  a simple walk from  particle to particle,  while following the
gradient until convergence: at each step of the walk, the SPH density of the
particle  is compared  to its  $N_{\rm HOP}$  closest neighbors  (which were
stored during the SPH smoothing step), the particle for the next step of the
walk being the  one with the largest SPH density.  We take $N_{\rm HOP}=16$,
as advocated by \citet{Hop}.
\item For each leave of the  tree, the connections with the other leaves are
created by searching  the saddle points on the  intersecting surfaces ${\cal
S}_{ij}$ between peak  patches $i$ and $j$. Each  surface ${\cal S}_{ij}$ is
made of particles  belonging to one of the peak patches  and having at least
one of their closest neighbors among  $N_{\rm HOP}$ in the other peak patch,
and vice versa. If the set ${\cal S}_{ij}$ contains only particles belonging
to $i$  or only particles belonging  to $j$, the connection  between $i$ and
$j$ is considered as non significant (because non symmetric) and eliminated,
${\cal S}_{ij}=\null$. Saddle points are local maxima in ${\cal S}_{ij}$. To
establish the  connectivity as a function  of a density  threshold, only the
highest saddle  point matters, when there  are several. The  search for this
saddle point involves finding the maximum of the SPH density among particles
belonging to $S_{ij}$.  We proceed as follows to estimate accurately the SPH
density in  $S_{ij}$.  For each particle  $A$ in $S_{ij}$,  say belonging to
peak  patch $i$  and with  density  $\rho_A$, we  consider the  list of  its
closest  neigbors among  $N_{\rm HOP}$  belonging  to peak  patch $j$,  with
density $\rho_k$, $k=1,\cdots,N_j \leq N_{\rm HOP}$.  The density associated
to    this   particle    in    ${\cal   S}_{ij}$    is    then   given    by
$\rho=\min(\rho_A,\rho_k)$. By applying this procedure, we locate accurately
$S_{ij}$ and  avoid slight overestimation of  the SPH density  at the saddle
point.

\item It  is possible  to build the  tree of structures  and sub-structures,
when the list  of neighboring leaves to which a given  leave is connected is
given,  as well as  the corresponding  saddle points.  This  is performed
recursively  by increasing progressively  a threshold  parameter, $\rho_{\rm
t}$, from  an initial value,  $\rho_{\rm TH}$, corresponding to  the typical
overdensity used to select galaxy  haloes, here called structures. A typical
choice  for  $\rho_{\rm  TH}$   is  $\rho_{\rm  TH}=81$,  which  corresponds
approximately to  friend-of-friend haloes selected with  a linking paremeter
$b=0.2$ (e.g., \citet{Hop}).  Suppose we are at step $n$  of the process and
let us compute step $n+1$. At this  point, we are sitting on a branch of the
tree --a  structure or a sub-structure-- and  we aim to draw  the details of
this  branch.   This (sub-)~structure  contains  a  number  of peak  patches
connected by saddle  points of densities $\rho_{\rm s}$.  For the considered
value of  $\rho_{\rm t}$, the  connections inside that  (sub-)~structure are
examined   and  destroyed   when  $\rho_{\rm   s}  <   \rho_{\rm   t}$.  The
(sub-)~structure is then broken into as many components as necessary. During
the  process,   the  particles  above  $\rho_{\rm  t}$   belonging  to  each
sub-component are tagged,  which allows us to determine  at any time various
properties of  a given (sub-)~structure,  namely the number of  particles it
contains,  its mass,  its  average  and maximum  SPH  density, for  possible
application  to  various morphological  criterions  of  selection. One  such
criterion  is Poisson  noise. In  order to  asses if  a  given sub-structure
containing $N$  particles should be considered  as statistically significant
compared  to  Poisson  noise,  its  average  density  must  be  sufficiently
significant compared to $\rho_{\rm t}$:
\begin{equation}
\langle \rho \rangle_{\rm sub-structure} > \rho_{\rm t} \left[1+
\frac{f_{\rm Poisson}}{\sqrt{N}} \right],
\label{eq:poissonsel}
\end{equation}
where  $f_{\rm  Poisson}$  is   a  ``$f_{\rm  Poisson}  \sigma$''  detection
parameter, typically  a few unities.  A good choice is  $f_{\rm Poisson}=4$.
If the sub-structure is below this  threshold, it disappears, i.e. it is not
considered in next  step of the recursion. At the end  of the selection, two
situations are possible: (i) two sub-structures or more are detected and new
nodes are created  in the tree (ii) the (sub)-structure  was not broken into
multiple components  and nothing happens at  this step. The  process is then
repeated  on the  new  sub-structures by  increasing  locally the  threshold
$\rho_{\rm t}$:
\begin{equation}
\rho_{\rm t} \rightarrow \rho_{\rm t} \times \left[ 1
+\frac{f_{\rm Poisson}}{\sqrt{N}}\right],
\end{equation}
until there  is only one peak  patch in the (sub-)structure.   Note that the
Poisson  noise selection,  eq.~(\ref{eq:poissonsel}) is  not applied  to the
haloes, when $\rho_{\rm t}=\rho_{\rm TH}$.
\end{enumerate}

At  the  end  of  the  process,   one  obtains  a  tree  of  structures  and
sub-structures,   where  each   node  of   the  tree   corresponding   to  a
(sub-)structure, with its position, its  number of particles, is mean square
radius, its average and maximum  SPH density, and the density $\rho_{\rm s}$
of the  highest saddle point which  connects it to  another substructure. In
addition, a flag  is given to each  particle. This flag is a  pointer to the
node the closest possible  to a leave (if not a leave),  which allows one to
find recursively the list of  particles belonging to any (sub-)structure and
thus perform some  more ellaborate post-treatement, such as  some relying on
dynamical  prescriptions (boundness).   The difficulty  in that  case  is to
estimate  accuratly  the gravitational  potential.  Its  computation can  be
rather costly, since ``pealing'' the (sub-)haloes requires iterating several
forward and backward walks in the tree of structures and sub-structures with
corresponding calculations of the gravitational potential.  Our prescription
is therefore at  the present time purely morphological  and does not involve
the estimate of the  gravitational potential.  The current implementation is
rather fast, most of the CPU time being taken by the SPH smoothing, e.g. 1-3
hours on 16 millions particles on current fast scalar processors.

Our  algorithm   is  called  ADAPTAHOP   since  we  aim  at   improving  HOP
(\citet{Hop}): the first two step above  are exactly the same as in HOP, but
the  last  two  are different.  Indeed,  in  HOP,  the  idea is  to  combine
informations on  the saddle  point densities, $\rho_{\rm  s}$, on  the local
maxima, $\rho_{\rm max}$,  inside a connected set of  peak patches to decide
whether it has to be broken  into multiple {\em disjoint haloes}. The aim of
HOP  is indeed  to improve  standard  friend-of-friend methods  in order  to
obtain more  compact and  spherical haloes. The  goal of ADAPTAHOP  is quite
different since it focuses on sub-structure detection.

In spirit, ADAPTAHOP  is in fact very similar to  the substructure finder of
\citet{Springel}: SUBFIND (see also \citet{SWTK2001}). Of course, there is a
major difference,  since SUBFIND has  in addition a  sophisticated dynamical
prescription   involving    exact   calculation   of    the   gravitationnal
potential. Springel  uses also a  slightly more elegant method  to construct
the   tree   of   structures    and   substructures   prior   to   dynamical
post-treatment. After step  one above, the idea is to  rank the particles by
decreasing  density  and  treat   them  in  this  order.  Investigating  the
distribution  of  particules  is  such  a way  is  equivalent  to  examining
isocontours of decreasing  density.  It uses (as in  ADAPTAHOP) the closests
neighbors  of a  particle  to decide  if  the particle  examined during  the
process (i) creates a new  (sub-)structure since it is isolated (ii) belongs
to an existing substructure or (iii) connects two substructures, which makes
the construction  of the tree  of structures and substructures  much simpler
than  in ADAPTAHOP  and more  accurate, since  there no  need for  using the
threshold parameter  $\rho_{\rm t}$. In  SUBFIND, no treatment is  made to
account for the local Poisson noise: it is not necessary because of the dynamical
post-processing, which destroys unbounded structures.

It  is   important  to  note   that  since  ADAPTAHOP  has   no  dynamical
post-treatment,  it gives  slightly different  results compared to  SUBFIND  in its
present form.  In particular, for  a given sufficiently massive  dark matter
halo , SUBFIND  (\citet{SWTK2001}) describes it in terms  of a large, smooth
central  component,  and  a  bunch  of  much  less  massive  sub-haloes.  In
ADAPTAHOP, the result is quite similar, except that the central component is
much  less spatially extended  (it is  extended up  to the  isocontour level
corresponding to  the saddle point connecting  it to a sub-halo),  and it is
therefore less massive.

\begin{figure}
\centerline{
\resizebox{8cm}{8cm}{\includegraphics{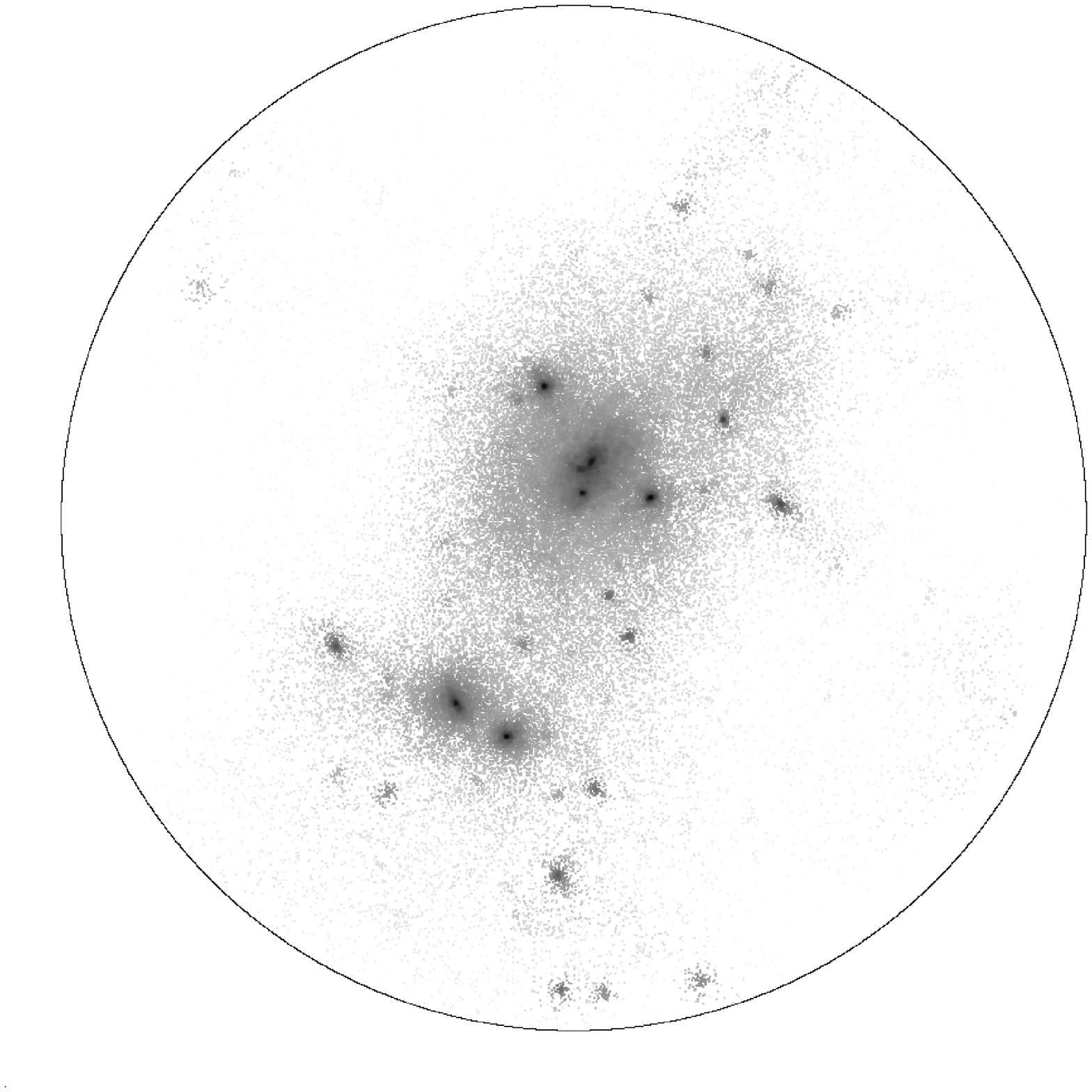}}
}
\centerline{
\resizebox{8cm}{8cm}{\includegraphics{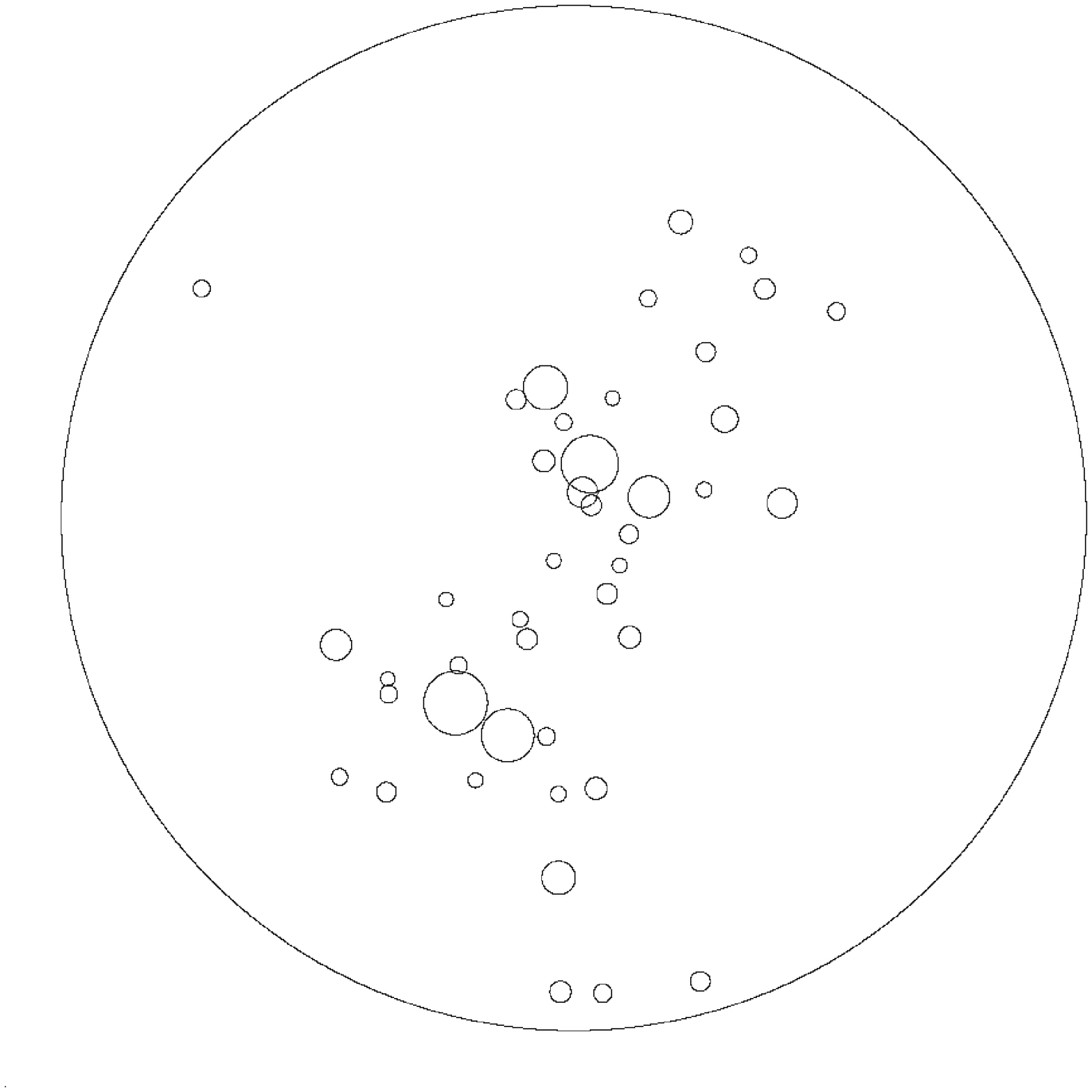}}
}
\caption[]{Illustration  of   the  output  of  ADAPTAHOP  for   one  of  the
simulations of  this work.  A sphere of  radius 5  Mpc centered on  the most
massive halo is  represented.  In the upper pannel,  the dark matter density
is shown  using a  logarithmic scale.  Darker  regions correspond  to higher
density contrasts.  The lower pannel  displays the detected  subhaloes (i.e.
the  most elementary  structures corresponding  to the  peak patches  or the
leaves of the tree). The size of the circle scales with $M^{1/3}$, where $M$
is the mass of the subhalo. Most  of the subhaloes seen on the figure belong
to  the most  massive  halo.  Clearly, ADAPTAHOP  is  rather successfull  at
detecting all the significant substructures.}
\label{f:figureada}
\end{figure}

Fig.~\ref{f:figureada} illustrates how well ADAPTAHOP performs in one of the
simulations we realized for this work, for the most massive halo detected in
this realization.

\section{ Statistics on the sphere}
\label{s:statsphere}
When dealing with spherical fields,  there are different ways to characterize
their angular  structure.  In  the present paper,  we essentially  deal with
centered statistics,  i.e. we describe the angular  structuration of scalar
or vector  fields relative to a  specific direction, defined by  the halo or
satellite's spin  $\bf S$.  Let us  first formally introduce   filtering on the
sphere, statistical  and angular averages, and  present one-point
statistics  (probability distribution  functions) while postponing
  two-point statistics
(correlation functions, or excess probability of joint events) to \citet{Aubert1}.

\subsection{One point statistics}
\label{s:statistics}
For any field, $x$, on the sphere, let us introduce the smoothed field, $( x
)_{\alpha}$ (filtered on scale $\alpha$) , as
\begin{eqnarray}
 (  x )_{\alpha} (  \MG{\Omega} )  &\equiv& \frac{1}{\int  \Theta_{\alpha} (
   \MG{\Omega}' ) \mathd \MG{\Omega}'} \int \Theta_{\alpha} ( \MG{\Omega}' -
   \MG{\Omega} )  x (  \MG{\Omega}' ) \mathd  \MG{\Omega}' \\  &\equiv& \int
   w_{\alpha}  ( \MG{\Omega}  - \MG{\Omega}'  )  x (  \MG{\Omega}' )  \mathd
   \MG{\Omega}',
\end{eqnarray}
where $\Theta_{\alpha}$stands for the top hat function,
\begin{equation}
 \Theta_{\alpha} (  \MG{\Omega} )  = 1 \quad  \textrm{if \quad}  | \vartheta
 |\leqslant \alpha,
\end{equation}
and $w_{\alpha}$  is defined by \ref{defxa}  the standard top hat filter on
the sphere.

  Consider now the centered  top-hat-filtered (on scale $\alpha$) field, $[ x
  ]_{\alpha}$, defined by
\begin{eqnarray} 
[  x ]_{\alpha}  &\equiv& (  x )_{\alpha}  ( \frac{\mathpi}{2}  ),\\ &\equiv&
\frac{1}{\int  \Theta_{\alpha} (  \MG{\Omega}' )  \mathd  \MG{\Omega}'} \int
\Theta_{\alpha} (  {\vartheta}' - \mathpi  / 2 )  x ( \MG{\Omega}'  ) \mathd
\MG{\Omega}', \\ &\equiv& \int W_{\alpha} ( \MG{\Omega}' ) x ( \MG{\Omega}' )
\mathd \MG{\Omega}' . \EQN{defW}
\end{eqnarray} 
Note that \Eq{defW}  defines $W_\alpha$.  Our filtering is  now centred, in
that the average is carried on  a window which is centred at the equatorial
plane  (since  we  are in  this  paper  interested  in the  polarisation  of
accretion processes with respect to  that plane).  Let us also introduce the
average of $x$ on the sphere, as
\begin{equation}
{ \bar  x} \equiv (  x )_{\mathpi /  2} = \frac{1}{\int  \mathd \MG{\Omega}}
   \int x ( \MG{\Omega} ) \mathd \MG{\Omega}, \label{defxa}
\end{equation}
We may also for a given $x$ define its contrast as
\begin{equation}
 \delta_x  \equiv \frac{x}{\bar x} - 1 . 
\end{equation}
Note that, in  contrast to standard cosmology, we expect  that $\bar x \neq
\langle x \rangle$, (i.e. no  ergodicity) since the angular average over one
virial sphere is not representative of the whole cosmological set, and since
$ \langle x  \rangle$ depends on $\vartheta$ whereas $\bar  x$ doesn't. As a
consequence,
\[ \langle \delta_x \rangle = \langle \frac{x}{\bar x} \rangle - 1 \neq
  \frac{ \langle  x \rangle}{ \langle \bar  x \rangle} - 1  \,.  \]
 Consider
  now   the   top-hat-filtered-centred   flux   density   contrast,
  $[\delta_\varpi]_\alpha $, defined by
\begin{equation}
 [\delta_{\varpi} ]_{\alpha} \equiv ( \delta_{\varpi} )_{\alpha} ( \mathpi /
   2 )  = \frac{1}{\bar  \varpi} \int W_{\alpha}  ( \MG{\Omega} )  \varpi \,
   \mathd \MG{\Omega} - 1 \, . 
\label{defdelta}
\end{equation} 
Since  by  construction,
   $[\delta_\varpi]_\alpha $, is a filtered version of $\delta_{\varpi}$, it
   inherits some  of it  statistical properties. In  particular, the  PDF of
   $\delta_{\varpi} ( \mathpi / 2 ) $ and $[\delta_\varpi]_\alpha$ should be
   quite similar provided $\alpha$ is small enough.

In the  main text, we  consider the anisotropic parameter,  $\delta_m \equiv
[\delta_{\rho v_r}]_ { \mathpi / 8}$, which therefore corresponds formally to
the centered top-hat-smoothed  (on scales of $\mathpi / 8$) mass flux density
contrast.  Following the same spirit, we could also consider quantities such
as  $[\delta_{\rho  v_r v^2}]_{  \mathpi  /  8}$,  which would  measure  the
anisotropy in the  accreted kinetic energy: the excess  of accreted kinetic
energy should allow us to track the excess of incoming virialised objects in
the equatorial plane without  performing their explicit identification.  One
should also consider $[\delta_{\rho v_r  L}]_{ \mathpi / 8}$, the anisotropy
in the  accreted momentum,  since this quantity  is directly related  to the
torque applied onto  the system by the infall.  More  generally still we could
investigate $( \delta_{\varpi} )_{\alpha} (  \vartheta ) $, the flux density
contrast  top-hat-smoothed  on   a  ring  of  size  $\alpha$  centred  on
$\vartheta$.

Note  that  we  can  think  of  the  harmonic  coefficients,  $a^{m}_{\ell}$
introduced  in section \ref{s:expansion}  as a  specific type  of filtering,
where  the  window  function,  $W_\alpha$,  is replaced  by  an  axisymmetric
spherical harmonic, $Y_{\ell}^0 (\MG{\Omega})$:
\begin{equation}
[\delta_\varpi]_{\ell}=     \frac{1}{\bar     \varpi}\int     Y^{0     \star
}_\ell(\MG{\Omega})            \varpi(\MG{\Omega})           \,           \d
\MG{\Omega}=\frac{a^{0}_{\ell}}{\bar \varpi}.
\end{equation}
We can also write ${\bar  \varpi}$ in terms of spherical harmonics:
\begin{equation}
{\bar         \varpi}\equiv\frac{1}{4\pi}\int         \varpi         \mathd
\MG{\Omega}=\frac{1}{\sqrt{4\pi}}\int   Y^{0  \star  }_0(\MG{\Omega})\varpi
\mathd \MG{\Omega}=\frac{a^{0}_{0}}{\sqrt{4\pi}}.
\end{equation}
Therefore we obtain :
\begin{equation}
[\delta_\varpi]_{\ell}=\frac{a_\ell^0}{{\rm sign}(a_0^0) \sqrt{C_0}},
\label{defalmtilde}
\end{equation}
where  $C_0=|a^{0}_{0}|^2/4\pi$ is  the  $\ell=0$ component  of the  angular
power spectrum  $C_\ell$.  

Since a step function can be expanded along spherical harmonics as
\begin{equation}
\Theta_\alpha(\vartheta-\frac{\pi}{2})=\sum_\ell                       b_\ell
Y_\ell^0(\vartheta,0)\,;
\end{equation}
therefore, $[\delta_\varpi]_\alpha$ defined by Eq.~\ref{defdelta} obeys
\begin{equation}
[\delta_\varpi]_{\alpha}=\sum_\ell b_\ell \,[\delta_\varpi]_{\ell}-1.
\end{equation}

Taking $x=\rho  v_r$ for example, we have :
\begin{equation}
\delta_{[\rho  v_r]}(  \vartheta, \varphi  )=\sum_{\ell,  m} d_\ell^m  Y_{\ell}^m  (  \vartheta,
\varphi  )=\frac{\rho  v_r(  \vartheta, \varphi  )}{\overline {\rho  v_r}} -1,
\label{eq:contrast}
\end{equation}
where
\begin{equation}
\overline {\rho  v_r}=\frac{1}{4\pi}\int\d{\vartheta}\d{\varphi}\rho  v_r(
\vartheta, \varphi  )\sin \vartheta= \frac{a_0^0}{\sqrt{4\pi}}.
\end{equation}
Since \[
\int\d{\vartheta}\d{\varphi} Y_{\ell}^m(  \vartheta,
\varphi  ) \sin \vartheta=\sqrt{4\pi}\delta_{l0}\delta_{m0}\]
(e.g. \citet{Moment}), we find :
\begin{equation}
d_\ell^m={\tilde  a^m_{\ell}}-\sqrt{4\pi}\delta_{l0}\delta_{m0}.
\end{equation}
We finally obtain :
\begin{equation}
\delta_{[\rho  v_r]}(  \vartheta, \varphi  )=\sum_{\ell,  m}
 {\tilde  a^m_{\ell}}  Y_{\ell}^m  (  \vartheta, \varphi  )  -1.
\end{equation}

\section{Convergence issues}
\label{s:convergence}

\subsection{Substructures \& the haloe's spin}

\begin{figure}          
 \centering
\resizebox{5cm}{5cm}{\includegraphics{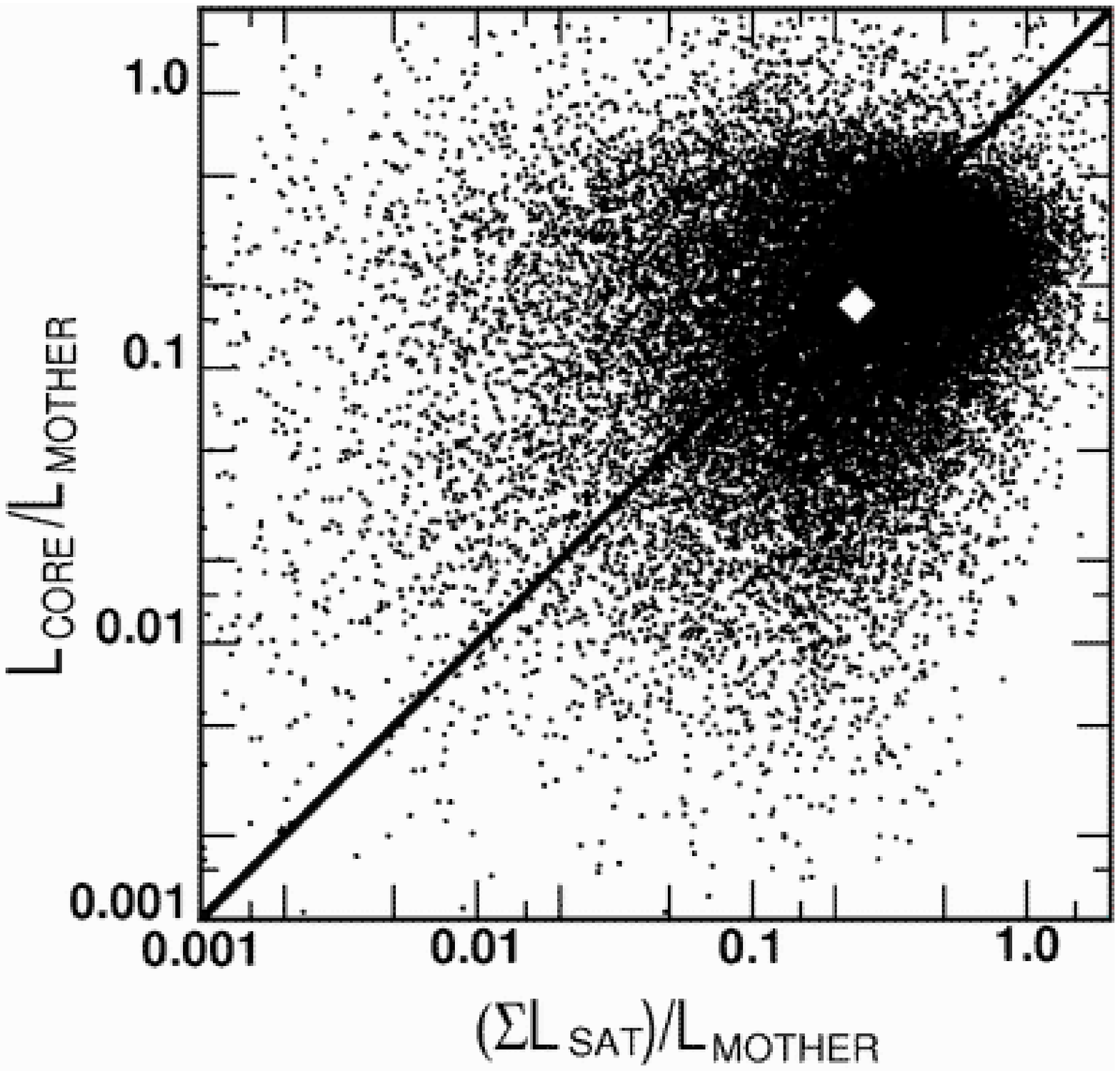}}
\resizebox{5cm}{5cm}{\includegraphics{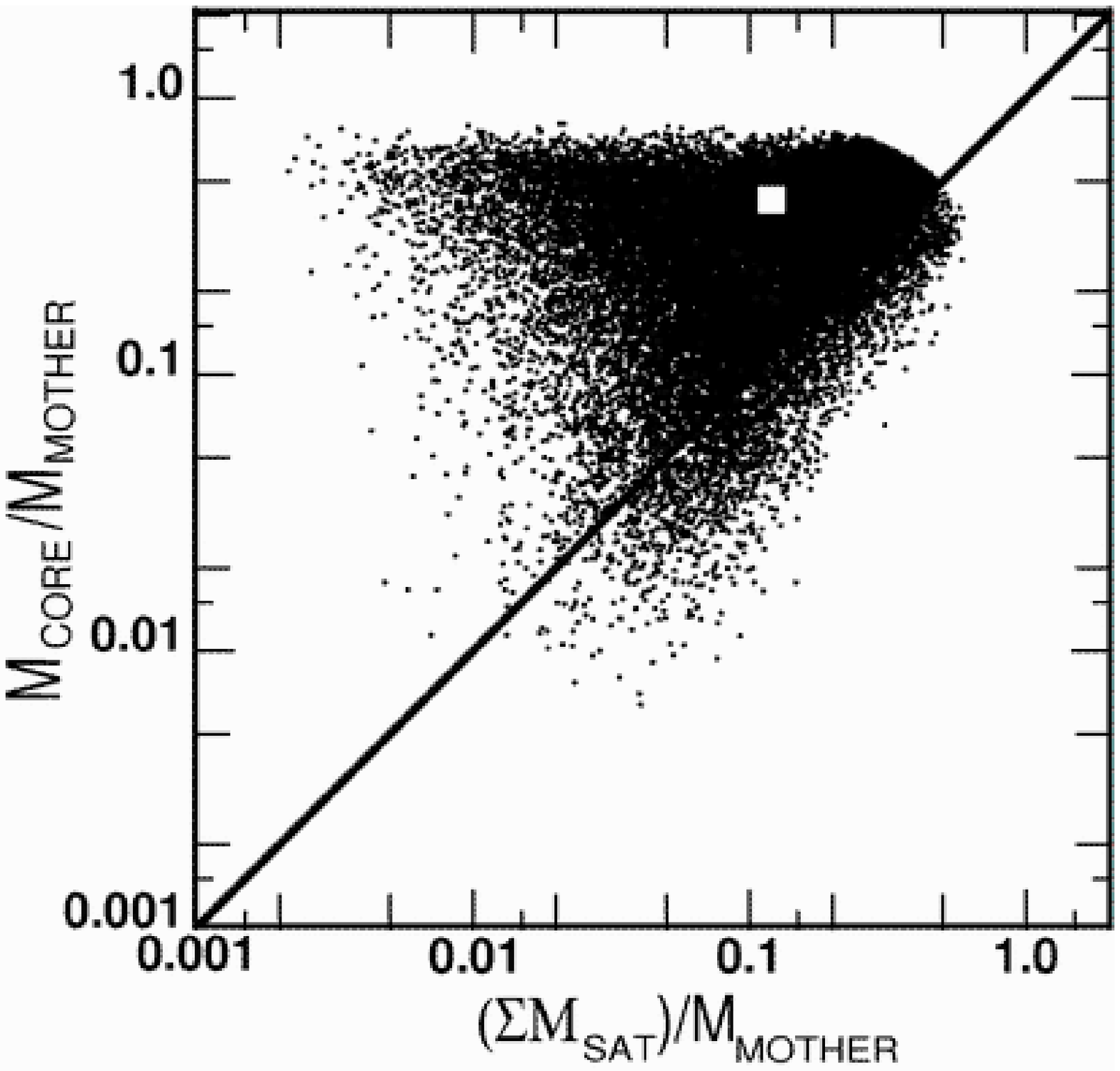}}
\caption{Comparison of the substructure's and the core's contribution to the
amplitude of the mother's  spin  and to  the mother's mass.  \textit{Top}: Comparison of
the core's contribution  to the mother's spin compared  to the contribution of
all   the  satellites  for   each  mother   detected  in   our  simulations.
\textit{Bottom}: same  comparison but for  the core's and  satellites' mass
relative to the mother's total  mass.  In both figures, the symbol indicates
the  barycenter of  the cloud  of  points while  the thick  line's slope  is
unity. While  the total  mass is dominated  by the core's  contribution, the
mother's spin is dominated by satellites showing that their specific orbital
momentum is more important than that of the core.  }
\label{contrib}
\end{figure}

For each tree  of substructure-satellites, we computed the  total spin inside
the  mother structure,  $\M{S}_M$,  and the  momentum  of each  substructure
inside  the  mother  structure,   $\M{L}_s$.  Then  we  compared  the  inner
satellites'  and the  contribution of  the core  to the  mother's  spin. The
comparison  is only  made on  the components  of the  substructures momentum
parallel to  $\vec S_M$.  The  results are shown in  Fig.~\ref{contrib}.  We
plotted the  \textit{total} contribution of satellites to  the mother's spin
versus  the core's contribution.   From the  barycenter of  the distribution
shown in Fig.~\ref{contrib}, it  appears that substructures contain about 80
$\%$ of the  total host's spin with a satellites  contribution of 50$\%$ and
about 30 $\%$  for the core.  The bottom panel  shows the total contribution
of  substructures  to the  mother's  mass  versus  the contribution  of  the
core.  As expected  given the  definition  of the  core, we  found that  the
relative proportion  are almost  reversed compared to  the previous  plot. A
core contains about half of  the total mass while satellites represent about
40$\%$ of the  total mass.  Clearly the specific  angular momentum is larger
in satellites than in the core.   The distance of satellites relative to the
mother's centre and their  velocities induce  a `lever  arm' effect.  Even if
satellite remnants  are light  in terms  of mass they  are important  if not
dominant for the spin of the galactic system.  This effect also suggests that the
mother's  spin is  aligned  with the  orbital  momentum  of infalling  satellites
because they determine the direction of the halo's spin.

\subsection{the mass dependence of $\langle \delta_m \rangle$}
\label{sec:appmasdep}
\begin{figure}          
 \centering
\resizebox{5cm}{5cm}{\includegraphics{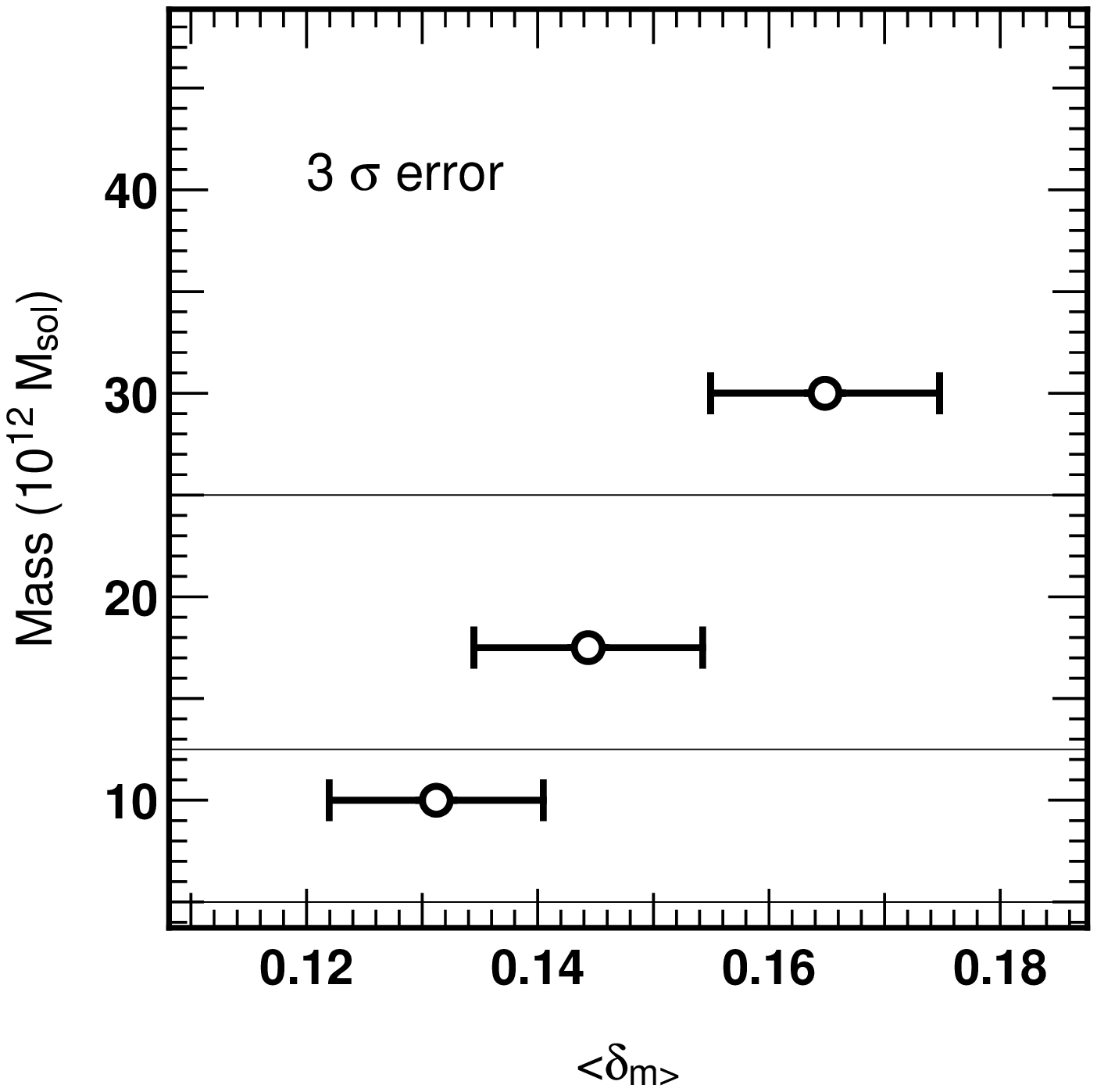}}
\caption{ Comparison of $\langle  \delta_m \rangle$ for different classes of
halo mass  at z=0. The  error bars stand  for the 3$\sigma$ error.  The thin
lines  separate the  three classes  of mass: $5\cdot10^{12}  M_\odot<  m <
1.25\cdot10^{13}  M_\odot$, $1.25\cdot10^{13}  M_\odot< m  < 2.5\cdot10^{13}
M_\odot$ and  $m >  2.5\cdot10^{13} M_\odot$. Each  class contains 16500
haloes.
}
\label{Imass}
\end{figure}
We measured the average excess of accretion $\langle \delta_m \rangle$ ( see
section \ref{s:one-point})  for three different class of  masses at redshift
z=0:    $5\cdot10^{12}    M_\odot<    m   <    1.25\cdot10^{13}    M_\odot$,
$1.25\cdot10^{13}   M_\odot<  m   <  2.5\cdot10^{13}   M_\odot$  and   $m  >
2.5\cdot10^{13}  M_\odot$.   Each  class  contains  approximatively  16  500
haloes.   The results  are shown  in Fig.  \ref{Imass}. It  is  found that
$\langle  \delta_m  \rangle$  increases   with  mass  but  does  not  change
significantly  even if  the three  classes cover  different  mass magnitudes.
Consequently, no class of mass dominates  when all the haloes are being used
in the computation of $\langle \delta_m \rangle$.

\vfill

\end{document}